\documentclass[11pt]{aastex63}
\usepackage{amsmath,amssymb,amstext}
\usepackage{appendix}
\usepackage{enumitem}
\usepackage{graphbox}
\usepackage{pdfpages}
\usepackage{etoolbox}
\usepackage{comment}
\makeatletter
\patchcmd{\@outputpage@head}{\@ifx{\LS@rot\@undefined}{}{\LS@rot}}{}{}{}
\makeatother
\graphicspath{{./}{figures/}}

\definecolor{darkgreen}{RGB}{0,100,0}
\definecolor{darkred}{RGB}{175,0,0}

\begin{document}

\title{Open-source Flux Transport (OFT). \\I. HipFT {--} High-performance Flux Transport}

\correspondingauthor{Ronald M. Caplan}
\email{caplanr@predsci.com}
\author[0000-0002-2633-4290]{Ronald M. Caplan}
\affil{Predictive Science Inc., 9990 Mesa Rim Road, Suite 170, San Diego, CA 92121, USA}
\author[0000-0003-0939-1055]{Miko M. Stulajter}
%\email{miko@predsci.com}
\affil{Predictive Science Inc., 9990 Mesa Rim Road, Suite 170, San Diego, CA 92121, USA}
\author[0000-0003-1662-3328]{Jon A. Linker}
%\email{linkerj@predsci.com}
\affil{Predictive Science Inc., 9990 Mesa Rim Road, Suite 170, San Diego, CA 92121, USA} 
\author[0000-0003-1759-4354]{Cooper Downs}
%\email{cdowns@predsci.com}
\affil{Predictive Science Inc., 9990 Mesa Rim Road, Suite 170, San Diego, CA 92121, USA} 
\author[0000-0003-0621-4803]{Lisa A. Upton}
%\email{lisa.upton@swri.org}
\affil{Southwest Research Institute, 6220 Culebra Road, San Antonio, TX 78238, USA}
\author[0000-0003-3191-4625]{Bibhuti Kumar Jha}
%\email{maitraibibhu@gmail.com}
\affil{Southwest Research Institute, 6220 Culebra Road, San Antonio, TX 78238, USA}
\author[0000-0003-4312-6298]{Raphael Attie}
%\email{rattie@gmu.edu}
\affil{NASA Goddard Space Flight Center, 8800 Greenbelt Road, Greenbelt, MD 20771, USA}
\author[0000-0001-9326-3448]{Charles N. Arge}
%\email{charles.n.arge@nasa.gov}
\affil{NASA Goddard Space Flight Center, 8800 Greenbelt Road, Greenbelt, MD 20771, USA}
\author[0000-0002-6038-6369]{Carl J. Henney}
%\email{carl.henney.1@spaceforce.mil}
\affil{Air Force Research Laboratory, Space Vehicles Directorate, Kirtland AFB, NM 87117, USA}

\date{\today}


%%%%%%%%%%%%%%%%%%%%%%%%%%%%%%%%%%%%%%%%%%%%%%%%%%%%%%%%%%%%
%%%%%%%%%%%   ABSTRACT
%%%%%%%%%%%%%%%%%%%%%%%%%%%%%%%%%%%%%%%%%%%%%%%%%%%%%%%%%%%%

\begin{abstract}
Global solar photospheric magnetic maps play a critical role in solar and heliospheric physics research. Routine magnetograph measurements of the field occur only along the Sun-Earth line, leaving the far-side of the Sun unobserved.  Surface Flux Transport (SFT) models attempt to mitigate this by modeling the surface evolution of the field.  While such models have long been established in the community (with several releasing public full-Sun maps), none are open source. The Open Source Flux Transport (OFT) model seeks to fill this gap by providing an open and user-extensible SFT model that also builds on the knowledge of previous models with updated numerical and data acquisition/assimilation methods along with additional user-defined features.  In this first of a series of papers on OFT, we introduce its computational core: the High-performance Flux Transport (HipFT) code (\url{github.com/predsci/hipft}).  HipFT implements advection, diffusion, and data assimilation in a modular design that supports a variety of flow models and options.  It can compute multiple realizations in a single run across model parameters to create ensembles of maps for uncertainty quantification and is high-performance through the use of multi-CPU and multi-GPU parallelism.  HipFT is designed to enable users to easily write extensions, enhancing its flexibility and adaptability.  We describe HipFT's model features, validations of its numerical methods, performance of its parallel and GPU-accelerated code implementation, analysis/post-processing options, and example use cases.
\end{abstract}

%%%%%%%%%%%%%%%%%%%%%%%%%%%%%%%%%%%%%%%%%%%%%%%%%%%%%%%%%%%%
%%%%%%%%%%%   INTRODUCTION
%%%%%%%%%%%%%%%%%%%%%%%%%%%%%%%%%%%%%%%%%%%%%%%%%%%%%%%%%%%%

\section{Introduction}
\label{sec:intro}
The magnetic field of the Sun plays a key role in solar and heliospheric physics, as it is a major driver of the structure and dynamics of the solar corona, and is the energy source for solar activity.  The field is measured most easily in the photosphere, and global observations are provided in the form of full-disk magnetograms by instruments such as the Helioseismic and Magnetic Imager (HMI) onboard the Solar Dynamics Observatory (SDO) \citep{scherrer2012helioseismic}, the NSO Global Oscillation Network Group (GONG) \citep{harvey1996global}, and the {\bf National Solar Observatory/Synoptic Optical Long-term Investigations of the Sun Vector Spectromagnetograph (NSO/SOLIS/VSM) \citep{Keller03a,Keller03b}}.    Most regular observations of the surface field are along the Earth-Sun line (although the PHI imager on Solar Orbiter \citep{solankietal2020} now provides intermittent measurements from other vantage points).  This creates large data gaps in the full surface field.
Observatories  create so-called ``synoptic'' maps by combining portions of magnetograms over the course of a solar rotation to produce full-sun maps of the magnetic field, often in Carrington coordinates.  These Carrington Rotation (CR) maps are really diachronic in nature  \citep[e.g.,][]{linkeretal2017}, as they are built up over time and do not attempt to represent  the Sun's magnetic field at any given instant.  By their nature, synoptic CR maps necessarily contain older data, and do not reflect the surface flux evolution and emergence that occurred after data ingestion.  From Earth's vantage point, one or both of the Sun's poles are either poorly observed or obscured throughout the year, and so filling the polar regions of the maps with values requires extrapolation or other methods. 

Full-Sun magnetic maps are usually created from measurements of the line-of-sight (LOS) field, as this component is most reliably measured, especially in weaker field regions.  They can be used to to predict solar irradiance and activity indices \citep{Chapman1986,Henney2015,Warren2021}, and they have a long history of use as boundary conditions in models of the solar corona and heliosphere, such as in potential field \citep[e.g.,][]{Wiegelmann2021} and magnetohydrodynamic (MHD) \citep[e.g.,][]{Riley2011, Mackay2012,Gombosi2018,feng2019magnetohydrodynamic} models.  The first potential field source-surface (PFSS) models  to use photospheric measurements \citep[e.g.,][]{altschulernewkirk1969} used the LOS field ($B_{LOS}$) directly as a boundary condition, and the first such global MHD model \citep{usmanov1993} used the radial field ($B_r$)  derived from a PFSS model using this specification. \citet{wang1992potential} argued that the the LOS field is predominantly radial where it is measured in the photosphere, and specifying $B_r$ by employing this assumption is the more appropriate boundary condition for coronal models.  \citet{mikiclinker96} 
used this approach to directly specify $B_r$ in a global MHD model.  This approach has now become standard for most models (potential field or MHD), to the point that observatories typically provide CR maps of $B_r$ derived from $B_{LOS}$.  

While diachronic maps provide a useful average description of the Sun's field over the course of a solar rotation, for many applications, a representation of the global field at a particular time is desired.  Instantaneous global observations of the field are not presently available, but the processes by which the magnetic flux on the Sun evolves (primarily differential rotation, meridional flow, supergranular diffusion, and random flux emergence) have been studied for many years.  
Following the Babcock-Leighton description of the solar dynamo \citep{Babcock1961,Leighton1964}
Surface Flux Transport (SFT) models have been developed to describe these processes (see the reviews by  \citet{sheeley2005}, \citet{Jiang2014} and \citet{Yeates2023} for a comprehensive history).  These models treat the photospheric $B_r$ as a passive scalar quantity that is evolved on the solar surface.  SFT models first appeared in the 1980s  \citep[e.g.,][]{devoreetal1984,wangetal1989} to investigate the evolution of the surface field over the course of the solar cycle.  Assimilative SFT models \citep[e.g.,][]{worden00a,schrijver2003photospheric} ingest data from magnetograms (typically $B_r$ derived from $B_{LOS}$) on the observed portion of the Sun's surface, and evolve the field on the unobserved portions (including the poles).  They produce a continuous approximation of the state of the photospheric magnetic field as a sequence of ``synchronic'' maps - maps that attempt to represent the state of the Sun's magnetic field at a given instant in time. While the emergence of new flux on the far-side of the Sun (such as active regions (ARs)) will be missed in this approach, the evolved existing field is expected to be much closer to the true state.  

Assimilative SFTs have  used a variety of assumptions and calculation methods, that can lead to quantitative differences between the maps.  Three assimilative SFTs that have been used frequently to generate maps for coronal modeling and space weather applications are the Lockheed Martin Solar and Astrophysics Laboratory (LMSAL) Evolving Surface-Flux Assimilation Model (ESFAM) \citep{schrijver2003photospheric}, the Air Force Data Assimilative Photospheric flux Transport (ADAPT) model, \citep{Arge2010,Arge2013,hickmann2015data}, and the Advective Flux Transport (AFT) model \citet{upton2013predicting}.  While ESFAM and ADAPT release pre-computed full-Sun maps for public use\footnote{ADAPT: \url{https://nso.edu/data/nisp-data/adapt-maps}, LMSAL-ESFAM: \url{https://www.lmsal.com/forecast}}, none of these models are open source.  As a result, users of maps must rely on the assumptions and parameter choices of the model developers, which may have been tailored for specific purposes.   The ADAPT model has pioneered the use of multiple realizations to characterize possible variability, but  \citet{Barnes2023} found that greater differences were present between maps generated by different SFTs than amongst the ADAPT ensembles. The consequences of different SFT map properties for coronal/heliospheric models have not been investigated extensively, although \citet{Knizhnik2024} demonstrated similar performance of solar wind models using ADAPT and AFT for a single Carrington rotation.  

To foster the open use and development of SFT models, and allow community investigation of the effects of different model assumptions and parameters, we have created the Open-source Flux Transport (OFT) model.  OFT is patterned after AFT in that it solves the transport equation on an Eulerian mesh, allows higher resolution than other models, and can utilize a {\bf supergranular} flow model \citep{Hathaway_2010,Hathaway_2015} to better represent the quiet sun network and the decay of ARs.  Like ADAPT, OFT allows for the generation of an ensemble of maps, and like ESFAM and ADAPT, can incorporate the addition of random flux to maintain the quiet sun background magnetic flux away from the window where assimilation occurs.

\begin{figure}[htb]
\centering
\includegraphics[width=0.45\textwidth]{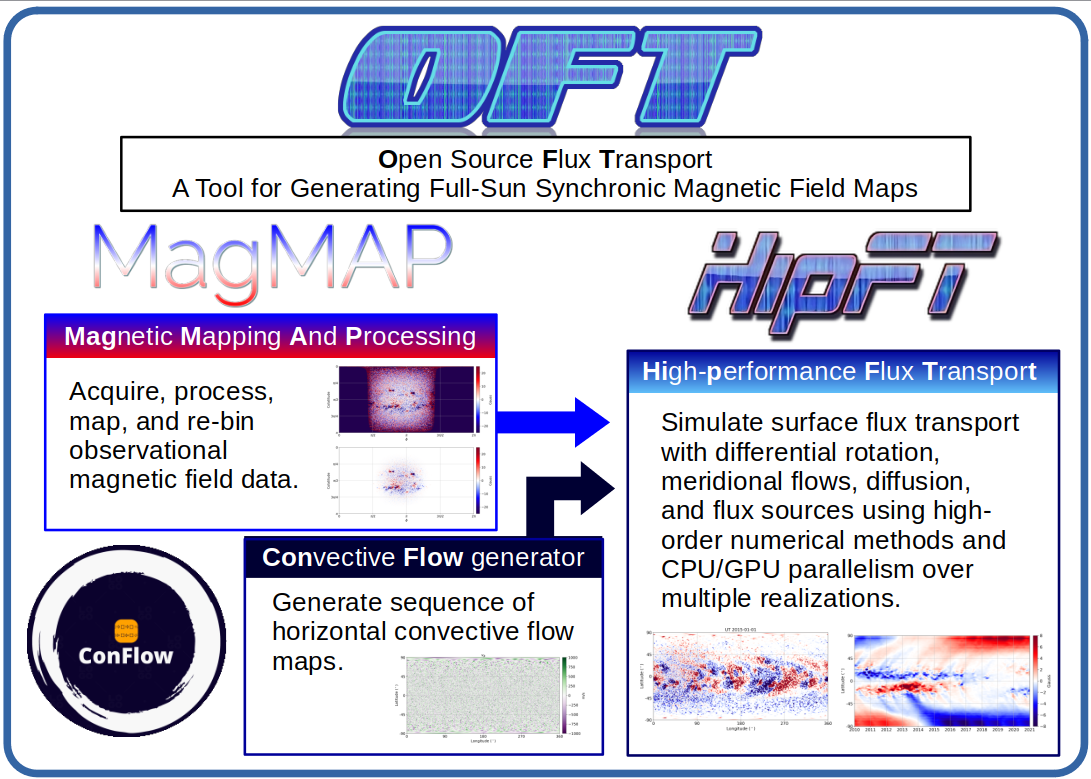}
\caption{Overview of the components of the Open-source Flux Transport model.\label{fig:oft}} 
\end{figure}
  
OFT is publicly hosted on github\footnote{\url{https://github.com/predsci/oft}}, allowing for community use, contributions, and development.  It blends several key state-of-the-art features from other SFT models into a modular, modern, and computationally efficient code base. OFT is broken up into three main components (shown in Fig.~\ref{fig:oft}):  Magnetic Mapping and Processing (MagMAP\footnote{\url{https://github.com/predsci/magmap}}), Convective Flow Generator (ConFlow\footnote{\url{https://github.com/predsci/conflow}}), and High-performance Flux Transport (HipFT\footnote{\url{https://github.com/predsci/hipft}}). Each component has its own independent public repository, which are all added to the OFT repository as git submodules\footnote{\url{https://git-scm.com/book/en/v2/Git-Tools-Submodules}}.  MagMAP is used to obtain and accurately remap line-of-sight and vector magnetograph observations into heliographic coordinates, as well as down-sample the resulting map to a desired resolution in an integral flux-preserving manner.  The resulting sequence of map files and meta data is then ready to be used by HipFT's data assimilation module.  Details on the data acquisition and mapping techniques used in MagMAP will be described in a forthcoming OFT paper \citep{oft3}.   ConFlow is used to generate a sequence of randomly-seeded {\bf supergranular} tangential surface flows which convect according to specified differential rotation and meridional flow models.  The resulting sequence of flows and metadata are then ready to be used in HipFT.  Details on the supergranular flow model used in ConFlow will be described in a forthcoming OFT paper \citep{oft2}.  HipFT (the focus of this paper) is the computational core of OFT, charged with integrating the surface flux transport model, and includes advection, diffusion, source terms, and data assimilation.  While loosely based on the AFT model, HipFT is a new code built from the ground up including multiple extensions and enhancements.  {\bf The version of HipFT used in this paper is {\tt v1.14.1} \cite{hipftcodezenodo}}.

The paper is organized as follows:  
In Sec.~\ref{sec:model} we describe the components of HipFT's flux transport model.  The numerical methods used to integrate the model and their validations are described in Sec.~\ref{sec:num}, while the code implementation and performance are shown in Sec.~\ref{sec:code}.  The use of the HipFT is described in Sec.~\ref{sec:codeuse} with selected example cases described in Sec.~\ref{sec:examples}.  We discuss the availability of the code in Sec.~\ref{sec:avail}, and summarize in Sec.~\ref{sec:summary}.

%%%%%%%%%%%%%%%%%%%%%%%%%%%%%%%%%%%%%%%%%%%%%%%%%%%%%%%%%%%%
%%%%%%%%%%%   MODEL
%%%%%%%%%%%%%%%%%%%%%%%%%%%%%%%%%%%%%%%%%%%%%%%%%%%%%%%%%%%%

\section{Surface flux transport model}
\label{sec:model}
As mentioned in Sec.~\ref{sec:intro}, HipFT's surface flux transport model is based on previous SFT models, especially the AFT model \citep{upton2013predicting}. 
SFTs generally solve some form of the advective-diffusion equation for $B_r$ with source terms (recently, an approach to SFT modeling using Physics-Informed Neural Networks has also been described \citep{athalathil2024surface}).  HipFT solves the following form of this equation:
\begin{equation}
\label{eq:main_model}
\dfrac{\partial B_r}{\partial t} = - \nabla_{s}\cdot\,(B_r\,{\bf v}) + \nabla_{s} \cdot (\nu\,\nabla_{s}\,B_r) + S + D,
\end{equation}
where $B_r({\bf x},t)$ is the surface radial magnetic field, ${\bf v}({\bf x},t,B_r)=(v_{\theta},v_{\phi})$ is the non-homogeneous surface flow velocity vector, $\nu({\bf x})$ is the diffusivity coefficient, $S({\bf x},t,B_r)$ is a source term, $\nabla_{s}$ and $\nabla_{s}\cdot$ are the two-dimensional spherical surface gradient and divergence operators respectively, $D({\bf x}, t, B_r, B_{r;d}(t))$ is the application of data-assimilation where $B_{r;d}(t)$ is the data being assimilated, and ${\bf x}=(\theta,\phi)$, where $\theta$ and $\phi$ are the co-latitudinal and longitudinal directions respectively.

\subsection{Surface Flows}
\label{sec:model_flows}

The surface flows are implemented using the advection term
\begin{equation}
\label{eq:advection}
\nabla_{s}\cdot\,(B_r\,{\bf v}) = \dfrac{1}{R_{\odot}\,\sin\theta}\dfrac{\partial}{\partial\,\theta}(\sin\theta\,B_r\,v_{\theta}) + \dfrac{1}{R_{\odot}\,\sin\theta}\,\dfrac{\partial}{\partial \phi}(B_r\,v_{\phi}),
\end{equation}
where $v_{\theta}(\bf{x},B_r)$ and $v_{\phi}(\bf{x},B_r)$ are the flow velocities in the co-latitude and longitudinal directions respectively, and we take the solar radius to be $R_{\odot}=6.69\times 10^{8}\,\mbox{m/s}$.  HipFT allows flexibility in specifying these flows, including allowing fully custom flows to be read from file(s).  It also includes built-in common analytic descriptions of differential rotation and meridional flows.  Here we describe these flow options, as well as a magnetic field flow attenuation.

\subsubsection{Analytic models for differential rotation and meridional flows}
\label{sec:flows_analytic}

\begin{figure}[htb]
\centering
\includegraphics[width=0.3\textwidth]{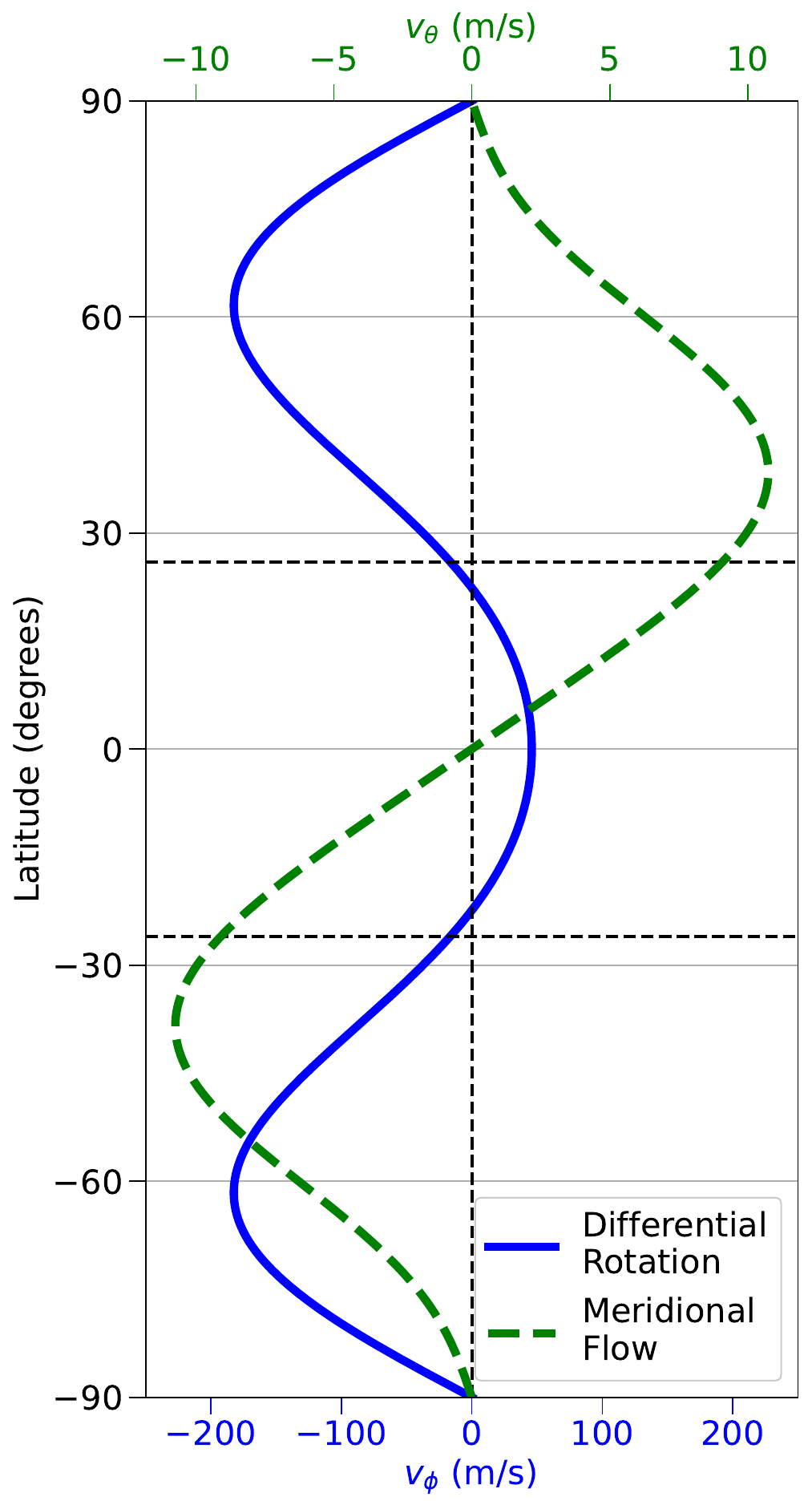}
\caption{Analytic flow models for differential rotation (blue solid line) and meridional flows (green dashed line) used in HipFT.  The profiles are shown {\bf relative to the Carrington rotational frame} using HipFT's default parameter values.\label{fig:model_flows_analytic}} 
\end{figure}

A commonly used model for differential rotation (DR) in the Carrington frame can be expressed as 
\begin{equation}
\label{eq:dr}
v_{\phi}(\theta)=\left[d_0 + d_2\,\cos^2(\theta) + d_4\,\cos^4(\theta)\right]\,\sin\,\theta,
\end{equation}
where the parameters $d_0$, $d_2$, and $d_4$ are inputs in units of $\mbox{m/s}$ and chosen based on analysis of solar observations.  Several studies have produced various values for these parameters \citep{howard1970spectroscopic,snodgrass1983magnetic,snodgrass1990rotation,upton2014effects}.  HipFT allows the user to specify these values and vary them over multiple realizations.  The default values are set to $d_0=46\,\mbox{m/s}$, $d_2=-262\,\mbox{m/s}$, and $d_4=-379\,\mbox{m/s}$, which are updated parameters used in the AFT model based on the data and procedures described in \citet{hathaway2022variations}.  In Fig.~\ref{fig:model_flows_analytic}, we show the profile of the default HipFT DR model.

For meridional flows (MF), several models have been used \citep{wang1989evolution,mackay2006models,sft_farm}.  For HipFT, we implement the form used in AFT \citep{hathaway2022variations}, given {\bf in the Carrington frame} by
\begin{equation}
\label{eq:mf}
v_{\theta}(\theta)=-\left[m_1\,\cos\,\theta + m_3\,\cos^3\theta + m_5\,\cos^5\theta\right]\,\sin\,\theta,
\end{equation}
where, as in the DR model, the parameters $m_1$, $m_3$, and $m_5$ are input in units of $\mbox{m/s}$ and chosen based on analysis of solar observations.  In HipFT, the default values are $m_1=22\,\mbox{m/s}$, $m_3=11\,\mbox{m/s}$, and $m_5=-28\,\mbox{m/s}$, which are derived in the same manner as those for DR described above.  In Fig.~\ref{fig:model_flows_analytic}, we show the profile of the default HipFT MF model.

\subsubsection{Custom flow profiles and supergranular flows}
\label{sec:flows_conflow}

HipFT allows the user to specify a sequence of velocity map files that specify flows over time to be added to any already selected analytical flow profile such as those in Sec.~\ref{sec:flows_analytic}.  A motivating use case for this feature is adding {\bf supergranular} convective flow models \citep{Rincon2018}.  In Sec.~\ref{sec:examples_1yr}, we show an example of a production HipFT run utilizing such flows generated by ConFlow \citep{oft2}.

\subsubsection{Flow attenuation}
\label{sec:flows_attenuation}

Using flows that are independent of the value of the surface magnetic field can miss important dynamics.  For example, active regions (AR) are observed to have different flows profiles than DR and MF measured in the quiet Sun \citep{Stenflo1974,fattenpaper2}.  To account for this, we add the option to attenuate the flow based on the magnitude of the surface field.  This takes the form
\begin{equation}
\label{eq:flow_attenuation}
v_{\theta/\phi}\rightarrow v_{\theta/\phi}\,\left[1.0-\tanh\left(\dfrac{|B_r|}{B_0}\right)\right],
\end{equation}
where $B_0$ is an input saturation value (default $B_0=500$ Gauss).

\subsection{Diffusion}
\label{sec:model_diffusion}

The decay of ARs and flux cancellation caused by supergranular and granular flows require extremely high-resolution flow profiles to capture directly, and are subject to other processes not able to be directly modeled in SFT models.  Instead, it is common to add a diffusion term to the SFT model to try to capture the mean effect.  The diffusion term in HipFT appears as 
\begin{equation}
\label{eq:diffusion}
\nabla_{s} \cdot (\nu\,\nabla_{s}\,B_r) =
\dfrac{1}{R_{\odot}^2\,\sin\theta}\dfrac{\partial}{\partial\,\theta}\left( \nu(\theta,\phi,B_r)\,\sin\theta\,\dfrac{\partial\,B_r}{\partial\,\theta}\right) 
+ \dfrac{1}{R_{\odot}^2\,\sin^2\theta}\,\dfrac{\partial}{\partial \phi}\,\left(\nu(\theta,\phi,B_r)\dfrac{\partial B_r}{\partial \phi}\right),
\end{equation}
where $\nu(\theta,\phi,B_r)$ is the chosen diffusivity.

The value for $\nu$ required to obtain the desired amount of flux cancellation and field decay varies widely in the literature, ranging from $100\,\mbox{km}^2\!/\mbox{s}$ to over $600\,\mbox{km}^2\!/\mbox{s}$ \citep{Jiang2014,Yeates2023}.  Selecting a scalar value for $\nu$ is difficult, as the required diffusivity to match observations can depend on spatial scale and the chosen flow velocities \citep{Jiang2014}.  A further consideration is the (sometimes overlooked) contributions to diffusivity caused by the model's numerical method used to integrate advection.  For example, when using finite-difference or finite-volume Upwinding (see Sec.~\ref{sec:num_advect_time}), a significant amount of diffusivity is added to the solution.  As a result, the model requires a smaller value of $\nu$ to achieve the desired amount of diffusion; making the required value resolution dependent.  Another example is when using central differencing for advection, where the scheme is unconditionally unstable unless enough diffusion is added for stabilization.

Diffusion is also useful when running with {\bf supergranular} convective flows.  Since small regions of convergent flows can appear, low-diffusive advection schemes such as the default scheme used in HipFT (see Sec.~\ref{sec:num_advect}) can result in high-valued `spike' pixels (as there are no radial inflow or outflow in the model).  Although such convergent flows are often temporary (due to the rapid changes in {\bf supergranular} flows) large magnitude spikes can occasionally still occur.  Introducing some amount of diffusion helps to reduce/eliminate them, with the optimal amount of diffusion to use being scheme and use-case specific.

The implementation of the diffusion equation of Eq.~\ref{eq:diffusion} allows HipFT to be re-purposed as an efficient flux-preserving magnetogram smoother (see Sec.~\ref{sec:examples}) in which case the diffusivity can be set based on the grid cell size as: 
\begin{equation}
\label{eq:diffgrid}
\nu_g = \alpha_{\nu}\,\left[(\Delta\theta)^2 + (\Delta \phi\,\sin\theta)^2\right],
\end{equation}
where $\alpha_{\nu}$ is a user specified factor and $\Delta\theta$ and $\Delta \phi$ are the local grid cell sizes.

HipFT also allows a user to easily write their own diffusivity as any function of space and $B_r$. Additionally, it includes the ability to read an external file for supplying a custom (optionally spatially-dependent) diffusivity.  

\subsection{Data assimilation}
\label{sec:model_dataassim}

Data assimilation refers to ingesting available observational data into the model over time and is a key component for SFT models being used to generate synchronic full-Sun magnetic maps.  Sophisticated methods of accounting for data uncertainty in the assimilation (such as Kalman filtering) have been utilized in SFT models \citep{hickmann2015data,Dash2024}.  The data assimilation in the present version of HipFT uses a custom weighting between the observed and model data.  It is designed to ingest the output data of the MagMAP software package \citep{oft3}, where at each assimilation time $t$, the field is updated as
\begin{equation}
\label{eq:data_assim}
B_r({\bf x},t) =  F({\bf x},t)\,B_{r;d}(t) + (1 - F({\bf x},t))\,B_r({\bf x},t), 
\end{equation}
where $F\in[0,1]$ is an assimilation weight function.  While the weight layer in the MagMAP assimilation maps (see \citet{oft3}) can be customized by the user based on more sophisticated uncertainty methods (such as those mentioned above), the default weighting is a simple function of center-to-limb angle and latitude, described as
\begin{equation}
\label{eq:data_assim_f}
F = \begin{cases}
   \mu^{\alpha_{\mu}}, &(\mu > \mu_{\mbox{\scriptsize lim}}) \land (|\theta_{\mbox{\scriptsize l}}| < \theta_{\mbox{\scriptsize l,lim}}) \\
   0, &\text{o.w.}
\end{cases}
\end{equation}
where $\mu=\cos\theta_d\in[0,1]$, $\theta_d$ is the center to limb angle of the observed solar disk data, $\theta_l$ is the map latitude, and $\alpha_{\mu}$, $\mu_{\mbox{\scriptsize lim}}$, and $\theta_{\mbox{\scriptsize l,lim}}$ are chosen parameters.  

The data assimilation functionality is implemented by reading in 3D MagMAP files. In each file, the first layer contains the new data, the second layer is taken to be the default full weight function to use directly ($F$), and the third layer contains the $\mu$ values extracted from the data, which can be used with the user-specified assimilation function of Eq.~\ref{eq:data_assim_f}.  Weights can also be fully customized by modifying the weight layer of the data files from MagMAP, or by implementing a new weight subroutine into HipFT.  Details on the weight models and their effects are described in a forthcoming OFT paper \citep{oft3}.  MagMAP is designed to ingest data from magnetograms, but in principle, other proxy sources of magnetic field, such as EUV imaging \citep{hess2020using} and/or helioseismology \citep{liewer2014testing}, could be used to insert ARs observed on the far side of the Sun \citep{Arge2013,chen2022inferring,sft_farm}.  

HipFT can optionally enforce flux balance during the assimilation step by rewriting Eq.~\ref{eq:data_assim} as 
\begin{equation}
\label{eq:data_assim_deltab}
B_r^{\mbox{\scriptsize new}} = B_r^{\mbox{\scriptsize curr}}  + \Delta B_r,
\end{equation}
where $\Delta B_r = F({\bf x})\,[B_{r;d} - B_r^{\mbox{\scriptsize curr}}]$.  Before the data is assimilated, $\Delta B_r$  is {\bf multiplicatively} flux balanced as
\begin{alignat}{1}
\label{eq:fluxbal}
\Delta B_r &= \Delta B_r/\sqrt{|\Phi_{+}/\Phi_{-}}, \qquad \Delta B_r > 0, \\
\Delta B_r &= \Delta B_r*\sqrt{|\Phi_{+}/\Phi_{-}}, \qquad \Delta B_r \le 0, \notag
\end{alignat}
where $\Phi_{+}$ and $\Phi_{-}$ are total fluxes of the initial difference, computed as surface integrals of $\Delta B_r$ for positive and negative values separately.
By using this method, neutral lines are preserved, and the balance of the flux is distributed proportionally to the field.

\subsection{Flux emergence}
\label{sec:model_source}

The quiet sun network is made up of an ocean of small-scale magnetic flux that is observed to be completely replaced by emergence and cancellation over a couple of days \citep{schrijveretal1997}.  To account for this source of flux, earlier SFTs (e.g. ADAPT and ESFAM) have incorporated random flux emergence into the maps.  HipFT uses the source term, $S$, in Eq.~\ref{eq:main_model} to facilitate flux emergence.  A static source term can be read from a file and then applied each time step. Additionally, two models of adding small-scale (grid-level) random flux emergence (RFE) are implemented.  For each model, the user can specify the total unsigned flux per hour, $\Phi/\mbox{\small hr}$, of the source terms, as well as the desired time between source terms (representing the lifetime of the random flux elements).  

Each model uses values sampled from a normal distribution with mean 0 and standard deviation $\sigma$
\[
X\sim \mathcal{N}(0,\sigma^2) = \frac{1}{\sigma\,\sqrt{2\,\pi}}\,\mbox{exp}\left[-\frac{x^2}{2\,\sigma^2}\right].
\]

For the first model, we set the average source term to represent a uniform unsigned flux per hour per cell.  We set the flux per hour of each cell such that the mean unsigned flux per hour ($\langle|\varphi|\rangle$) of each cell is equal to the chosen total unsigned flux per hour divided evenly across the total number of grid cells $N_{\mbox{\scriptsize cells}}$:
\[
\langle|\varphi|\rangle = \frac{\Phi/\mbox{\small hr}}{N_{\mbox{\scriptsize cells}}}.
\]
Since for a set of random variables $X\sim \mathcal{N}(0,\sigma^2)$,
\begin{equation}
\label{eq:mean}
\langle |X|\rangle = \dfrac{\int_{-\infty}^{\infty}|x|\,\mathcal{N}(0,\sigma^2)\,dx}{\int_{-\infty}^{\infty}\mathcal{N}(0,\sigma^2)\,dx} = \sqrt{\frac{2}{\pi}}\,\sigma,
\end{equation}
we find
\[
\sigma =  \frac{\Phi/\mbox{\small hr}}{N_{\mbox{\scriptsize cells}}}\,\sqrt{\frac{\pi}{2}}.
\]
Since if $X\sim \mathcal{N}(0,\sigma^2)$ and $Y\sim \mathcal{N}(0,1)$, then $X = \sigma\,Y$, we can use a standard normal distribution to generate the flux per hour of each cell $i$ as
\[
\varphi_i = \frac{\Phi/\mbox{\small hr}}{N_{\mbox{\scriptsize cells}}}\,\sqrt{\frac{\pi}{2}}\,Y_i.
\]
Dividing by the local cell area ($\Delta A_{j,k} = R_{\odot}^2\,\sin\theta_j\,\Delta \theta_j\,\Delta \phi_k$) yields the source $B_r/\mbox{\small hr}$ value for each cell:
\begin{equation}
\label{eq:rfe_m1}
S_{j,k} = \frac{\Phi/\mbox{\small hr}}{\Delta A_{j,k}\,N_{\mbox{\scriptsize cells}}}\,\sqrt{\frac{\pi}{2}}\,Y_i.
\end{equation}
This method creates larger source values in regions with finer grid resolution. {\bf The motivation is to mimic the behavior of magnetograph} observations where a finer image resolution will often reveal larger values of {\bf small-scale} unsigned flux that would otherwise be averaged out over larger resolution elements. In practice for uniform grids in $\theta$ and $\phi$, this will lead to larger absolute values of $B_r$ in grid-cells near the poles due to the $\sin\theta$ weighting in the cell area. However, the contribution from each individual grid cell can be rapidly spread throughout the grid during the calculation via a combination of diffusion and flows, especially at the poles due to the $1/\sin\theta$ geometric factor in the longitudinal direction.

For the second model, we set the average source term to represent a uniform unsigned flux per hour per unit area:
\[
\langle|S|\rangle = \frac{\Phi/\mbox{\small hr}}{4\,\pi\,R_{\odot}^2}.
\]
Using Eq.~\ref{eq:mean}, we now get
\[
\sigma = \frac{\Phi/\mbox{\small hr}}{4\,\pi\,R_{\odot}^2}\,\sqrt{\frac{\pi}{2}},
\]
and therefore the source $B_r/\mbox{\small hr}$ value for each cell is computed as
\begin{equation}
\label{eq:rfe_m2}
S_i = \frac{\Phi/\mbox{\small hr}}{4\,\pi\,R_{\odot}^2}\,\sqrt{\frac{\pi}{2}}\,Y_i.
\end{equation}
This method results in a uniform distribution of $B_r$ values over the grid cells regardless of cell area or grid spacing. This may be desirable when the surface flux distribution is thought to be fully resolved and the amount of random flux emergence per unit area should be constant. In practice, for a run using a uniform grid in $\theta$ and $\phi$, the combination of diffusion and flows can lead to weaker, more diffuse contributions at the poles relative to the equatorial latitudes.

Due to their practical differences, the choice {\bf of which of the above methods to use will  depend on the particular application.  Because the second method (Eq.~\ref{eq:rfe_m2}) represents a uniform unsigned flux per unit area and is thus insensitive to the local grid resolution, it is the default method used in HipFT.}

{\bf In either} method, two source maps of random flux ($S_{\mbox{\tiny RFE};1}$ and $S_{\mbox{\tiny RFE};2}$) are generated and flux balanced in the same manner as Eq.~\ref{eq:fluxbal}.  At every step, a linear interpolation between the two current source maps is calculated for the current time $t$ and the resulting source term is integrated in Eq.~\ref{eq:main_model}:
\begin{equation}
\label{eq:rfe_s}
S_{\mbox{\tiny RFE}}(t) = (1-\alpha(t))\,S_{\mbox{\tiny RFE};1} + \alpha(t)\,S_{\mbox{\tiny RFE};2},
\end{equation}
where $\alpha(t) = (t - t_{\mbox{\tiny RFE},1})/\tau_{\mbox{\tiny RFE}}$.  When the time of the second source map is reached, a new source map is generated and the process repeats.  This ensures a smooth evolution of the random flux elements over time.

\begin{figure}[htb]
\centering
$\begin{array}{c}
\includegraphics[height=0.17\textwidth]{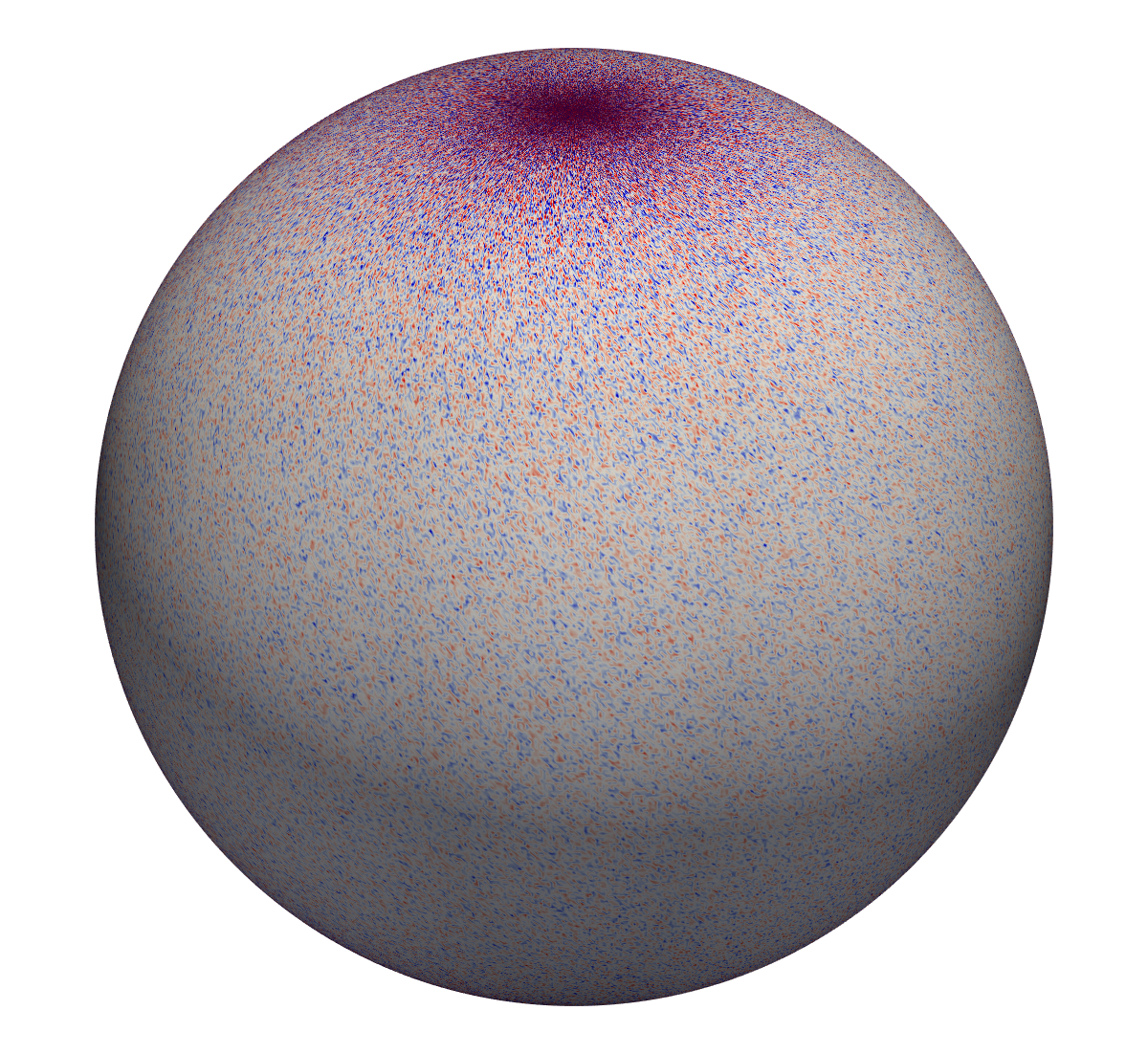}
\includegraphics[height=0.17\textwidth]{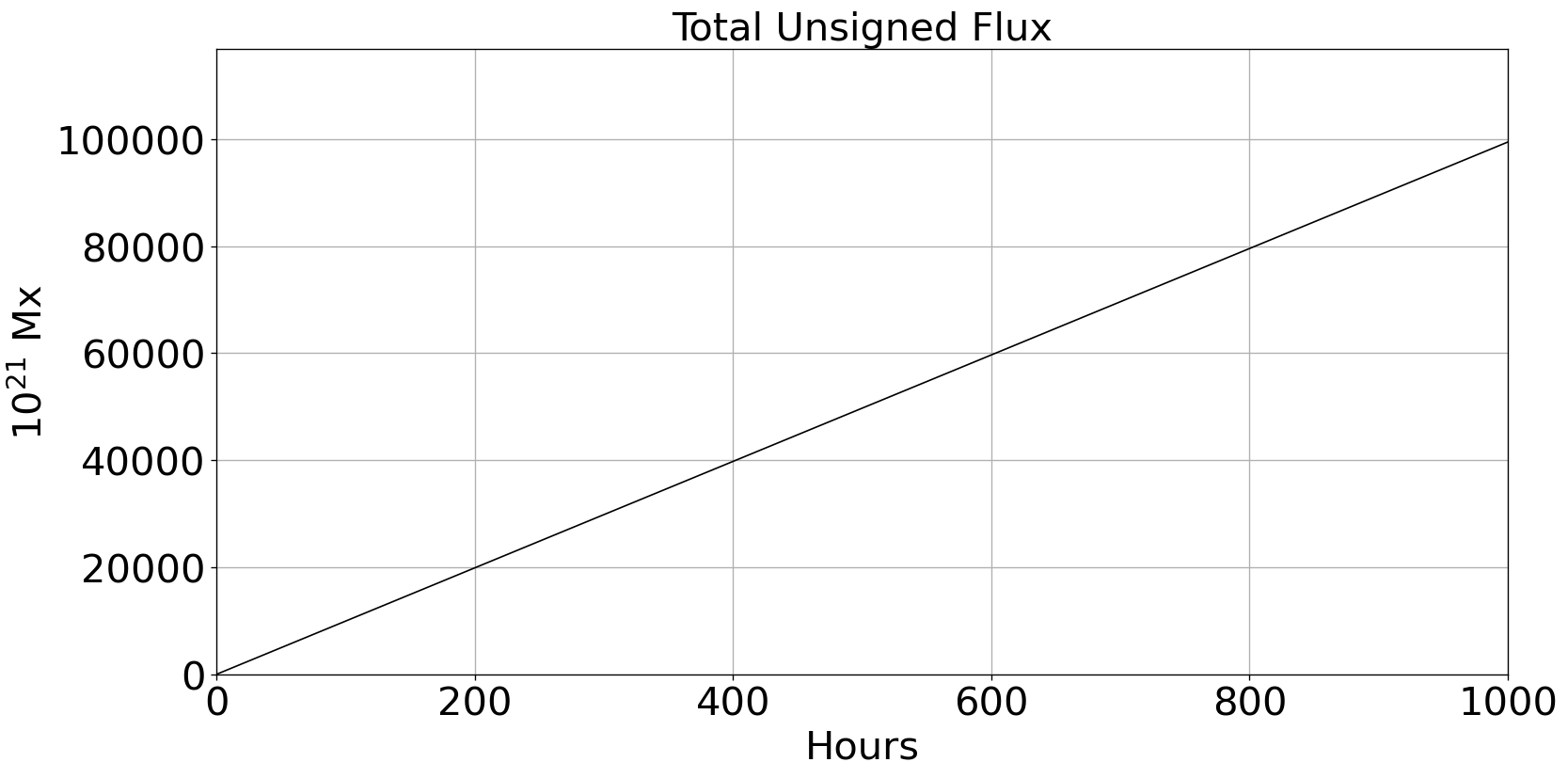}
\includegraphics[height=0.17\textwidth]{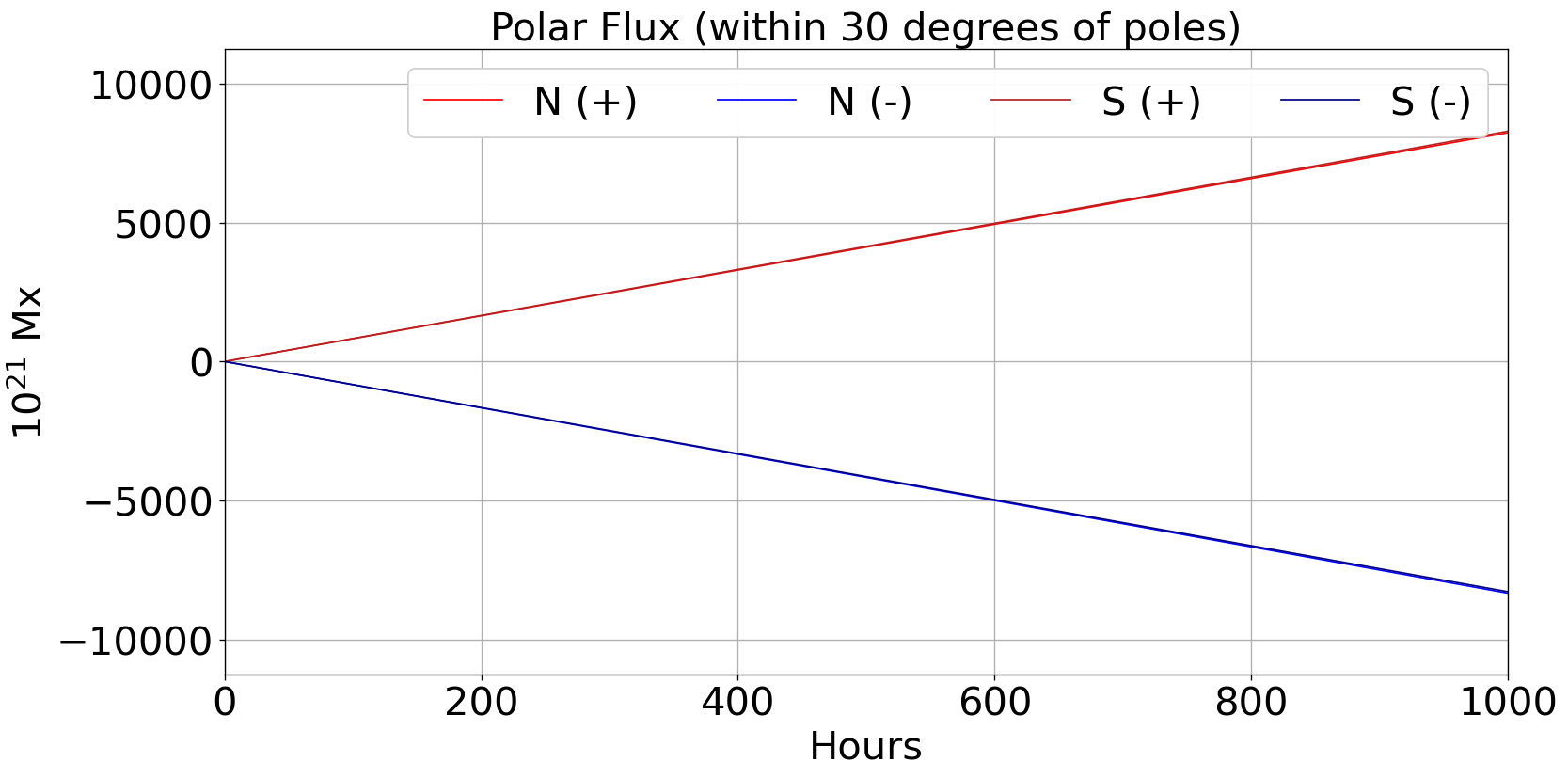}
\\
\includegraphics[height=0.17\textwidth]{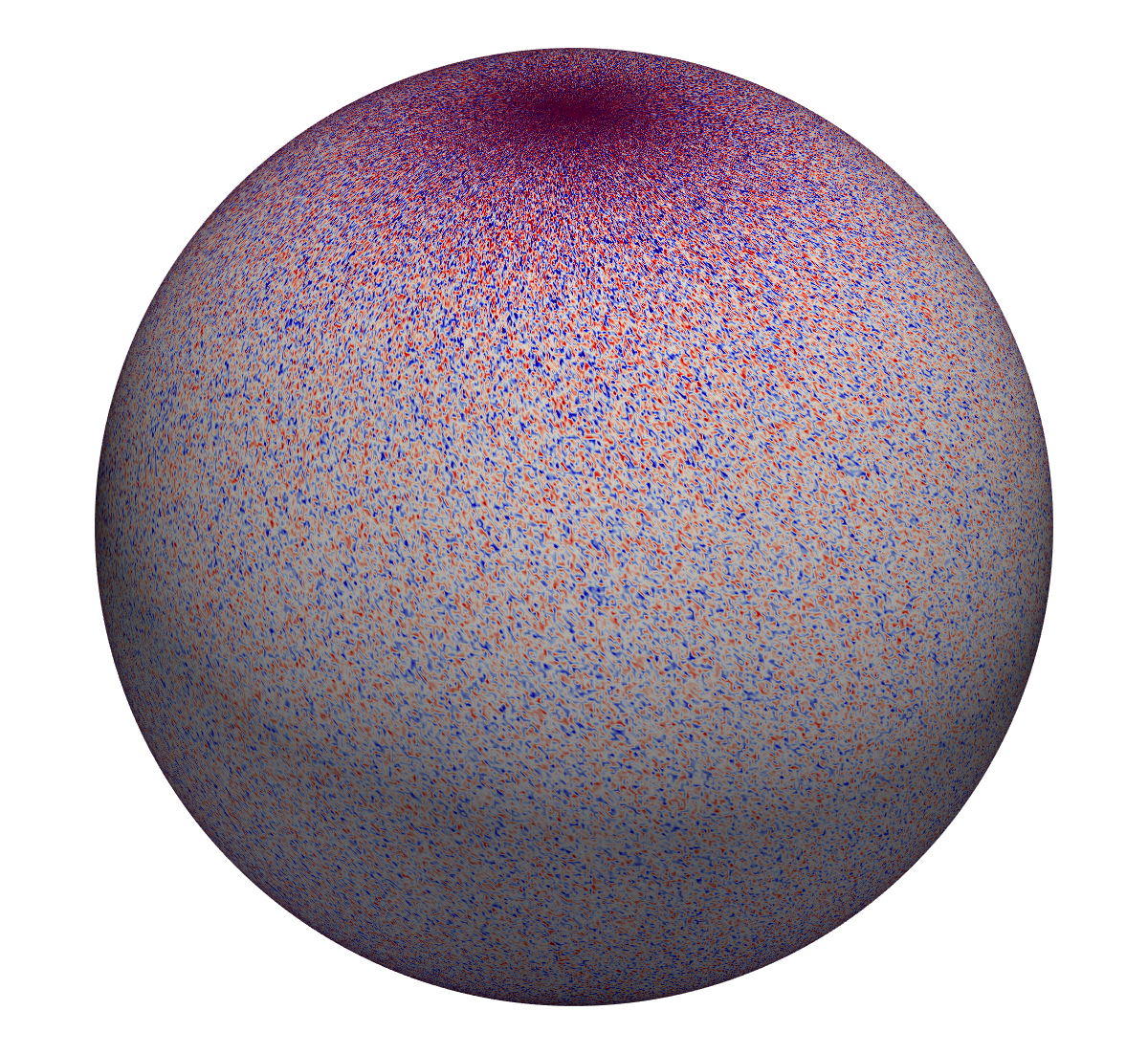}
\includegraphics[height=0.17\textwidth]{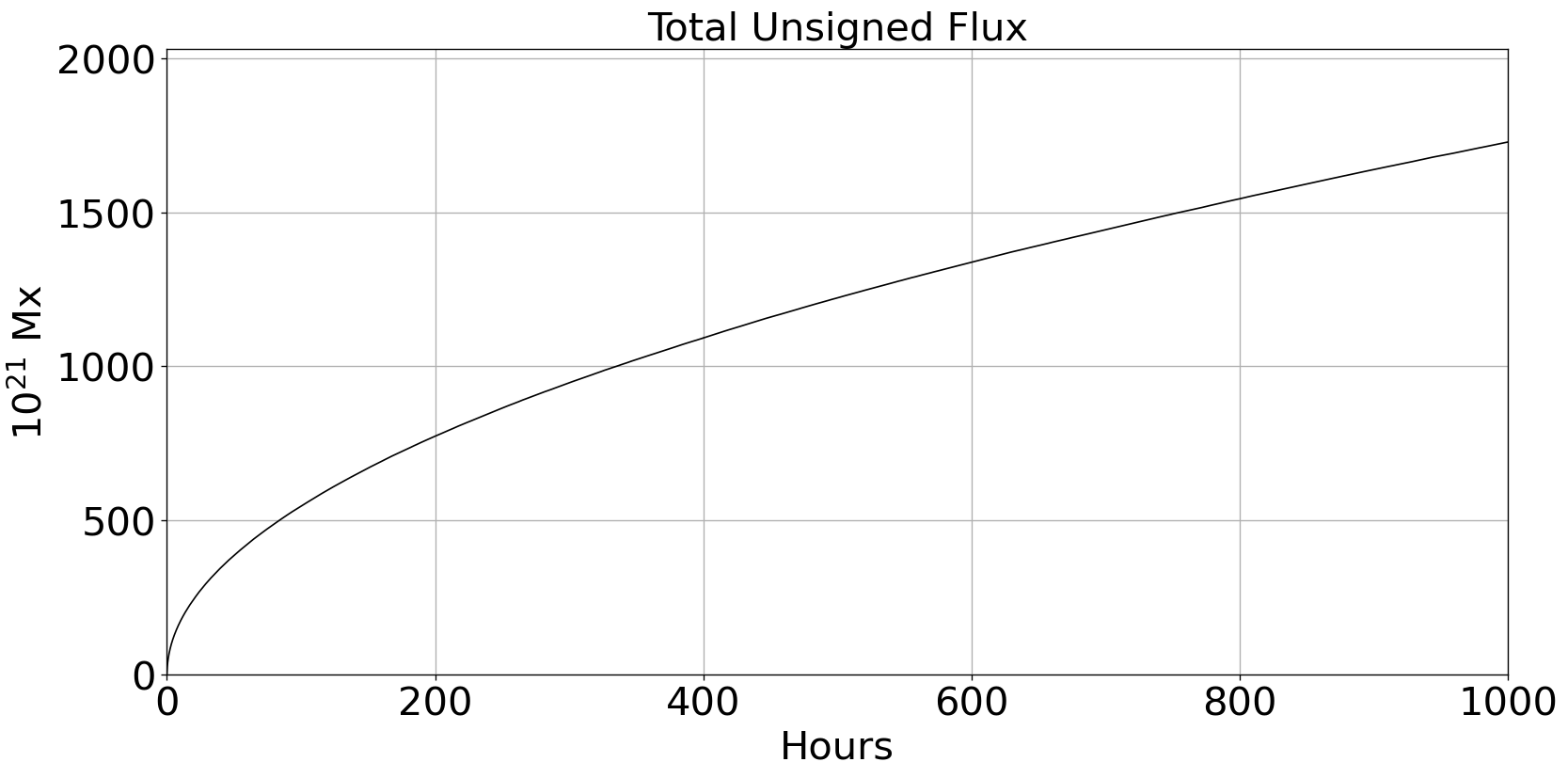}
\includegraphics[height=0.17\textwidth]{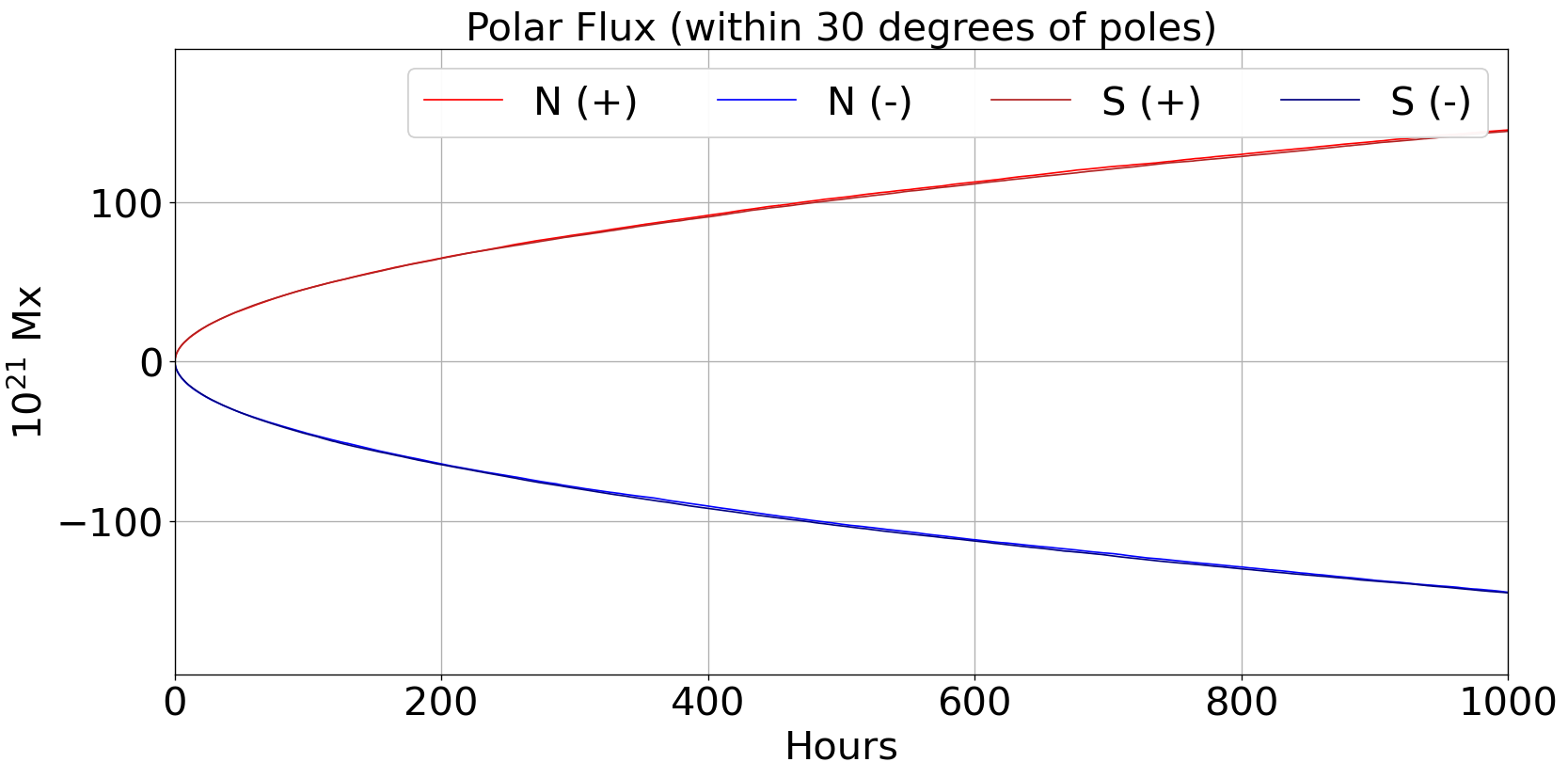}
\\
\includegraphics[height=0.17\textwidth]{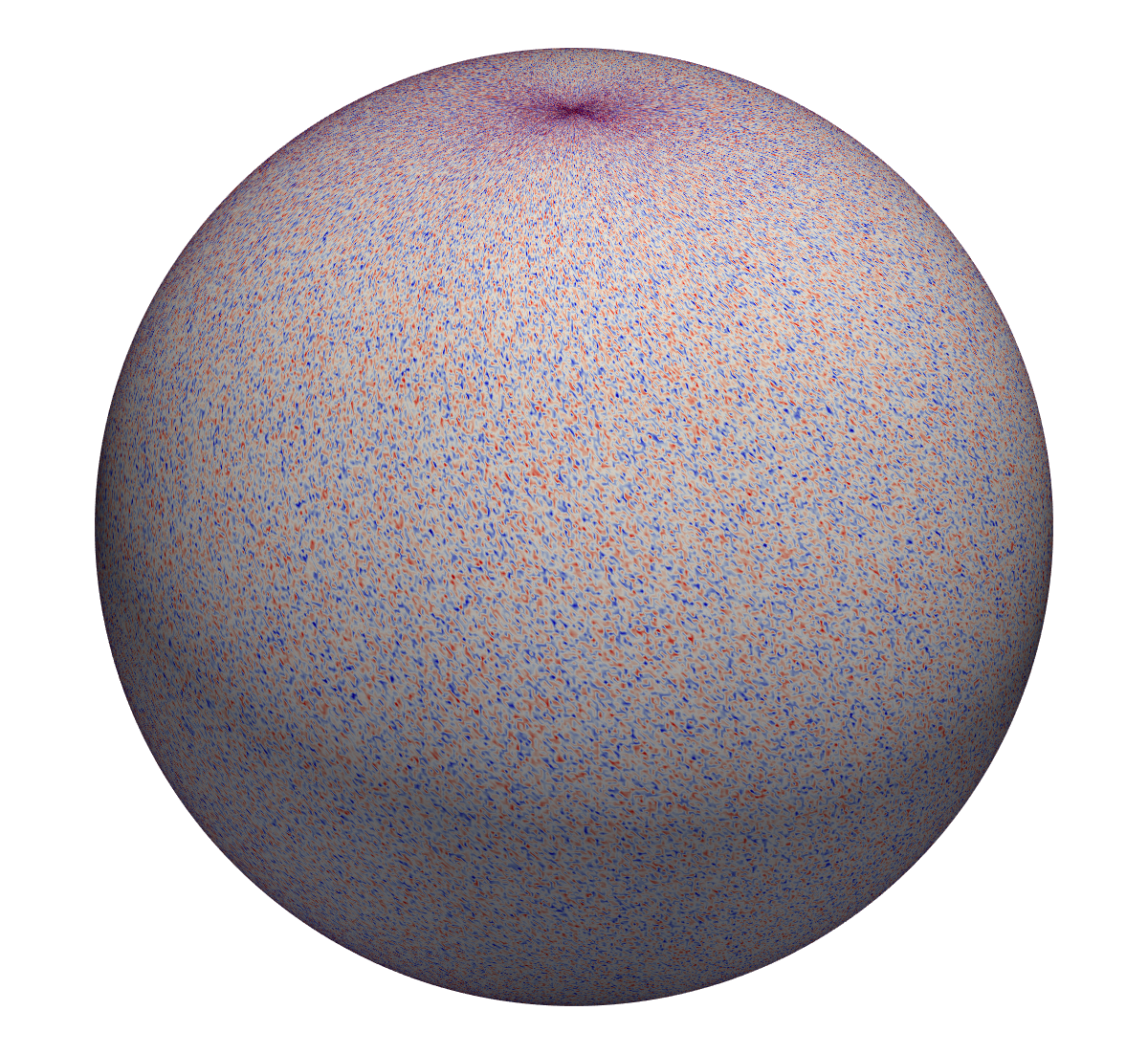}
\includegraphics[height=0.17\textwidth]{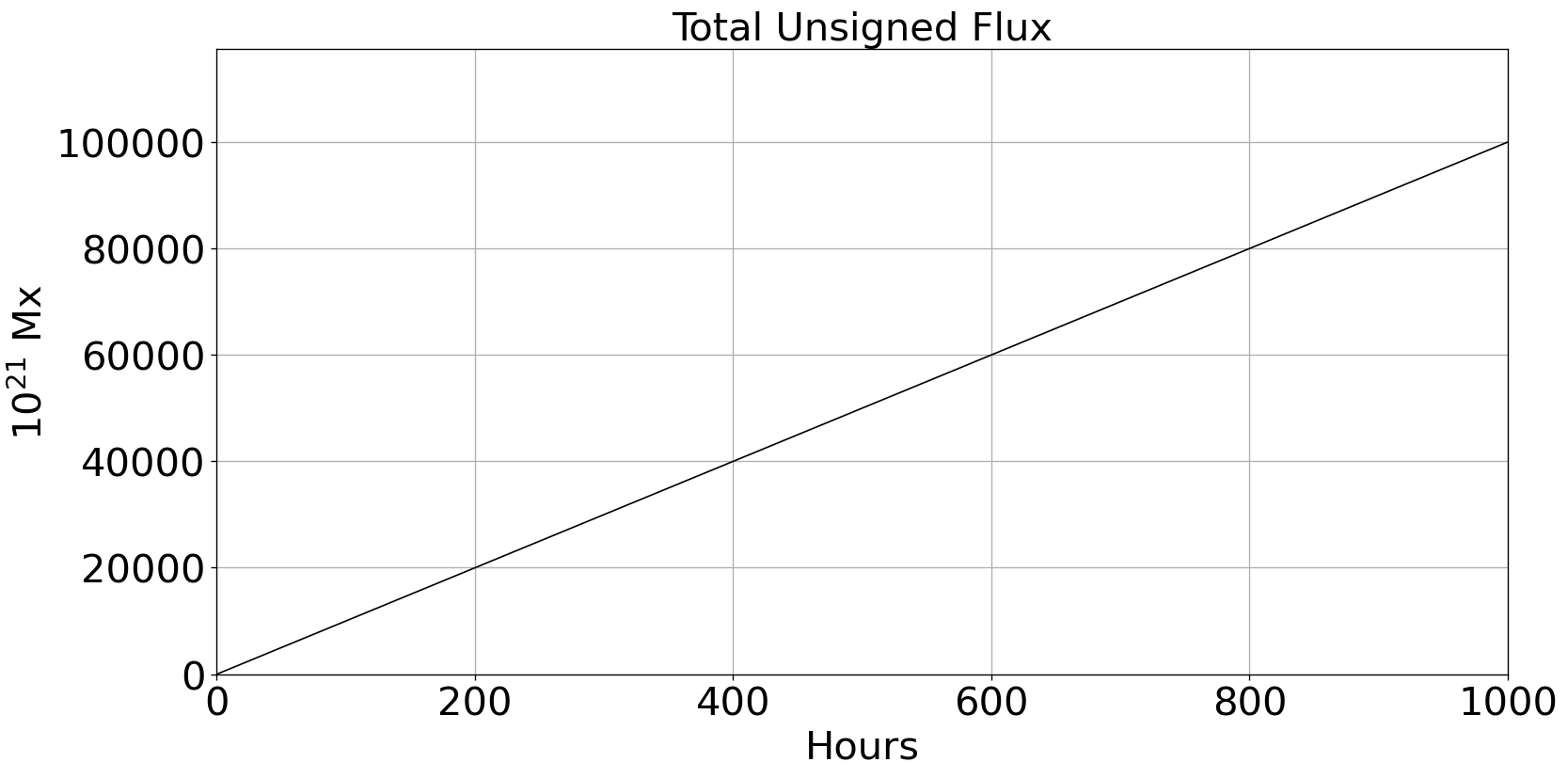}
\includegraphics[height=0.17\textwidth]{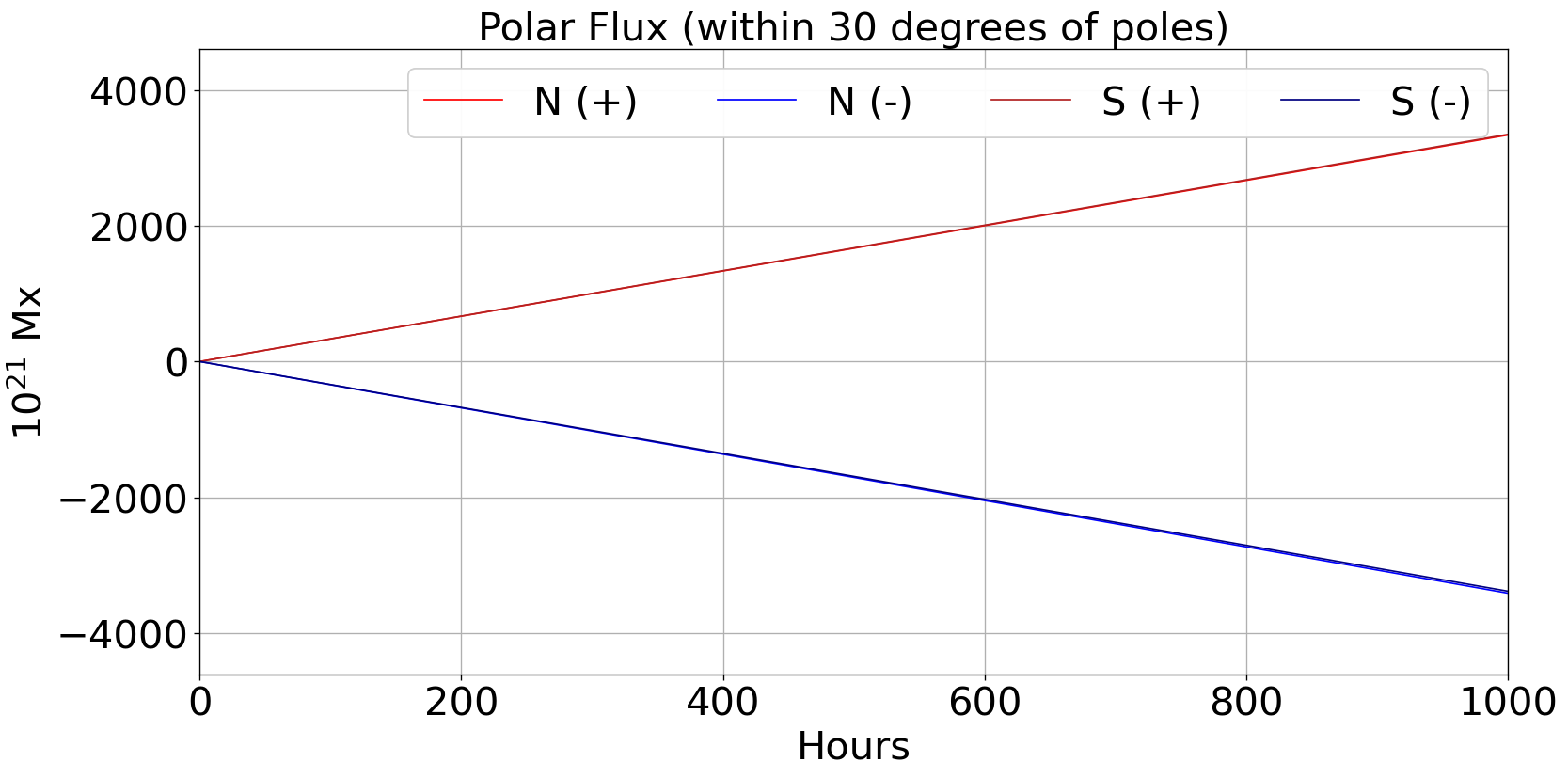}
\\
\includegraphics[height=0.17\textwidth]{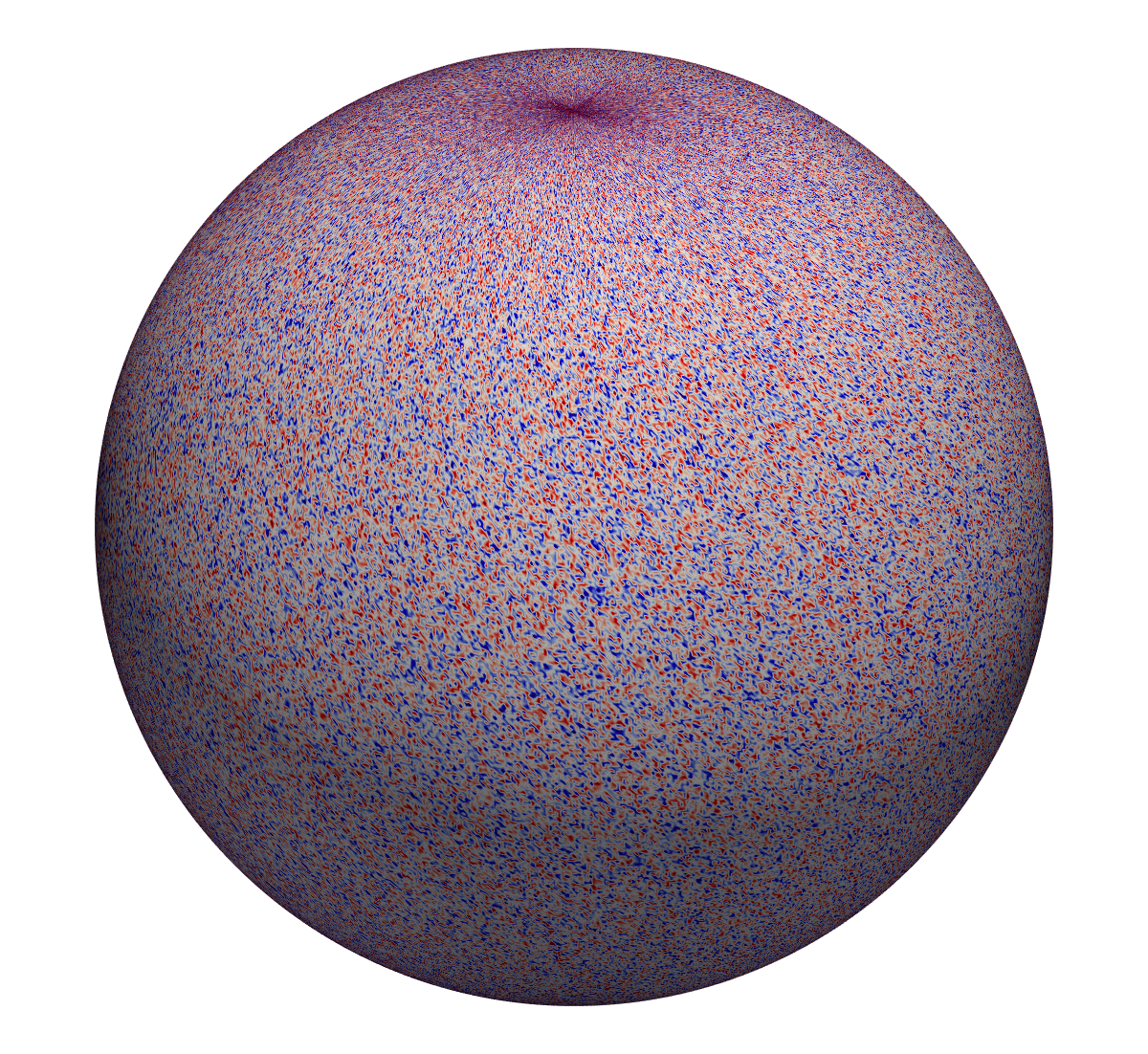}
\includegraphics[height=0.17\textwidth]{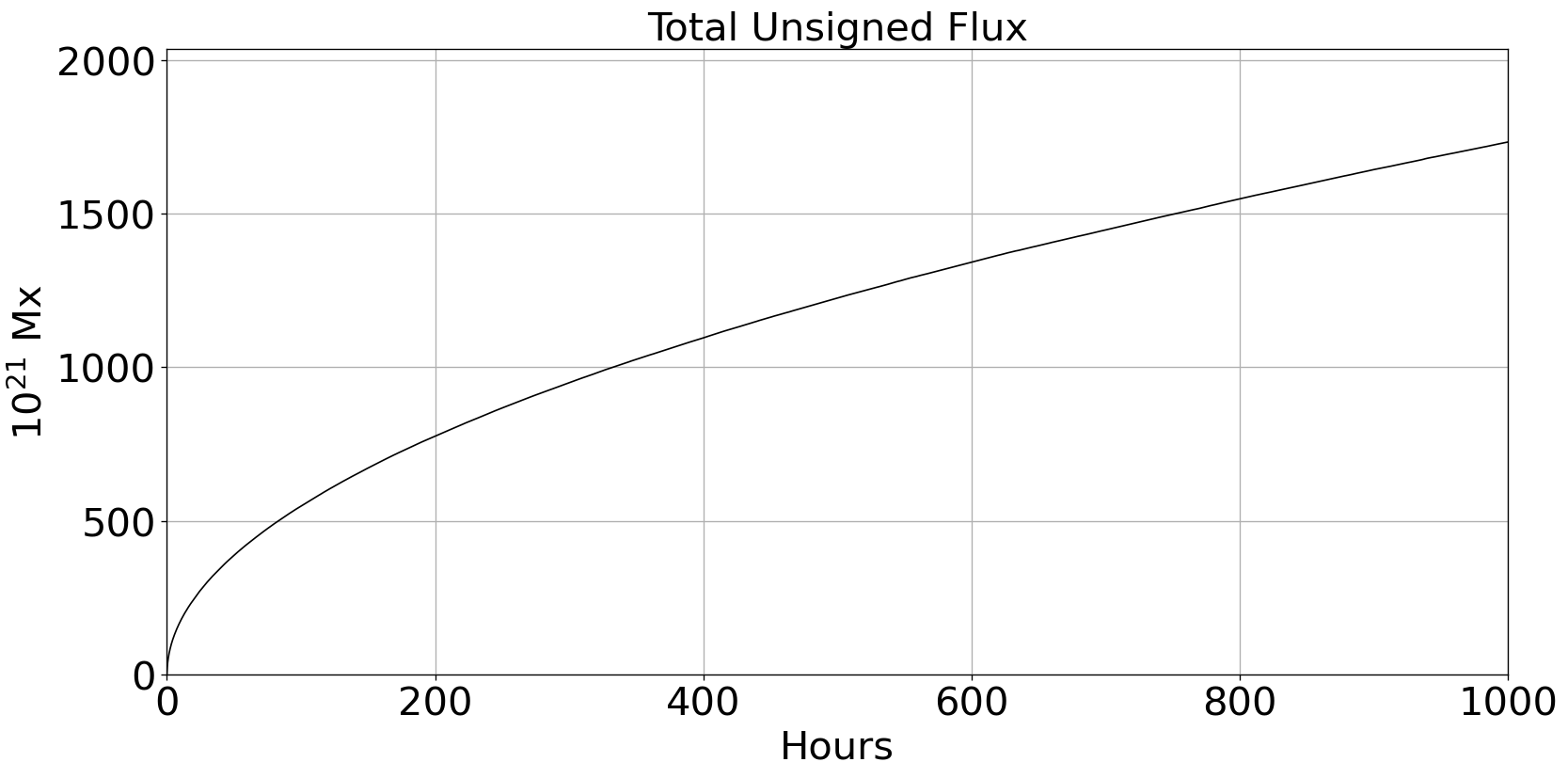}
\includegraphics[height=0.17\textwidth]{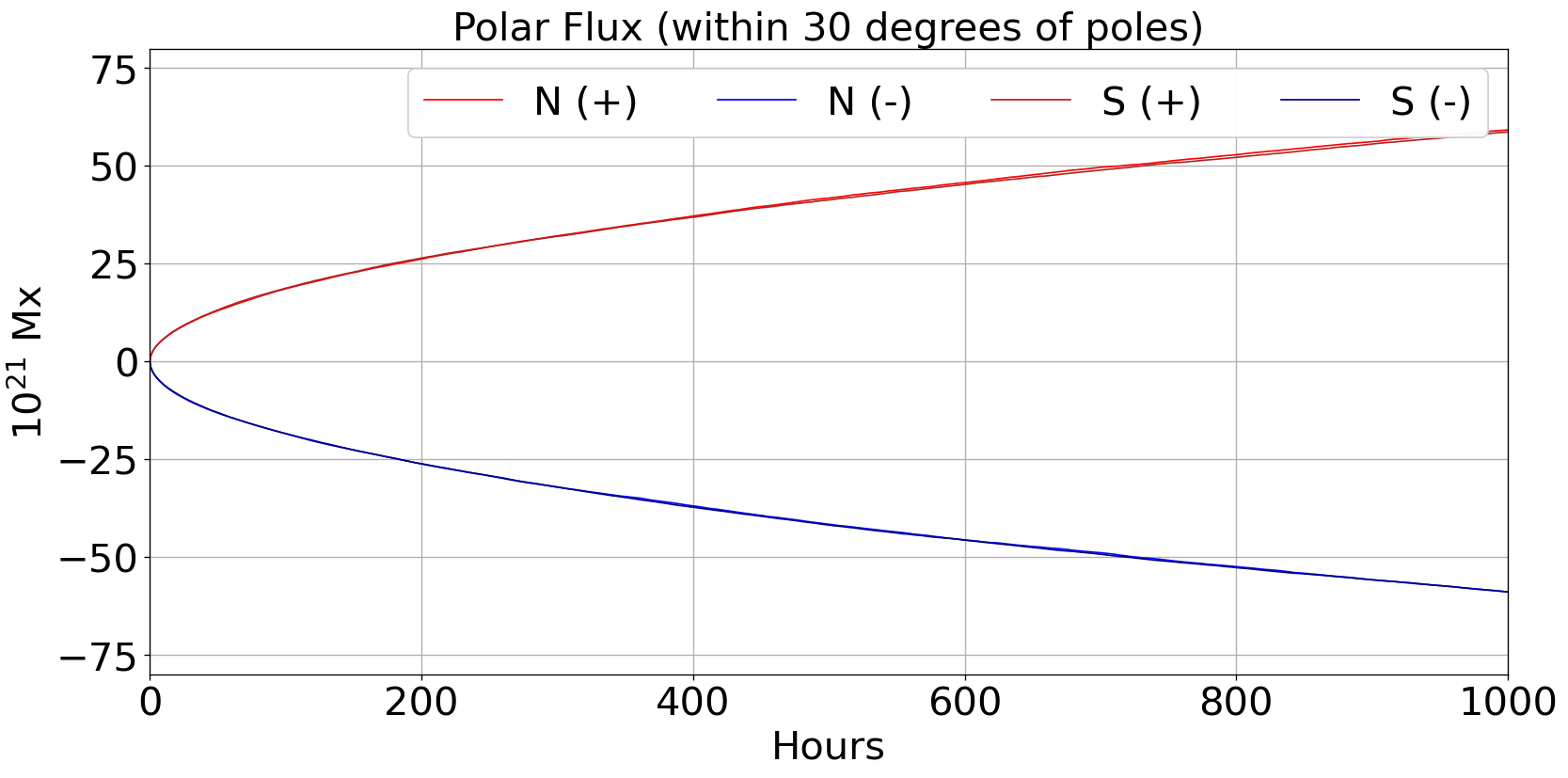}
\end{array}$
\caption{Random flux emergence source term runs for method 1 (M1) and method 2 (M2) with an end time of 1000 hours.  A total unsigned flux rate of  $100\times 10^{21}\,\mbox{Mx}/\mbox{\small hr}$ is set, and no other parts of the SFT model are active (no flows and no diffusion).  The results for M1 with lifetimes of infinity and 0.3 hours, and M2 with lifetimes of infinity and 0.3 hours are shown top to bottom.  For each run, the final map is shown (at a $b_0$ angle of 30 degrees) along with the total unsigned flux, and the polar flux within 30 degrees (left to right).  The final map for runs with infinite lifetimes are shown on a color scale from -1000 to 1000 Gauss, while those with 0.3 hour lifetimes are shown from -100 to 100 Gauss.   The latitudinal dependence of the resulting field for method 1 is apparent.\label{fig:rfe}} 
\end{figure}

To validate the implementation of RFE in HipFT, in Fig.~\ref{fig:rfe} we show the result of running with only the RFE source term activated (i.e. no flows or diffusion) starting with a zero map for each method shown above with both an infinite and finite ($0.3$ hour) RFE lifetime.
We see that the specified total unsigned flux per hour (here, $100\times 10^{21}\,\mbox{Mx}/\mbox{\small hr}$) is realized in the case of infinite lifetime, while with a finite lifetime, it grows much more slowly due to flux cancellation between the pairs of RFE samples.  In both cases, the flux remains balanced.  The latitudinal dependence of the field when using method 1 can also be seen.  See Sec.~\ref{sec:examples_1yr} and \citet{oft3} for more details about the effects of the RFE model choices on production runs.

\subsection{Multiple Realizations}
\label{sec:model_realizations}

When using phenomenological models like SFT that have a multitude of uncertainties including data quality, data-derived model parameters, model choices, multiple resolutions, etc., it is very important to try to identify and quantify the uncertainties (QU).  A key tool in QU is generating multiple realizations of the model solutions by varying the model parameters, leading to an ensemble of solutions.  Ideally, these ensembles should properly sample the span of the farthest reasonable boundaries in parameter space of the current model while keeping the number of realizations as low as possible.  

HipFT has been designed to compute many realizations of maps within a single run efficiently.  This makes it straight forward to set a range of model parameters to generate many realizations.  The python post processing scripts included with HipFT are also designed to process the multiple realizations together.  Currently, HipFT allows the user to specify an array of parameters for diffusion coefficient values, flow attenuation values, the model coefficients for DR and MF flow profiles, data assimilation latitude limits, $\mu$ limits, and weight exponents.  The mechanism to provide model parameters across realizations is modular, making it straight forward for users to add new cross-realization parameters.  In Sec.~\ref{sec:code} we show how the realizations are distributed across the computational units of a HipFT run, while in Sec.~\ref{sec:examples} we show an example run with multiple realizations.

\subsection{Analysis}
\label{sec:model_analysis}

There are a variety of derived quantities that are useful when analyzing full-Sun magnetic maps.  In HipFT we compute several of these quantities (described below) and output them at a chosen cadence in a text file per realization we refer to as a `run history'.  Python scripts are also provided to compute these quantities on maps that have already been written to disk, which can be used to generate history files on observational data products or other SFT models after converting their maps to the HipFT file and grid format.  

The derived quantities written to the history files are the total positive and negative flux given by 
\[
\int_{\Omega} B_r\,d\Omega, \qquad (B_r > 0) \qquad \mbox{and} \qquad \int_{\Omega} B_r\,d\Omega, \qquad (B_r < 0),
\]
respectively, where $\Omega=\sin\theta\,d\theta\,d\phi$, the total fluxes limited to the polar regions as defined by a user selected latitude cutoff, the area of the selected polar regions, the minimum, maximum, and minimum magnitude of $B_r$ over the whole map, and the strengths of the equatorial and axial dipole moments \citep{WangSheeley1991}.  The strength of equatorial dipole moment is calculated as
\begin{subequations}\label{eq:eqdipole1}
\begin{equation}
H(t) = \sqrt{h_1^2(t) + h_2^2(t)}, \tag{\ref{eq:eqdipole1}}
\end{equation}
where
\begin{align}
h_1(t) &= \dfrac{3}{4\pi}\,\int_{\Omega} B_r\,\sin\theta\,\cos\phi\,d\Omega, \label{eq:eqdipole1a} \\
h_2(t) &= \dfrac{3}{4\pi}\,\int_{\Omega} B_r\,\sin\theta\,\sin\phi\,d\Omega, \label{eq:eqdipole1b}
\end{align}
\end{subequations}
while the strength of the axial dipole moment is calculated as
\begin{equation}
D(t) = \dfrac{3}{4\pi}\,\int_{\Omega} B_r\,\cos\theta \; d\Omega.
\end{equation}
When running validation tests, we compute the difference between the HipFT solution and the true analytic solution using an absolute value version of the Hanna and Heinold (HH) metric \citep{HHbook}
\begin{equation}
\label{eq:hh}
\mbox{HH}_{||} = \sqrt{\dfrac{\sum_{i=1}^N\,|X_i-Y_i|^2}{\sum_{i=1}^N\,|X_i||Y_i|}},
\end{equation}
where $N$ is the total number of grid points, $X$ is the computed final map, and $Y$ is the analytic solution map.  The $\mbox{HH}_{||}$ metric has been shown to be better for testing numerical solutions than the more commonly used normalized root mean-square error \citep{HHgtNRMSD}.  

HipFT contains several post-processing scripts that utilize these analysis outputs, including using them to compute additional derived quantities as described in Sec.~\ref{sec:code_postproc}.

%%%%%%%%%%%%%%%%%%%%%%%%%%%%%%%%%%%%%%%%%%%%%%%%%%%%%%%%%%%%
%%%%%%%%%%%   NUMERICAL METHODS
%%%%%%%%%%%%%%%%%%%%%%%%%%%%%%%%%%%%%%%%%%%%%%%%%%%%%%%%%%%%

\section{Numerical Methods}
\label{sec:num}
HipFT uses a variety of numerical methods for each part of the integration of Eq.~\ref{eq:main_model}.  In this section, we describe the implemented schemes, starting with a description of the initial conditions for the tests that will be used to validate the schemes. We culminate with the validations shown in Sec.~\ref{sec:num_valid}.

\subsection{Test cases}
\label{sec:testcases}
We use two test initial conditions, each of which have analytic time-dependent solutions.  We can also set a constant angular velocity in $\phi$ corresponding to a full rotation, allowing us to test the advection schemes with any initial condition, including full-Sun magnetic maps.  To normalize the tests, each is run to an end time of $672$ hours (approximately one Carrington rotation) and, for analytic solutions, use a function amplitude of one.  The $\phi$ velocity for rigid rotation tests is set to $v_{\phi} = 1.80766\,\mbox{km/s}\,\sin\theta$.

The first test case is known as the "soccer ball" function\footnote{\url{https://www.chebfun.org/examples/sphere/SphereHeatConduction.html}}.  It is an analytic time-dependent solution for the diffusion operator on the surface of a sphere, which can be combined with a full rigid rotation $\phi$ velocity to test advection-diffusion schemes.  It is described as 
\begin{equation}
\label{eq:soccerball}
u(\theta,\phi,t)=e^{-42\,\nu\,t}\left(\,Y_6^0(\theta,\phi)+\sqrt{\dfrac{14}{11}}\,Y_6^5(\theta,\phi)\right),
\end{equation}
where $\nu$ is the chosen diffusion coefficient (in our tests we use $\nu=500\,\mbox{km}^2\!/\mbox{s}$), and $Y_l^m(\theta,\phi)$ are the tesseral spherical harmonics. To compute the spherical harmonics, we have implemented the First Modified Forward Column Recursion method of \citet{2002_legendre_poly_alg_holmes} based on the MATLAB code {\tt spherefun.sphharm} from the ChebFun package \citep{2014_ChebFun_Driscoll}.

The second test solution is a pair of modified Gaussians of opposite polarity described as
\begin{eqnarray}
\label{eq:testblob}
B_r(\theta,\phi,t) =
&-\dfrac{1}{\sin\theta}\exp\left[
-\dfrac{\left(\theta-\pi/4-v_{\theta}\,t\right)^2}{\sigma}
-\dfrac{\left(\phi-\pi/2-v_{\phi}\,t\right)^2}{\sigma}\right] 
\\
&+
\dfrac{1}{\sin\theta}\exp\left[
-\dfrac{\left(\theta-\pi/4-v_{\theta}\,t\right)^2}{\sigma}
-\dfrac{\left(\phi-3\pi/2-v_{\phi}\,t\right)^2}{\sigma}\right]. \notag
\end{eqnarray}
The $1/\sin\theta$ term in front ensures an exact solution in $\theta$ over time (see Appendix~\ref{a:thetasol} for details).  Although, in general, the term creates a divergence at the pole, our test has two equal and opposite polarities approaching the pole, which should cause the flux to cancel, mitigating the issue.  We use $\sigma = 0.03$ and set $v_{\phi}$ such that the Gaussian profiles travel a full rotation.  For $v_{\theta}$, we set its value such that the Gaussians travel an angular distance of $\Delta\theta=\pi/2$, making an end solution that should be the same as the initial solution flipped in latitude.  By setting (or not) the velocities, we can test advection in the $\phi$ and $\theta$ directions independently or together.

\begin{figure}[htb]
\centering
$\begin{array}{c}
\includegraphics[width=0.2\textwidth]{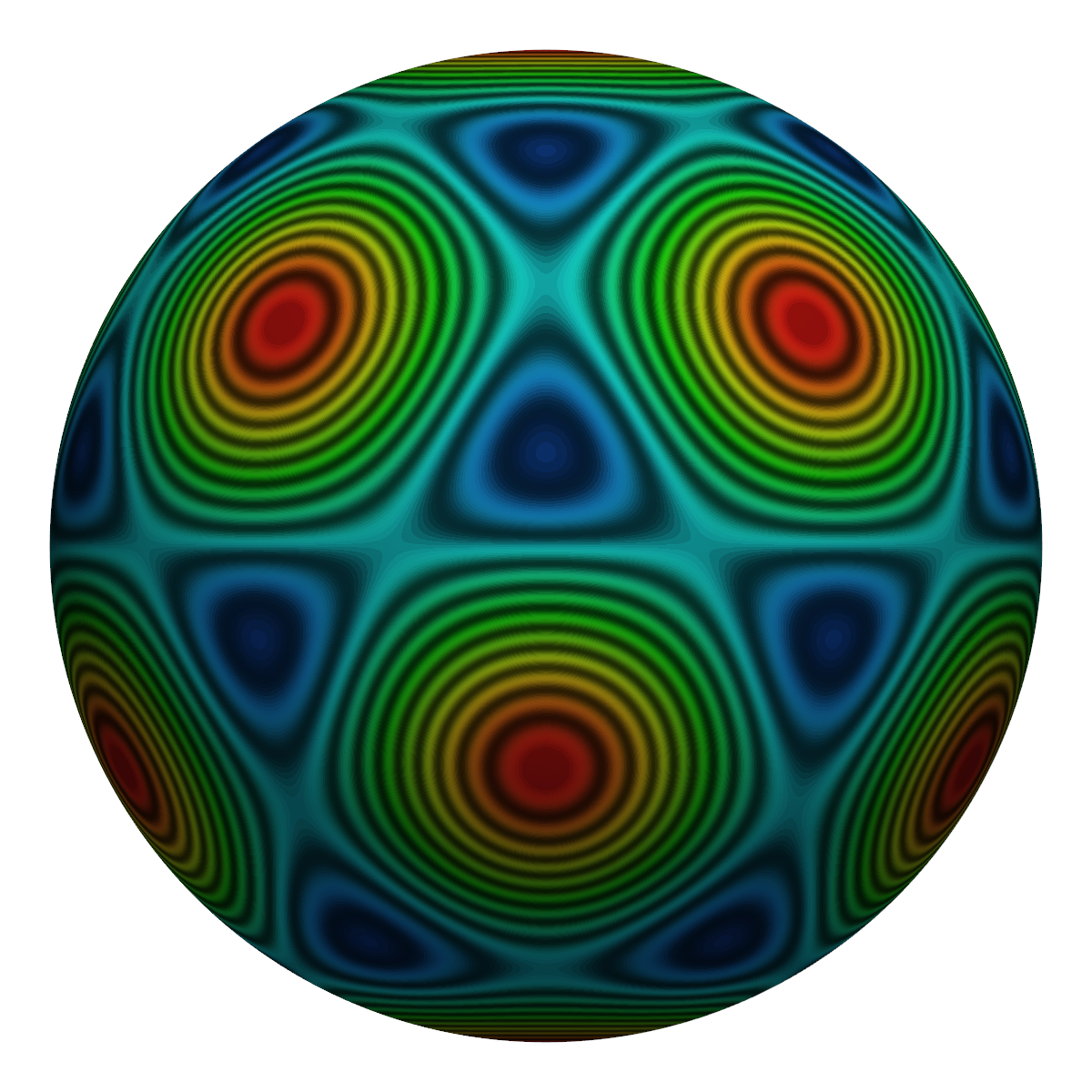}
\includegraphics[width=0.2\textwidth]{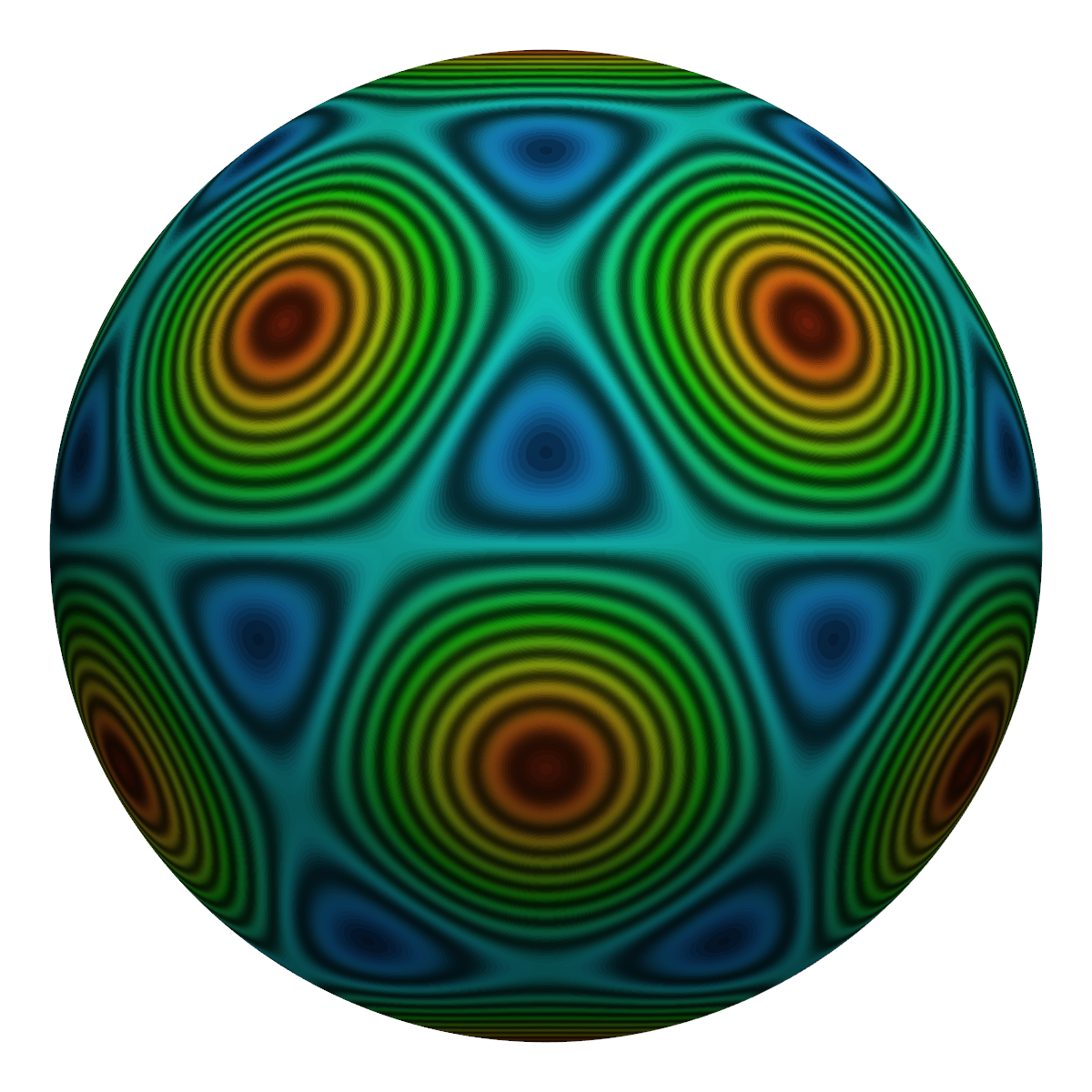}
\\
\includegraphics[width=0.2\textwidth]{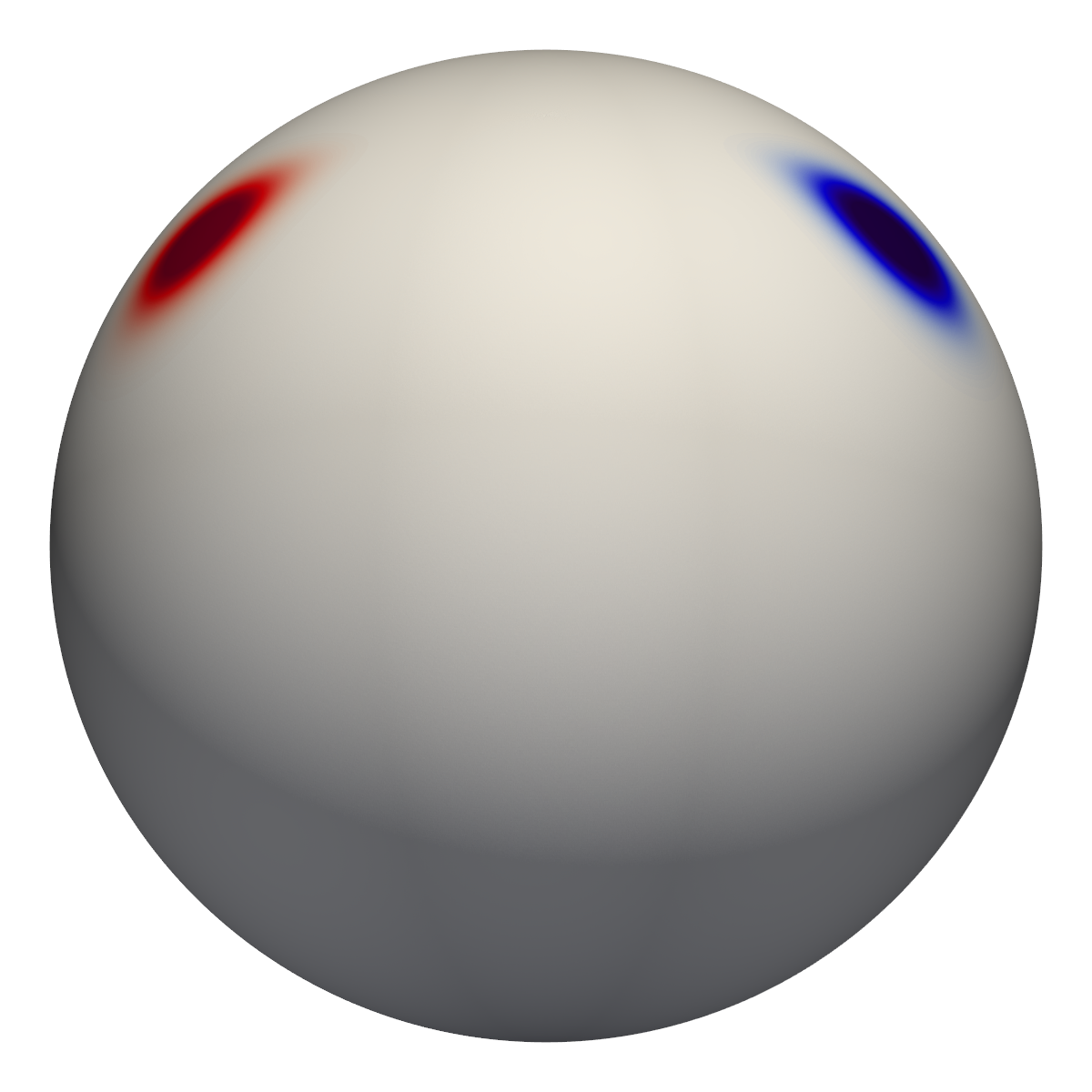}
\includegraphics[width=0.2\textwidth]{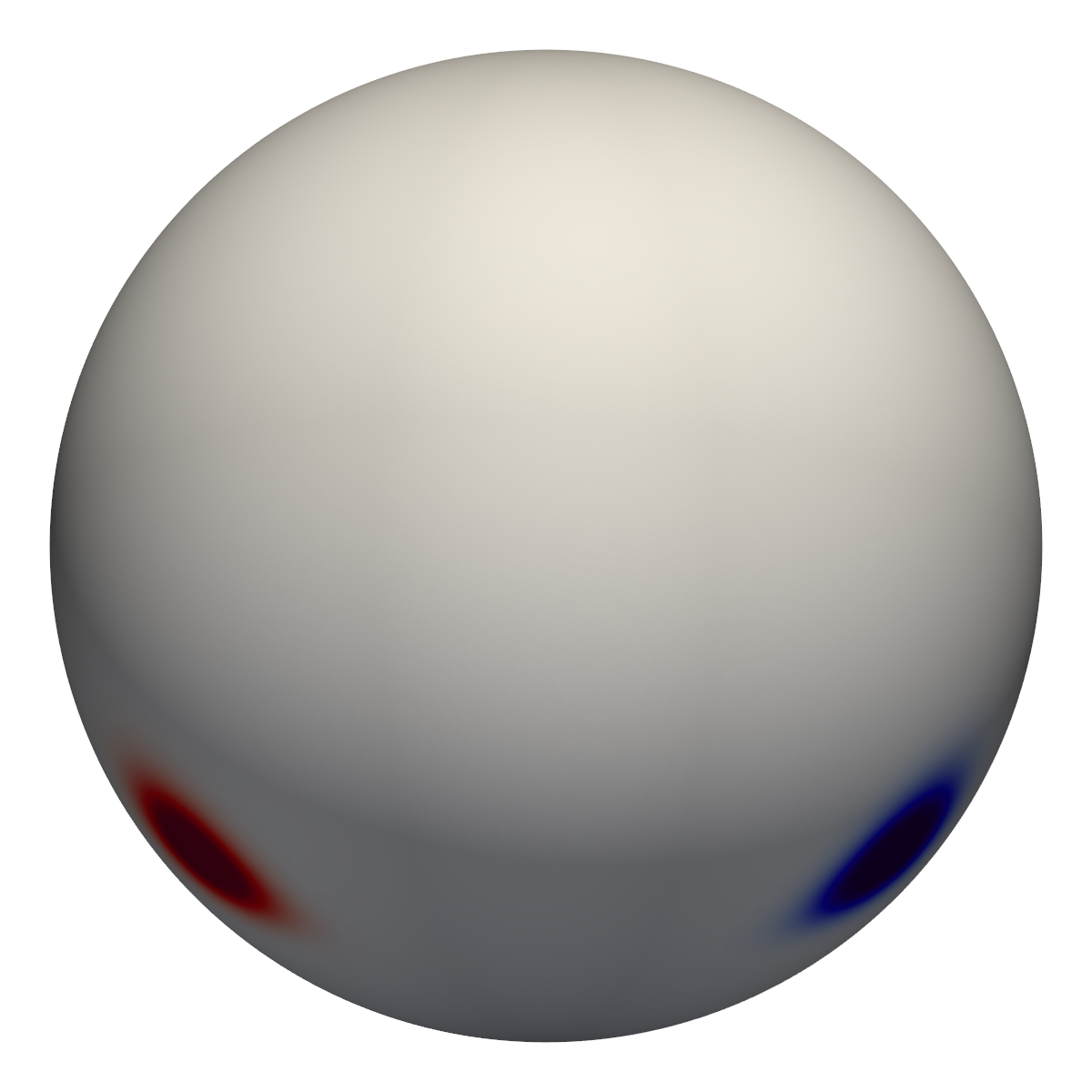}
\end{array}$
\caption{Test cases used for HipFT validations.  The top row is the soccer ball test of Eq.~\ref{eq:soccerball}, while the bottom row is the Gaussians test of Eq.~\ref{eq:testblob}.  For each row, the initial condition is shown on the left, and the solution after 672 hours is shown on the right (for the Gaussians test, the initial condition is shown at a $b_0$ angle of $30^{\circ}$, while the end solution is shown at a $b_0$ of $-30^{\circ}$).  The solutions are shown for the default resolution of $512 \times 1024$.\label{fig:testcases}} 
\end{figure}

The initial and analytic solution maps for the two test cases are shown in Fig.~\ref{fig:testcases}.
For all test cases, we use the $\mbox{HH}_{||}$ metric of Eq.~\ref{eq:hh} to compare the final computed output map with the analytic solution.  

\subsection{Grid}
\label{sec:num_grid}
\begin{figure}[htb]
\centering
\includegraphics[height=0.35\textwidth]{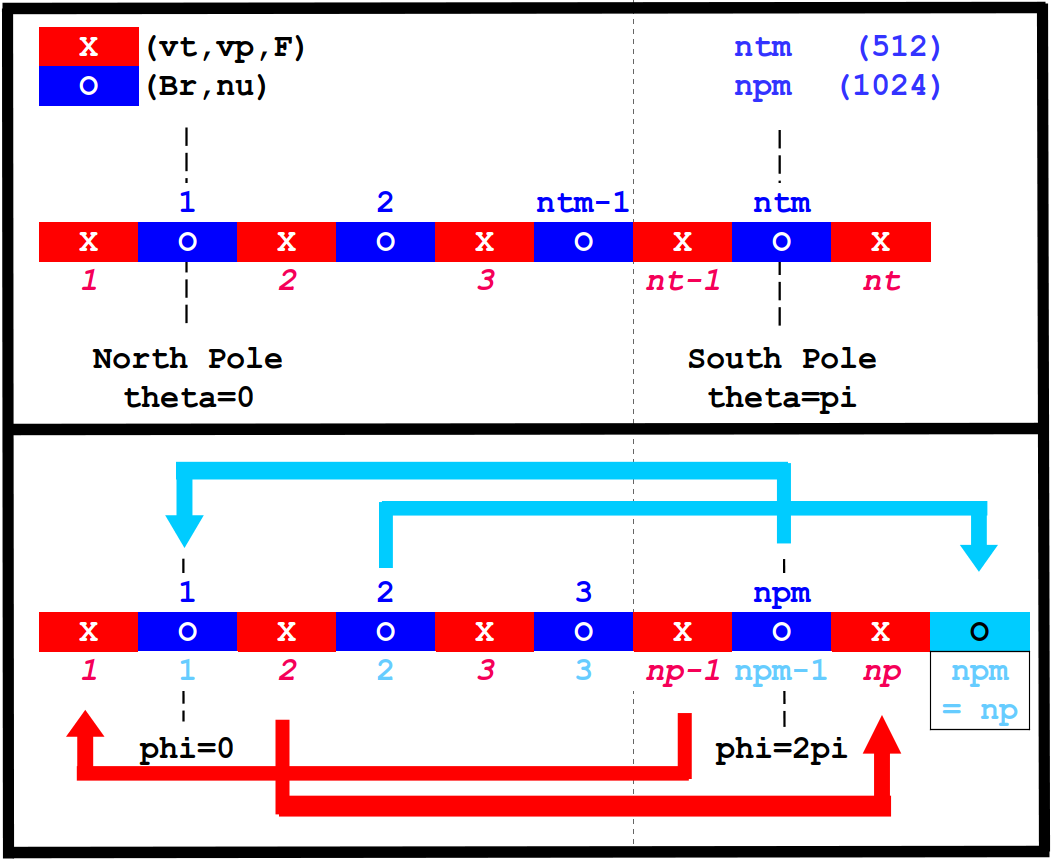} \qquad
\includegraphics[height=0.35\textwidth]{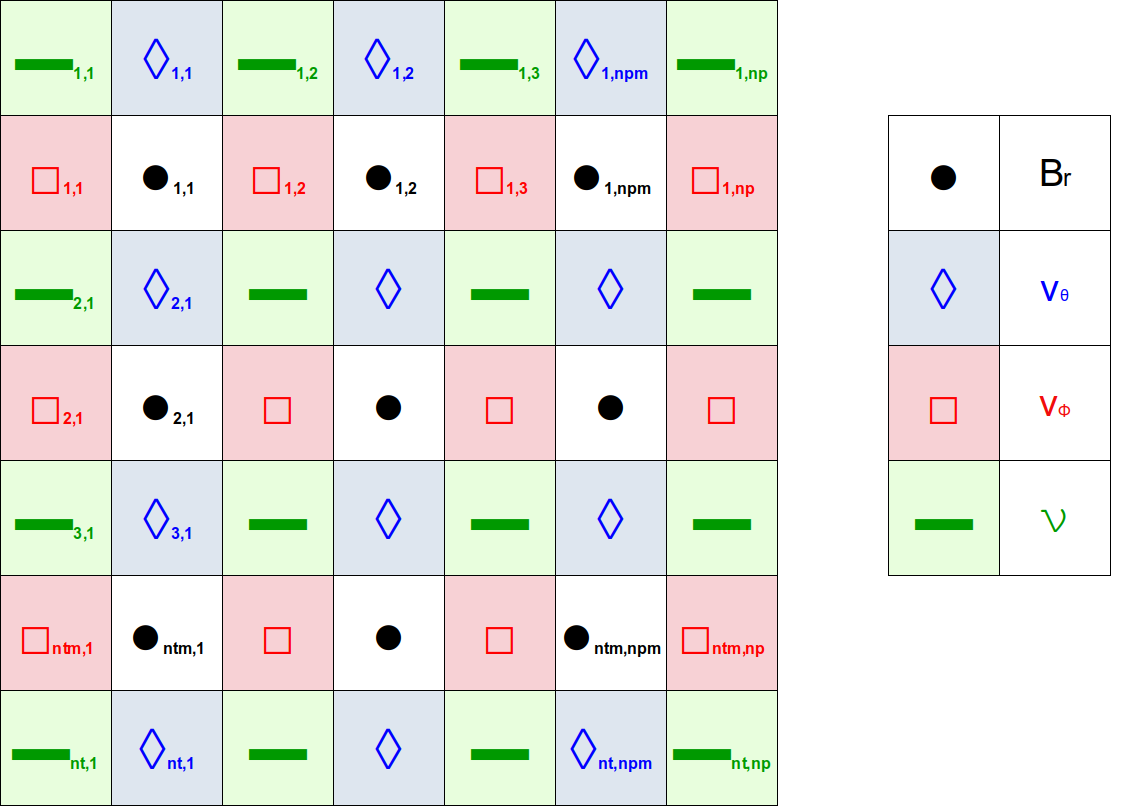}
\caption{Computational grid layout of HipFT. In the left image, the light blue $\phi$ grid limit is used within the code, while the dark blue limits are used for the input and output maps.  The two-point overlap and seam directions are indicated by the light blue arrows.   The staggering of the field, velocity components, and diffusivity are shown in the right image.
\label{fig:num_grid_details}}
\end{figure}

HipFT uses a logically-rectangular non-uniform spherical surface grid shown in Fig.~\ref{fig:num_grid_details}.  
The main grid (where $B_r$ resides) is defined over $\phi\in[0,2\pi]$ and $\theta\in[0,\pi]$, which includes a one-point overlap in the $\phi$ periodic boundary (i.e. both the $\phi=0$ and $\phi=2\pi$ values are in the files).  For internal calculations, an additional $\phi$ point is added, yielding a two-point overlap in the $\phi$ periodic boundary (shown in the lower left panel of Fig.~\ref{fig:num_grid_details} as the light blue square). The flow velocities and the advection fluxes (see Sec.~\ref{sec:num_advect_space}) are placed on `half' grids that are staggered to the $B_r$ field in their corresponding direction (as shown in the right panel of Fig.~\ref{fig:num_grid_details} as the blue and red shaded squares).  The diffusion viscosity (see Sec.~\ref{sec:num_diff}) is defined on the `half-half' grid (green shaded squares in the right panel of Fig.~\ref{fig:num_grid_details}).  

During the computation, when needed, the $\phi$ direction is `seamed' by copying the overlapping point values as shown by the arrows in the left panel of Fig.~\ref{fig:num_grid_details}.

For production runs of OFT, we currently use a default resolution of $\phi\,\times\,\theta$ = $1024 \times 512$.  The resolution in HipFT is set based on the initial input map, and any data assimilation and/or flow maps must match that resolution (the OFT repository contains tools needed to re-bin and process maps to any desired resolution and can be used prior to running HipFT).  Alternatively, the resolution can be set using input parameters for the case of a zero-value starting map or for validation test cases.
 
\subsection{Units}
\label{sec:num_units} 
 
HipFT internally uses spatial units of $R_{\odot}=6.96\times 10^{10}\,\mbox{cm/s}$ and time units of hours. It includes a module with preset constants to easily convert inputs into the internal code units.  For instance, to convert the flow velocities from \mbox{m/s} to $R_{\odot}/\mbox{hr}$ we have $\mbox{m\_s\_to\_rs\_hr}=5.172413793103448\times 10^{-6}$, and for converting the diffusivity $\nu$ from $\mbox{km}^2\!/\mbox{s}$ to $R_{\odot}^2/\mbox{hr}$ we have $\mbox{km2\_s\_to\_rs2\_hr}=7.43162901307967\times 10^{-9}$.

\subsection{Time steps}
\label{sec:num_dt} 

The time steps in HipFT are determined by a combination of stability limits, input/output times, and user choices.  At each step, the time step is initially set to the remaining time for the simulation.  It is then reduced through a series of checks which include the advection flow CFL stability condition (see Sec.~\ref{sec:num_advect_time}), the time to the next requested map output, the time to the next data assimilation, the time to the next input flow, and the time to the next random flux generation (depending on which features in the code are being used).  This method ensures that features like data assimilation and map output are performed exactly at the times proscribed, while also maintaining stable integration of the model.  Additionally, various user options (such as a minimum and maximum time step) are available for additional control.

\subsection{Operator splitting}
\label{sec:num_opsplit}

At each time step, the various components of Eq.~\ref{eq:main_model} are integrated separately (i.e. operator split) in the following sequence:
\begin{alignat}{2}
\label{num_opsplit_nostrang}
B_r^{*} &= A(B_r^n,\Delta t), \\
B_r^{**} &= D(B_r^*,\Delta t), \notag \\
B_r^{n+1} &= S(B_r^{**}), \notag
\end{alignat}
where $A$ is the advection term, $D$ is the diffusion term, $S$ is the source term, $n$ is the current step number, and $\Delta t$ is the current step's time step (the data assimilation occurs at the end of the advance).  This sequence of operator splitting can introduce a $O(\Delta t)$ error \citep{OPSPLIT}.  Another option is to use Strang splitting \citep{MacNamara2016} defined as 
\begin{alignat}{2}
\label{num_opsplit_strang}
B_r^{*}    &= S(B_r^{n},   \Delta t/2)  \\
B_r^{**}   &= A(B_r^{*},   \Delta t/2), \notag \\
B_r^{***}  &= D(B_r^{**},  \Delta t),   \notag \\
B_r^{****} &= A(B_r^{***}, \Delta t/2), \notag \\
B_r^{n+1}  &= S(B_r^{****},\Delta t/2), \notag
\end{alignat}
which has a lower $O(\Delta t^2)$ splitting error.  Using Strang splitting can be computationally more expensive due to the additional advection and source advances in each step.  However, this is only when using a constant time step (not set by the flow CFL).  When using the maximum allowed flow CFL time step, the CFL stable time step is now twice as large (due to the $\Delta t/2$ step sizes), cutting the total number of steps in half, negating the cost of the extra advection and source advances.  Moreover, if diffusion is active, there are now half as many diffusion advances (each with a time step twice as large).  Since the computation time of the super time stepping PTL scheme does not grow linearly with the time step (see Sec.~\ref{sec:num_diff_time}), this can make the overall run faster, especially since diffusion can be a large portion of the total run time.  Therefore, the time cost (or savings) of using Strang splitting is problem specific.  In terms of accuracy, we have found that often the spatial errors dominate the total error, so Strang splitting does not significantly improve the accuracy overall.  Therefore, the choice to use Strang splitting in HipFT is typically based on computational cost.  As an example, in Table~\ref{table:strang}, we show the results of running the $\phi$-rotation soccer ball advection-diffusion test case of Eq.~\ref{eq:soccerball} with and without Strang splitting, using both a small fixed time step and a time step set by the stable flow CFL.
\begin{table}[htb]
\begin{center}
\begin{tabular}{|l|r|r|r|}
\multicolumn{4}{c}{Fixed Time Step} \\
\hline 
          & $\Delta t$ & Wall Clock & Error\\
\cline{2-4}
No Strang  & 0.50 hr & 23.8 sec & $4.4\times 10^{-5}$\\
Strang     & 0.50 hr & 26.7 sec & $4.0\times 10^{-5}$\\
\hline 
\multicolumn{4}{c}{Maximum Stable Flow CFL Time Step} \\
\hline 
          & $\Delta t$ & Wall Clock & Error\\
\cline{2-4}   
No Strang   & 0.62 hr & 21.1 sec & $4.8\times 10^{-5}$ \\       
Strang       & 1.25 hr & 15.6 sec & $4.8\times 10^{-5}$\\ \hline 
\end{tabular}
\caption{Effects of Strang splitting in HipFT for the $\phi$-rotation diffusion-advection test case of Eq.~\ref{eq:soccerball}.  The run was performed with and without Strang splitting using a fixed time step, and using the maximum allowed stable flow CFL time step. The wall clock time (run on an NVIDIA RTX 3090Ti GPU) and the final $\mbox{HH}_{||}$ solution error is shown.  Due to the reduction in the number of diffusion advances, using Strang splitting is faster when using the maximum stable time step, but slower when using a fixed time step.}
\label{table:strang}
\end{center}
\end{table}
We see that with a fixed time step, the run is slower using Strang splitting, but when using the maximum allowed CFL time step, the run is faster.  In all cases, the total errors are comparable.  As the advantage of using Strang splitting is problem specific, it is disabled by default in HipFT.

\subsection{Advection}
\label{sec:num_advect}

\subsubsection{Advection spatial schemes}
\label{sec:num_advect_space}

To {\bf discretize} the advection term in Eq.~\ref{eq:main_model}, we use a finite-difference of the form
\begin{align*}
\left[ \nabla_{s}\cdot\,(B_r\,{\bf v}_{s})\right]_{(j,k)} &\approx \dfrac{\sin\theta_{j+\frac{1}{2}}\,F_{\theta:j+\frac{1}{2},k} - \sin\theta_{j-\frac{1}{2}}\,F_{\theta,j-\frac{1}{2},k}}{\sin\theta_j\,\Delta\theta_j} + \dfrac{F_{\phi:j,k+\frac{1}{2}} - F_{\phi:j,k-\frac{1}{2}}}{\sin\theta_j\,\Delta\phi_k},
\end{align*}
where $F_{\theta}$ and $F_{\phi}$ are computed either with an Upwinding/Central or WENO3-CS(h) scheme as described below.

\noindent {\bf Upwinding (UW) and Central Difference (CD)} \newline

Here, we set
\[
F_{i-1/2} = v_{i-\frac{1}{2}}\,\dfrac{1}{2}\left[(1 - \mbox{uw})\,B_{r:i}+(1 + \mbox{uw})\,B_{r:i-1}\right],
\]
where
\[
\mbox{uw} = \alpha_{\mbox{\scriptsize uw}}\,\mbox{sign}(v_{i-\frac{1}{2}}),
\]
and $\alpha_{\mbox{\scriptsize uw}}\in[0.5,1]$. Setting $\alpha_{\mbox{\scriptsize uw}}=0.5$ makes the scheme a central difference scheme which is second-order accurate, but when combined with explicit time stepping schemes, is unconditionally unstable \citep{numpdebook}.  This can be countered by adding enough diffusion to the model to stabilize the method (as is done in the AFT model).  The default value in HipFT when using this method is $\alpha_{\mbox{\scriptsize uw}}=1$ (upwind) which results in a first-order accurate discretization ($O(\Delta \theta)$,$O(\Delta \phi)$) and adds a significant amount of diffusivity to the solution.

\noindent {\bf Weighted Essentially Non-Oscillatory, third-order scheme with grid-step epsilon (WENO3-CS(h))} \newline

In order to have a more accurate advection integration and avoid the implicit diffusion added by the upwind scheme, we have implemented a Weighted Essentially Non-Oscillatory (WENO) scheme \citep{liu1994weighted}.  We chose to implement a WENO3 scheme as a balance of solution quality and complexity.  Special consideration is needed due to the use of non-homogeneous velocities and a non-uniform grid \citep{smit2005grid,shadab2019fifth}.  For the flux, we use a localized Lax-Friedrichs (LLF) spitting \citep{shu2009high,li2006weno}.  The use of spherical coordinates poses a challenge for WENO implementations, specifically in the theta direction.  Here, we treat the theta direction in the same manner as a Cartesian dimension which is a commonly used option even though it may potentially reduce the accuracy \citep{li2006weno, mignone2014high}.  While WENO3 schemes are formally 3rd-order accurate in smooth solution regions, they can be less accurate in the presence of critical points and discontinuities without adding additional grid cells to the scheme \citep{Baeza2020}.  However, \citet{Cravero2015} showed that by using the cell spacing for the required `small' constant $\epsilon_w$ (used to avoid a division by zero) in the formulation, the WENO3 scheme can retain its 3rd-order accuracy even close to local extrema on nonuniform grids.  We denote this scheme as WENO3-CS(h) and is the one implemented in HipFT.  While using other values of $\epsilon_w$ can make the standard WENO3 scheme more accurate, knowing in advance what value is best is difficult, making the WENO3-CS(h) scheme easier to use and more robust.  A full description of the scheme is given in Appendix~\ref{a:weno3}, where in Fig.~\ref{fig:wenoeps_vs_wenoh}, the convergence for the Upwind, WENO3 with a constant $\epsilon_w$, and WENO3-CS(h) schemes are compared.

\begin{figure}[htb]
\centering
$\begin{array}{c}
\includegraphics[width=0.25\textwidth]{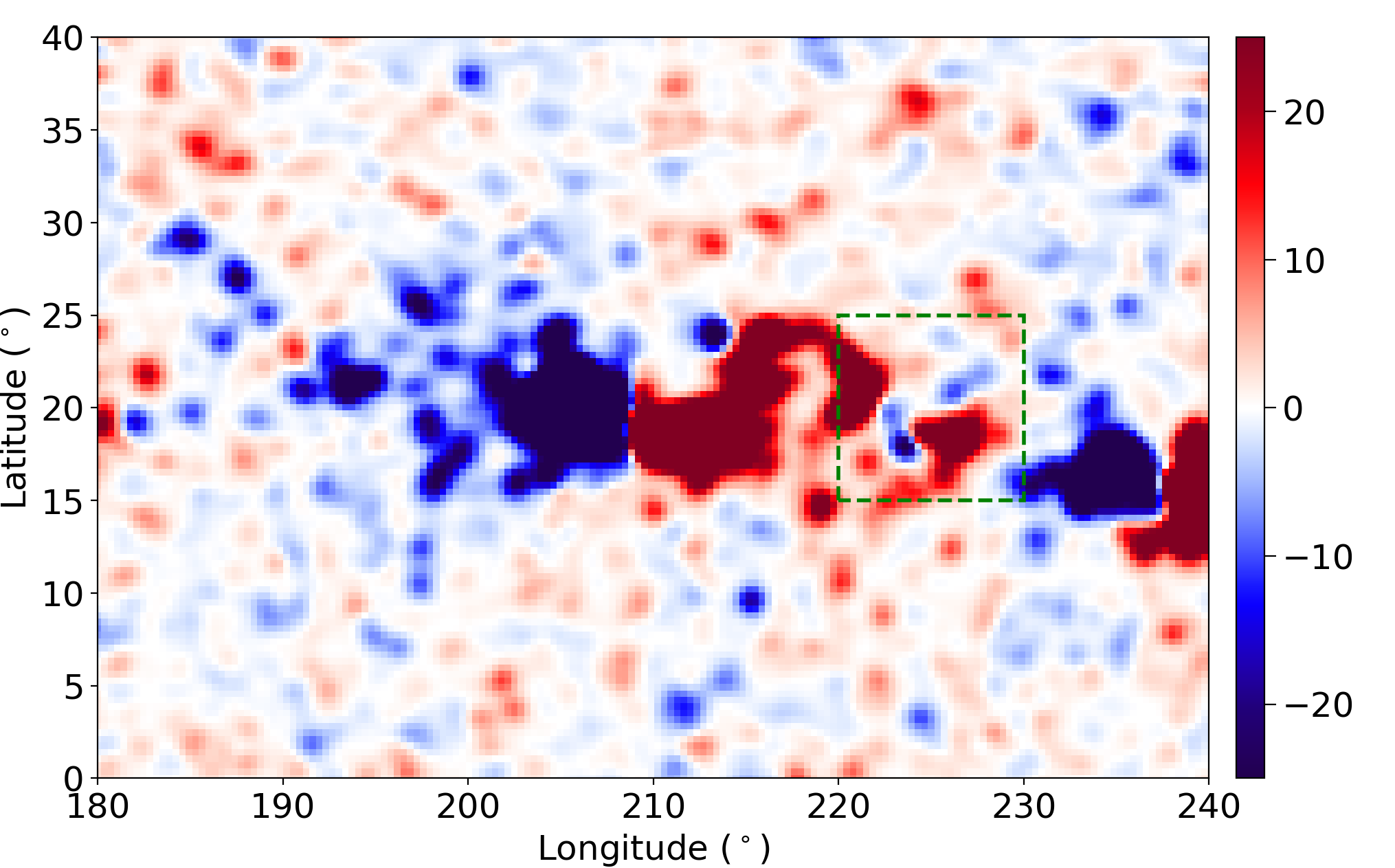}
\\
\includegraphics[width=0.25\textwidth]{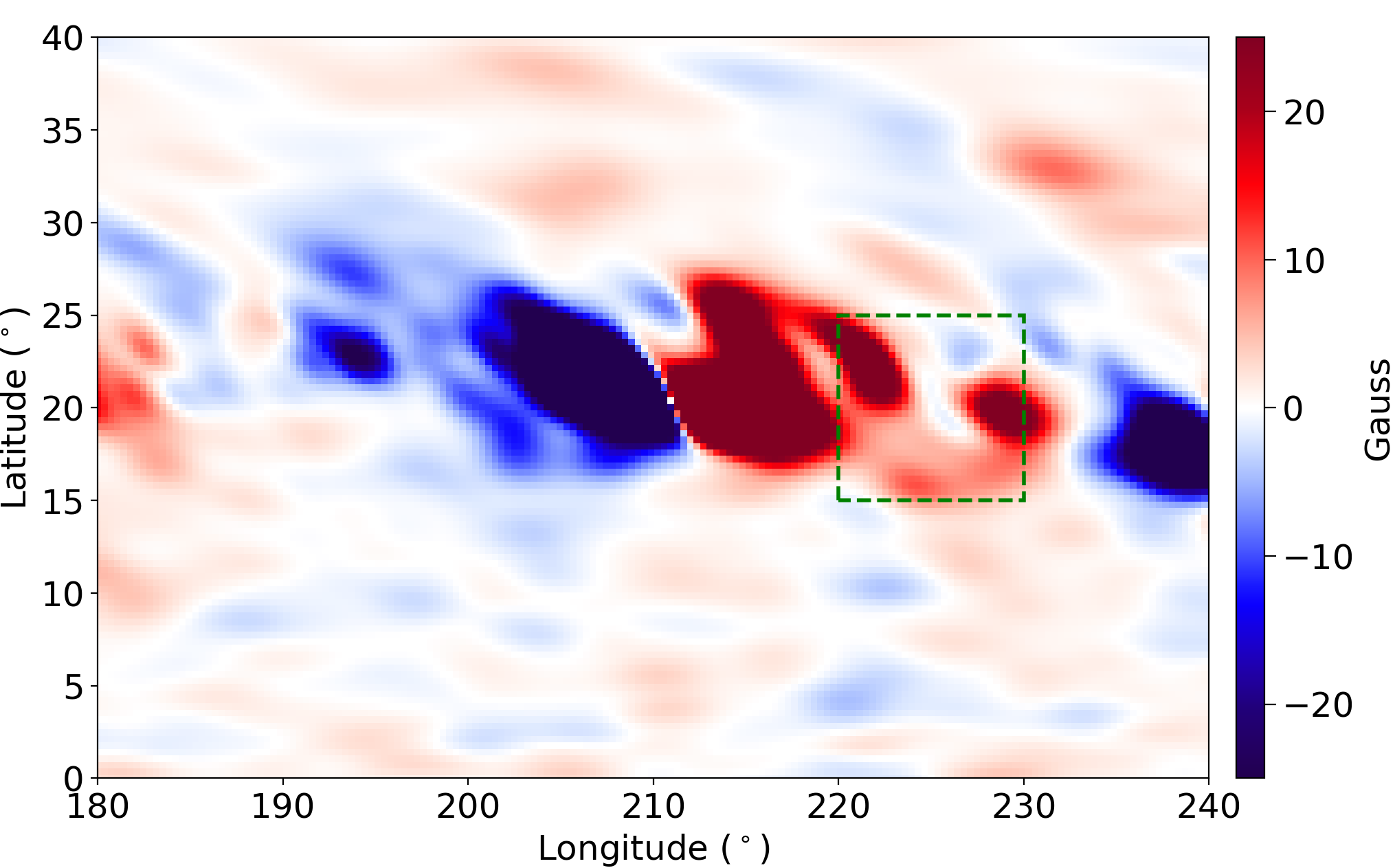}
\includegraphics[width=0.25\textwidth]{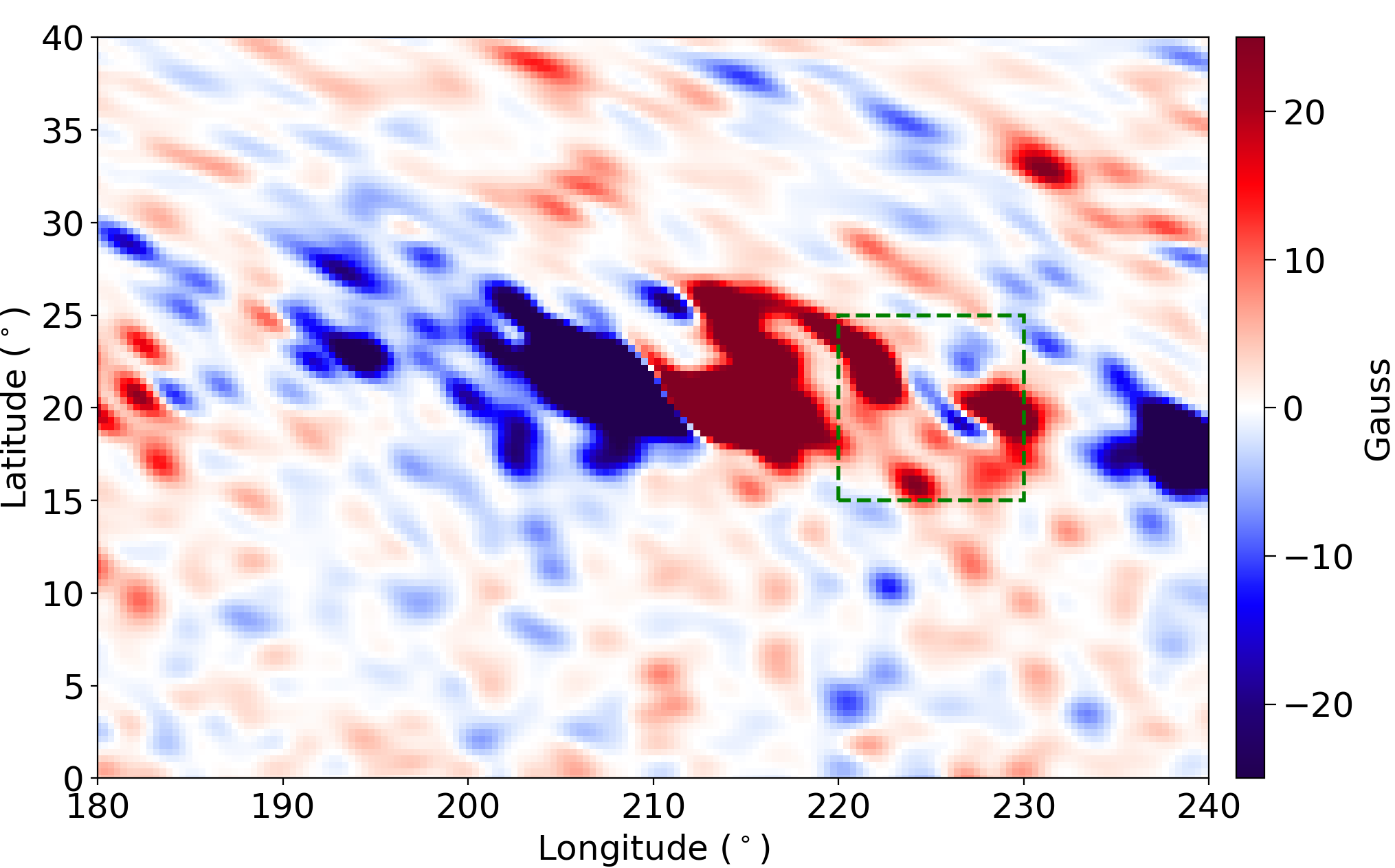}
\end{array}$
\caption{Qualitative comparison of the Upwind and WENO3-CS(h) scheme.  An initial magnetic map (top) was integrated in HipFT for 672 hours with default analytic differential and meridional flows (no diffusion) with the upwind scheme (bottom left) and the WENO3-CS(h) (bottom right).  A zoomed-in portion of the maps is shown, highlighting how the upwind scheme is much more diffusive than the WENO3-CS(h) scheme.  This can cause qualitative structural changes, such as the elimination of the small parasitic polarity highlighted in the green square. \label{fig:wenoeps_vs_wenoh2}} 
\end{figure}

In Fig.~\ref{fig:wenoeps_vs_wenoh2} we show the qualitative difference between the upwind and WENO3-CS(h) schemes in a realistic case of applying meridional and differential rotation flows to a magnetic map.
We see that the upwind scheme is much more diffusive than the WENO3-CS(h) scheme, leading to the washing out of detailed structures, while the WENO3-CS(h) retains them.

\noindent {\bf Boundary conditions} \newline

For transport on a spherical surface, there are no physical boundaries, but mathematically, boundaries appear because of the periodicity in $\phi$ and the geometric singularity at the poles in $\theta$. In the $\phi$ direction, we invoke a periodic boundary condition over the 2-cell overlap described in Sec.~\ref{sec:num_grid}.  Since the grid is not split across the $\theta$ or $\phi$ domain over MPI ranks (see Sec.~\ref{sec:code}), the periodic cells are readily accessible without needing inter-process communication.

In $\theta$, the poles are handled specially.  Since the WENO3-CS(h) scheme has a 5-point stencil, we use the upwind scheme at the cells adjacent to the poles.  Even though this can reduce the accuracy of the scheme in the $\theta$ direction, is it a simple solution that typically does not degrade the accuracy of solutions with little structure at the pole.  For the poles themselves, we utilize a finite-volume approach described in Appendix~\ref{a:polebc}. 

\subsubsection{Advection temporal schemes}
\label{sec:num_advect_time}

\noindent {\bf Forward-Euler (FE)} \newline
The forward Euler scheme is first-order accurate in time ($O(\Delta t)$) and described as
\begin{equation}
\label{eq:adv_euler}
B_r^{n+1} = B_r^n - \Delta t\,F_A(B_r^n)
\end{equation}
where
\[
F_A(B_r^n) = \nabla_{s}\cdot\,({\bf v}_{s}^n\,B_r^n).
\]
To ensure stability, the time step must be bound by the CFL condition, which in this case is approximated by 
\begin{equation}
\label{eq:advection_timestep}
\Delta t < \frac{1}{2}\,\left[ \frac{|v_{\theta}|}{\Delta \theta} + \frac{|v_{\phi}|}{\sin\theta\,\Delta\phi}  \right]^{-1},
\end{equation}
where the velocities are taken from the maximum of the flanking staggered values in each direction.  We also apply a `safety factor' of $\sim 0.9$ to the limit.

\noindent {\bf Strong Stability Preserving Runge-Kutta (SSPRK)} \newline
When using the WENO3-CS(h) scheme from Sec.~\ref{sec:num_advect_space}, the use of a higher-order time-stepping scheme is required for stability, and the use of a total value diminishing (TVD) scheme is typically used for its additional desirable properties \citep{weno_rk}.  HipFT provides two TVD scheme options: the third-order, three stage, strong stability preserving Runge-Kutta (SSPRK(3,3)) scheme and the third-order, four stage SSPRK(4,3) scheme \citep{Spiteri2002,ssprkbook}.  The SSPRK(3,3) (also know as RK3-TVD) and SSPRK(4,3) are described as
\begin{equation}
\begin{array}{c}
\hline
\mbox{SSPRK(3,3)} \\
\hline
\begin{array}{c}
f^{(1)} = f^n - \Delta t\,F_A(f^n) \\
f^{(2)} = \frac{3}{4}\,f^n + \frac{1}{4}\,f^{(1)} - \frac{1}{4}\,\Delta t\,F_A(f^{(1)}) \\
f^{n+1} = \frac{1}{3}\,f^n + \frac{2}{3}\,f^{(2)} - \frac{2}{3}\,\Delta t\,F_A(f^{(2)})
\end{array}
\\
\hline
\mbox{SSPRK(4,3)} \\
\hline
\begin{array}{c}
f^{(1)} = f^n - \frac{1}{2}\,\Delta t\,F_A(f^n) \\
f^{(2)} = f^{(1)} - \frac{1}{2}\,\Delta t\,F_A(f^{(1)}) \\
f^{(3)} = \frac{2}{3}\,f^n + \frac{1}{3}\,f^{(2)} - \frac{1}{6}\,\Delta t\,F_A(f^{(2)}) \\
f^{n+1} = f^{(3)} -  \frac{1}{2}\,\Delta t\,F_A(f^{(3)})
\end{array}
\end{array}
\end{equation}
The SSPRK(4,3) adds an additional evaluation stage, while remaining third-order accurate.  Its advantage lies in that, while the stable time step limit for the SSPRK(3,3) is the same as Eq.~\ref{eq:advection_timestep} used for the upwind scheme, the limit for the SSPRK(4,3) scheme is twice as large \citep{ssprkbook}, i.e.
\begin{equation}
\label{eq:ssprk43_dt}
\Delta t < \left[ \frac{|v_{\theta}|}{\Delta \theta} + \frac{|v_{\phi}|}{\sin\theta\,\Delta\phi}  \right]^{-1}.
\end{equation}
Therefore, adding $\sim30\%$ more computation is off-set by needing $50\%$ less time steps (if the overall calculation time step is being set by the flows).  This advantage is increased when combined with diffusion, as the diffusion term is expensive to calculate per time step.  For example, running the soccer ball diffusion test case of Eq.~\ref{eq:soccerball} at $512 \times 1024$ resolution with rigid rotation in $\phi$, using the maximum allowed time step with SSPRK(3,3) takes $38.5$ seconds on an RTX 3090Ti GPU, while for SSPRK(4,3), it takes $26.7$ seconds ($30\%$ faster). The $\mbox{HH}_{||}$ (see Eq.~\ref{eq:hh}) between the analytic solution and the simulation for the two runs is nearly the same ($4.3\times10^{-5}$ for SSPRK(3,3)  and $4.8\times10^{-5}$ for SSPRK(4,3)) showing that the speedup of the SSPRK(4,3) scheme does not significantly effect the solution accuracy.  Therefore, HipFT uses the SSPRK(4,3) scheme by default.

\subsection{Diffusion}
\label{sec:num_diff}

\subsubsection{Diffusion spatial schemes}
\label{sec:num_diff_space}

To {\bf discretize} the diffusion term in Eq.~\ref{eq:main_model}, we use the second-order central finite difference
\begin{align}
\label{eq:diffusion_op}
&\nabla_{s} \cdot (\nu({\theta,\phi})\,\nabla_{s}\,B_r) 
\approx 
\\
&\dfrac{1}{\sin\theta_j\,\Delta\theta_j}
\left[
\frac{\nu_{j+\frac{1}{2},k+\frac{1}{2}} + \nu_{j+\frac{1}{2},k-\frac{1}{2}}}{2}\,\sin\theta_{j+\frac{1}{2}}\,\dfrac{B_{r:j+1,k}-B_{r:j,k}}{\Delta\theta_{j+\frac{1}{2}}}
\right. \notag
\\
&\qquad \qquad -
\left.
\frac{\nu_{j-\frac{1}{2},k+\frac{1}{2}} + \nu_{j-\frac{1}{2},k-\frac{1}{2}}}{2}\,\sin\theta_{j-\frac{1}{2}}\,\dfrac{B_{r:j,k}-B_{r:j-1,k}}{\Delta\theta_{j-\frac{1}{2}}}
\right] \notag
\\
&
+\dfrac{1}{\sin^2\theta_j\,\Delta\phi_k}
\left[
\frac{\nu_{j+\frac{1}{2},k+\frac{1}{2}} + \nu_{j-\frac{1}{2},k+\frac{1}{2}}}{2}\,\dfrac{B_{r:j,k+1}-B_{r:j,k}}{\Delta\phi_{k+\frac{1}{2}}} 
\right. \notag
\\
&\qquad \qquad  \qquad -
\left.
\frac{\nu_{j+\frac{1}{2},k-\frac{1}{2}} + \nu_{j-\frac{1}{2},k-\frac{1}{2}}}{2}\,\dfrac{B_{r:j,k}-B_{r:j,k-1}}{\Delta\phi_{k-\frac{1}{2}}}
\right], \notag
\end{align}
where since the diffusivity $\nu$ is located on a staggered grid in both the $\theta$ and $\phi$ direction, it needs to be averaged to the correct location in the stencil.  
In order to apply the operator in a simple and efficient manner, and to allow the easy computation of the stable Euler time step size limit (see Sec.~\ref{sec:num_diff_time}), we represent the operator as applied to the inner cells (without boundary conditions) as a symmetric self-adjoint sparse matrix in a custom DIA format \citep{DIACSR}, where the matrix coefficients are given by
\begin{equation}
\label{eq:diffusion_matrix}
\begin{array}{|c|c|}
\hline
a_{j,k-1} & 
\dfrac{\nu_{j+\frac{1}{2},k-\frac{1}{2}} + \nu_{j-\frac{1}{2},k-\frac{1}{2}}}{2\,\Delta\phi_{k-\frac{1}{2}}\,\Delta\phi_{k}\,\sin^2\theta_{j}}
\\
\hline
a_{j-1,k} &
\dfrac{(\nu_{j-\frac{1}{2},k+\frac{1}{2}} + \nu_{j-\frac{1}{2},k-\frac{1}{2}})\,\sin\theta_{j-\frac{1}{2}}}{2\,\Delta\theta_{j-\frac{1}{2}}\,\Delta\theta_{j}\,\sin\theta_{j}}
\\
\hline
a_{j,k} &
-\left[a_{j-1,k}+a_{j+1,k}+a_{j,k-1}+a_{j,k+1}\right]
\\
\hline
a_{j+1,k} &
\dfrac{(\nu_{j+\frac{1}{2},k+\frac{1}{2}} + \nu_{j+\frac{1}{2},k-\frac{1}{2}})\,\sin\theta_{j+\frac{1}{2}}}{2\,\Delta\theta_{j+\frac{1}{2}}\,\Delta\theta_{j}\,\sin\theta_{j}}
\\
\hline
a_{j,k+1} &
\dfrac{\nu_{j+\frac{1}{2},k+\frac{1}{2}} + \nu_{j-\frac{1}{2},k+\frac{1}{2}}}{2\,\Delta\phi_{k+\frac{1}{2}}\,\Delta\phi_{k}\,\sin^2\theta_{j}}
\\
\hline
\end{array}
\end{equation}

\noindent {\bf Boundary conditions} \newline
In the $\phi$ direction, we invoke a periodic boundary condition over the 2-cell overlap as described in Sec.~\ref{sec:num_advect_space}.  For the poles, we utilize the same finite-volume approach used in the advection scheme as described in Appendix~\ref{a:polebc}. 

\subsubsection{Diffusion temporal schemes}
\label{sec:num_diff_time}

A major difficulty in integrating diffusion operators explicitly is the restrictive stable time step, making the computation expensive and often impractical.  One way to avoid this is to use implicit methods, such as backward Euler scheme solved with iterative Krylov solvers (e.g. the {\bf preconditioned} conjugate gradient (PCG) method) \citep{IterativeMethods_SAAD_Book} and/or multi-grid methods \citep{2000_Briggs_Multigrid}.  However, these methods can be complicated to implement and parallelize efficiently (especially on GPUs).  Additionally, when used in cases where the time step is very large compared to the explicit Euler time step limit, they can become too inaccurate for production use \citep{Dawes2021}.  

Extended stability Runge-Kutta methods (also known as `Super Time Stepping' (STS)) are a class of methods that are explicit and unconditionally stable \citep{verwer1996explicit}.  They are straight-forward to implement and easy to parallelize, while being competitive in computational performance to implicit methods.  For example, in \citet{ASTRONUM16}, we compared the solution and performance of the second-order RKL2 scheme \citep{RKL2_2014} compared to a PCG solver for a thermodynamic MHD model and found that the RKL2 method had equal or better performance to the PCG solvers, while yielding overall similar solution results.  However, some solution oscillations were found in regions of high gradients in small grid cells (where the explicit Euler time step was very small compared to the overall time step).  In \citet{Caplan2024}, a practical time step limit (PTL) was developed to be able to achieve much less oscillatory solutions for both STS (in that case the RKG2(3/2) scheme \citep{RKG2B}) and PCG methods for large time steps.  It was found that using the PTL with RKG2(3/2) has similar performance to the PCG method, but yielded better solutions.    

HipFT implements both RKL2 and RKG2(3/2) schemes for the diffusion operator (the explicit forward Euler scheme is also included for reference and testing).  See Section 3 and Appendix B of \citet{ASTRONUM16} for details on the implementation of RKL2, and Appendix A of \citet{Caplan2024} for RKG2(3/2).  Both schemes are combined with the PTL sub-cycling procedure described in \citet{Caplan2024}.  In practice, the PTL cycling is typically not triggered for most use cases of HipFT, with the notable exception of running HipFT as a map smoother (see Sec.~\ref{sec:examples_smooth}).

The RKL2 and RKG2(3/2) yield similar accuracy and computational efficiency. For example, running the test case 2 of Eq.~\ref{eq:soccerball} with diffusion only (no flows) using the RKL2 scheme took 0.45 seconds for the diffusion advance (on an NVIDIA RTX 3090Ti GPU) with a final $\mbox{HH}_{||}$ error of $8.0\times 10^{-5}$, while using the RKG2(3/2) scheme took 0.55 seconds with a final $\mbox{HH}_{||}$ error of $6.3\times 10^{-5}$.  So in this case, the RKL2 is a little faster than RKG2 but also a little less accurate.  However, here the PTL condition was not activated.  In some cases, the RKG2(3/2) scheme could activate less PTL cycles than RKL2, counterbalancing the performance differences.  For reference, the same run using explicit forward Euler took 320 seconds with a final error of $1.3\times 10^{-5}$.  The error is less due to the much smaller time step ($1.8\times 10^{-4}$ hr) where the STS methods ran in one full 672 hr time step.  In HipFT, the RKG2(3/2) scheme is used by default.

\subsection{Validation of default methods}
\label{sec:num_valid}
\begin{figure}[htb]
\centering
\includegraphics[width=0.25\textwidth]{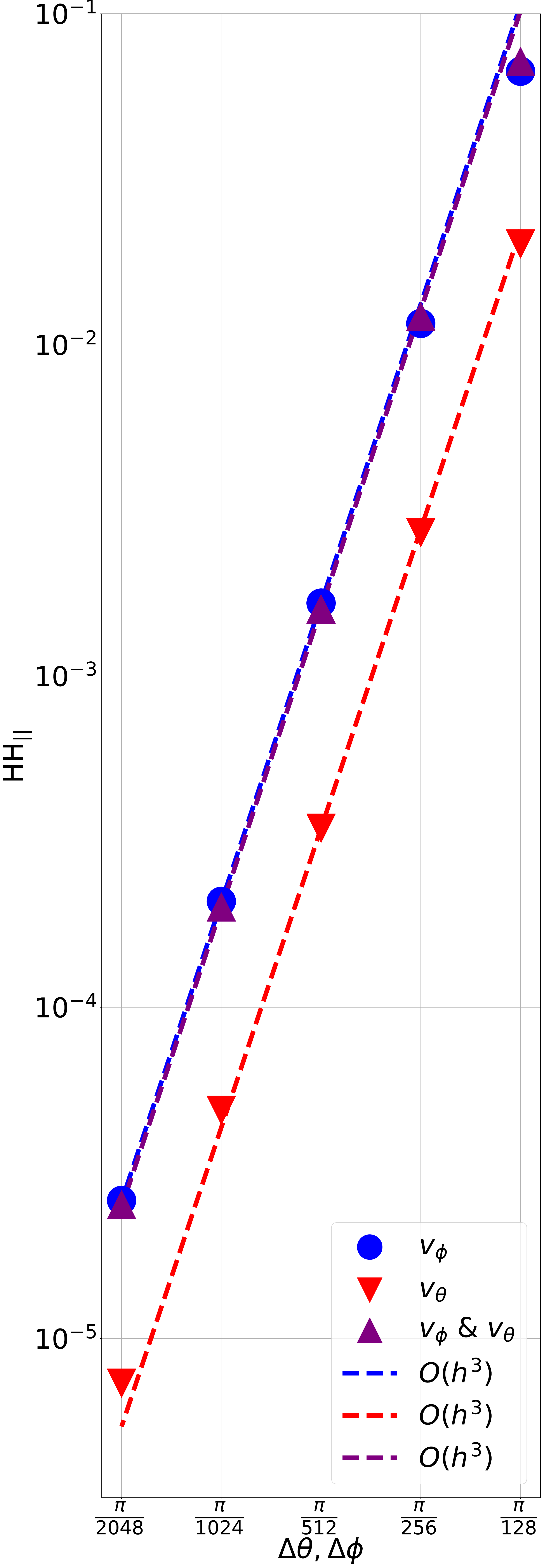}
\caption{Convergence of test case 2 (Eq.~\ref{eq:testblob}) using the default numerical schemes of HipFT.  The blue marks are for a $\phi$-rotation with $v_{\theta}=0$, the red marks are for a constant $v_{\theta}$, with $v_{\phi}=0$, and the purple marks are for runs using both $v_{\phi}$ and $v_{\theta}$.  For runs with a zero-value velocity component, the resolution of the zero-velocity direction is set to $512$ for $n_{\theta}$ and 1024 for $n_{\phi}$.  The third-order convergence of the WENO3-CS(h) scheme can be seen in all three runs. \label{fig:valid_advect}}
\end{figure}

Here, we show validations of the numerical schemes in HipFT using the test cases described in Sec.~\ref{sec:testcases}.  We focus on the default/recommended methods chosen based on the discussion and tests described in the previous sections.  The default advection method is the CS-WENO3(h) spatial scheme of Sec.~\ref{sec:num_advect_space} combined with the SSPRK(4,3) temporal scheme from Sec.~\ref{sec:num_advect_time}.  In Fig.~\ref{fig:valid_advect}, we show convergence results for running the test case 1 of Eq.~\ref{eq:testblob} for a resolutions spanning $n_{\phi}=256$ to $n_{\phi}=4096$.  For each resolution, we perform runs with a constant angular velocity in $\phi$, a constant velocity in $\theta$, and with both velocities active.  
We see that the code exhibits 3rd order accuracy as expected.  The errors in the $\phi$-direction are higher than those in the $\theta$ direction, due to the much higher velocities used in $\phi$.  In the $\theta$ direction, we see the convergence starting to diverge from third-order at the highest resolution tested.  This may be due to the use of first-order upwinding for the pole-adjacent cells in the $\theta$-direction or the second-order polar boundary conditions.

\begin{figure}[htb]
\centering
\includegraphics[width=0.25\textwidth]{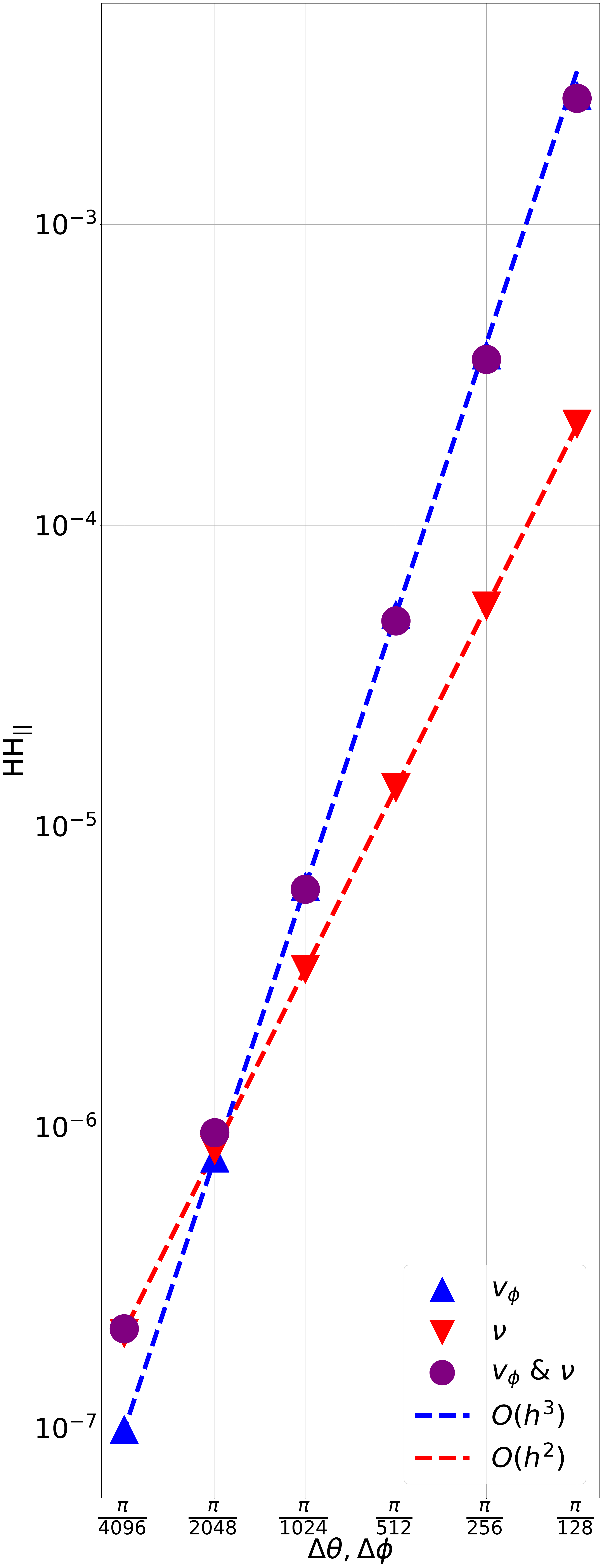}
\caption{Convergence of test case 1 (Eq.~\ref{eq:soccerball}) using the default numerical schemes of HipFT.  The purple marks are with diffusion and advection, the red marks are with diffusion only, and the blue marks are with advection only.  The advection error dominates the total error for most resolutions. However, since the diffusion error is second-order while the advection error is third-order, eventually at very high resolution, the diffusion error starts to dominate. \label{fig:valid_advect_diffuse}}
\end{figure}

The default diffusion method is the finite central difference spatial scheme of Eq.~\ref{eq:diffusion_op}, combined with the RKG2(3/2) temporal scheme adaptively cycled with the PTL condition described in Sec.~\ref{sec:num_diff_time}.  In Fig.~\ref{fig:valid_advect_diffuse}, we show convergence results for running the initial `soccer ball' profile of Eq.~\ref{eq:soccerball} for 672 hours with a diffusion coefficient of $500\,\mbox{km}^2\!/\mbox{s}$. We also show the same solution including advection with a constant angular velocity in $\phi$ that yields a full rotation (with no Strang splitting).  We also show the error for runs using only advection over a full rotation with the soccer ball initial condition.
We see that the errors associated with advection are significantly larger than those for diffusion.  However, since the CS-WENO3(h) scheme exhibits third-order accuracy, the errors intersect at a high resolution, resulting in the diffusion error dominating.   This transition happens at an extremely high resolution of $n_{\phi}=8192$, which is considerably larger than a typical HipFT run.  If such high resolutions eventually become needed, a 4th-order spatial diffusion operator would be straight-forward to implement.

The validation tests performed in this paper all used a uniform grid.  However, HipFT is capable of using a non-uniform grid, which can be useful to coarsen the grid near the poles to increase the flow CFL time step limit and for performing high-resolution evolutions of localized regions on the Sun, such as active regions or coronal holes.  Since the current default use cases of HipFT use a uniform grid, we leave the formal validation of running on non-uniform grids for a future publication.

%%%%%%%%%%%%%%%%%%%%%%%%%%%%%%%%%%%%%%%%%%%%%%%%%%%%%%%%%%%%
%%%%%%%%%%%   CODE IMPLEMENTATION AND PERFORMANCE
%%%%%%%%%%%%%%%%%%%%%%%%%%%%%%%%%%%%%%%%%%%%%%%%%%%%%%%%%%%%

\section{Code Implementation and Performance}
\label{sec:code}

HipFT is written in modern Fortran \citep{curcic2020modern} including features of the new Fortran 2023 standard \citep{ISO153912023}.  It is designed to be easily modified and expanded, with a modular structure and clear, descriptive names for variables, function, and subroutines.  It has been tested to work on a variety of compilers and hardware targets, including x86 and Arm CPUs, as well as NVIDIA and Intel GPUs.  

\subsection{Parallelism}
\label{sec:code_par}

HipFT uses a hybrid parallelization scheme, where MPI is used to spread groups of map realizations across compute units (within or across distributed nodes), while Fortran's {\tt do concurrent} (DC) is used to compute a group of realizations in parallel within a compute unit.  This can take the form of multi-threading on multi-core CPUs, or parallel computation on GPUs.  For example, when only modeling one realization, no MPI parallelism is used, and the code is run either multi-threaded across local CPU cores or on a single GPU.  When using MPI to run many realizations across multiple GPUs, the code assumes the code is launched with 1 MPI rank per GPU to set the GPU device number.  For multi-core CPUs, the number of MPI ranks and parallel threads should be set carefully using optimal affinity options, which can vary from system to system.  For example, on an AMD EPYC ROME CPU with 4 NUMA domains, it is optimal to set the CPU to 4xNUMA mode and run HipFT with 16 threads per NUMA domain with 4 MPI ranks (one per NUMA).  These kinds of affinity considerations are a common complication in running hybrid MPI+multi-threading applications \citep{Arul2015}, which is mostly avoided when running HipFT on GPUs.

A key issue when running codes on GPUs is data management.  Most GPUs have their own local memory separate from the CPU's memory.  Accessing/transferring data from one to the other is often much slower than the memory speed of the GPU, sometimes by orders of magnitude.  Two strategies to avoid this overhead are to either manually move the data to and from the GPU in a way that minimizes the number and size of the transfers, or use an automatic memory management system.  Automatic memory management can be a compiler/software feature, run-time feature, and/or driver-level feature.  Some GPUs and APUs have special hardware that allow the CPU and GPU to directly access each others' memory, often referred to as `unified memory', which can be used to greatly improve performance of automatic memory management.  

In order to keep HipFT as portable as possible, we manually manage the data movement between GPU and CPU using the OpenMP API, specifically OpenMP Target data directives \citep{deakin2023programming}.  This allows HipFT to run efficiently on compilers/hardware that do not support an automatic memory management system.  For more information on the GPU parallelization of HipFT and its portability across GPU vendors, see \citet{waccpd24}.

While we leave the details of building the code to the installation guide in the HipFT repository, it is important to point out that the ubiquitous GCC Fortran compiler {\tt gfortran} does not currently support directly multi-threading DC loops.  However, it does contain an automatic multi-threading option activated during compilation to parallelize HipFT.  We refer the user to the sample build scripts in the repository for more details.

\subsection{Performance}
\label{sec:code_perf}

Here, we test the performance of HipFT on both server and consumer CPUs and GPUs.  For the first test, we use the multi-realization example run provided in the github repository in the {\tt examples/flux\_transport\_1rot\_flowAa\_diff\_r8} folder.  This run uses analytic flows, flow attenuation, and diffusion to evolve a starting HMI Carrington map for 674 hours at the default $1024\times 512$ resolution.  Eight realizations are computed, varying the diffusivity and flow attenuation values, allowing us to run the code with up to 8 MPI ranks.

\begin{figure}[htb]
\centering
\includegraphics[width=\textwidth]{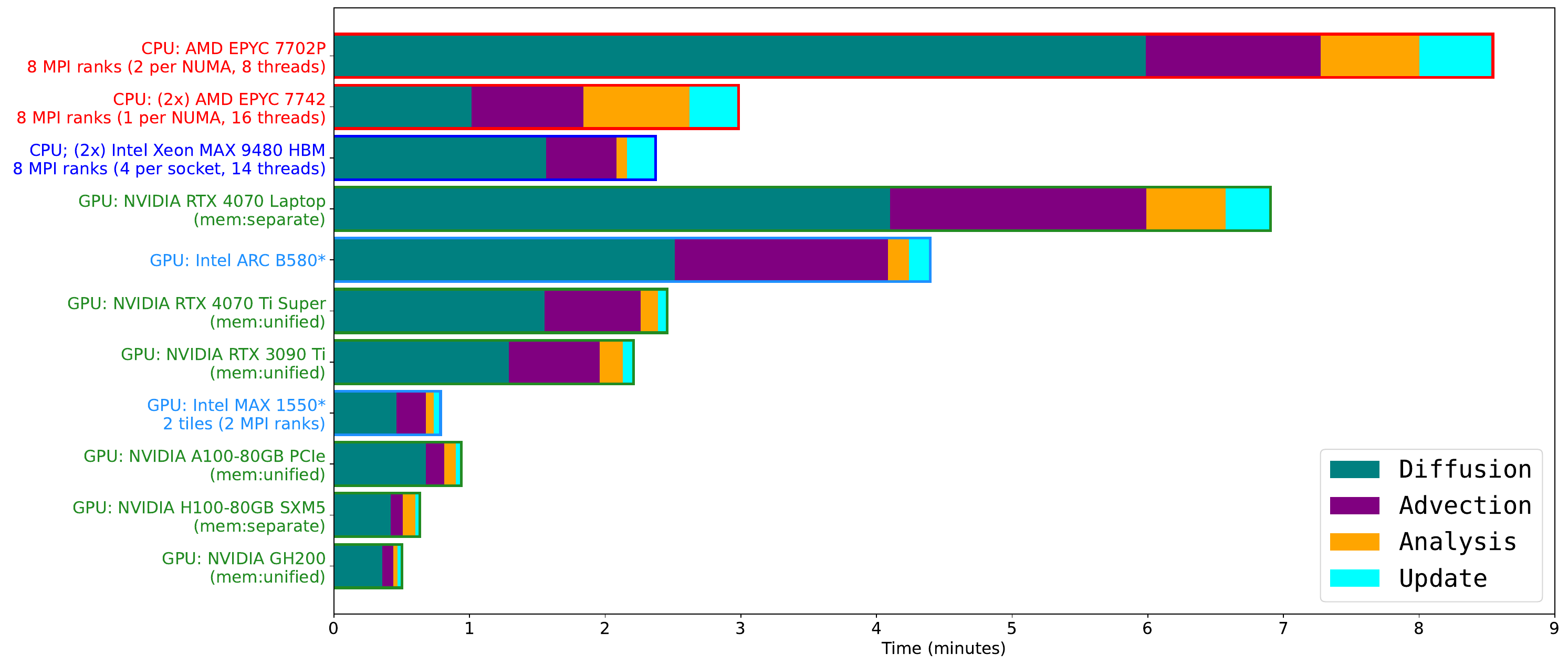}
\caption{Timing results for running HipFT on the example case {\tt flux\_transport\_1rot\_flowAa\_diff\_r8} provided in the git repository on a variety of CPU and GPU hardware.  The time taken for each part of the code is indicated (and I/O time is not included).  Only a single CPU node or single GPU is tested here.  For CPUs, the eight realizations of the run is spread across the sockets and/or NUMA domains using MPI, with the number of threads per MPI rank indicated.  On NVIDIA GPUs, the memory management method used for each GPU is indicated (see \citet{waccpd24} for details).   For the Intel MAX GPU, MPI is used to spread the realizations across compute tiles. \label{fig:perf1}}
\end{figure}

In Fig.~\ref{fig:perf1} we show timing results for various server CPUs and GPUs, as well as some consumer GPUs.  The portion of time each part of the HipFT calculation required is indicated.  Due to large variations in file system speeds across the systems, we omit the I/O time from the plots, but note that the total time for I/O on a locally-attached drive for these runs was small (less than 1 second).
We see that runs on a single professional GPU are faster than those on the multi-core CPUs.  We highlight the result of being able to run on both NVIDIA and Intel GPUs, demonstrating the performance portability of the Fortran standard implementation (see \citet{waccpd24} for details).  We also note that the inexpensive consumer GPUs ran as fast as a modern two-socket CPU server with over 100 cores.  For reference, the same run performed on a modern desktop CPU (Ryzen 9700X) with 8 dual-threaded cores (running with 8 MPI ranks, 2 threads per rank) took $\sim23$ minutes.  Therefore, adding an inexpensive consumer GPU to a modern personal desktop computer can yield over an $8\times$ speedup of HipFT (even more for an older desktop CPU). 

\begin{figure}[htb]
\centering
$\begin{array}[b]{cc}
\begin{array}[b]{r}
\includegraphics[width=0.25\textwidth]
{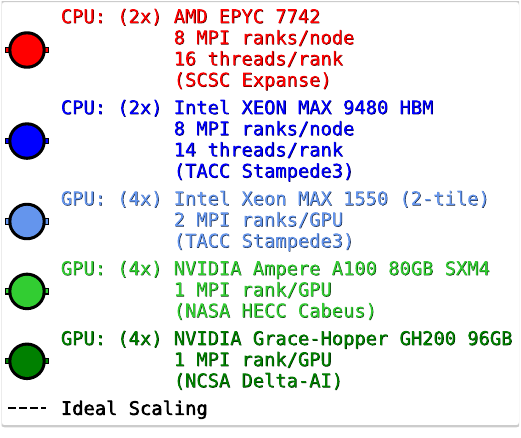} \\
\includegraphics[width=0.25\textwidth]
{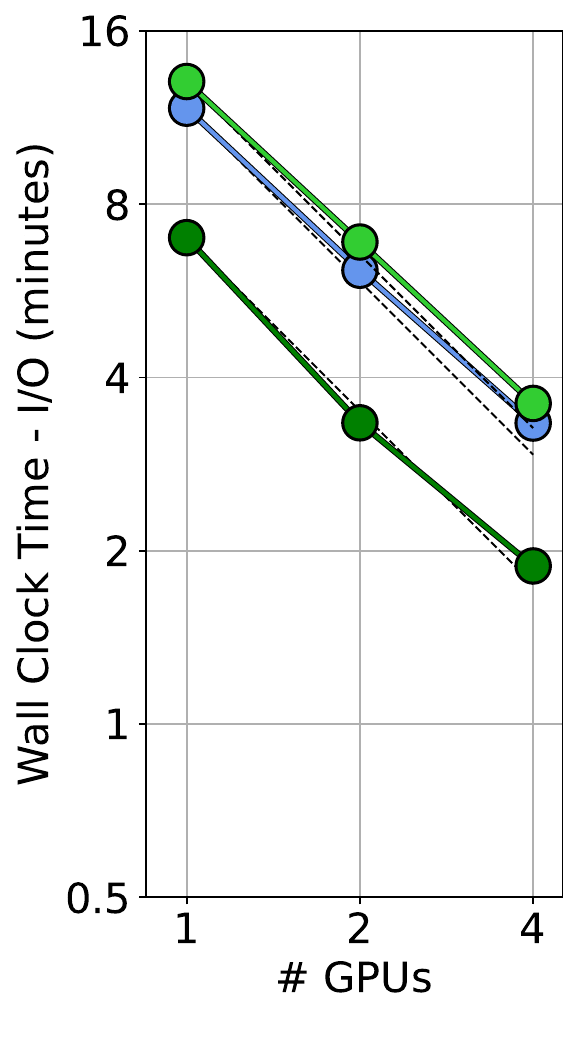}
\end{array}
&
\includegraphics[width=0.33\textwidth]{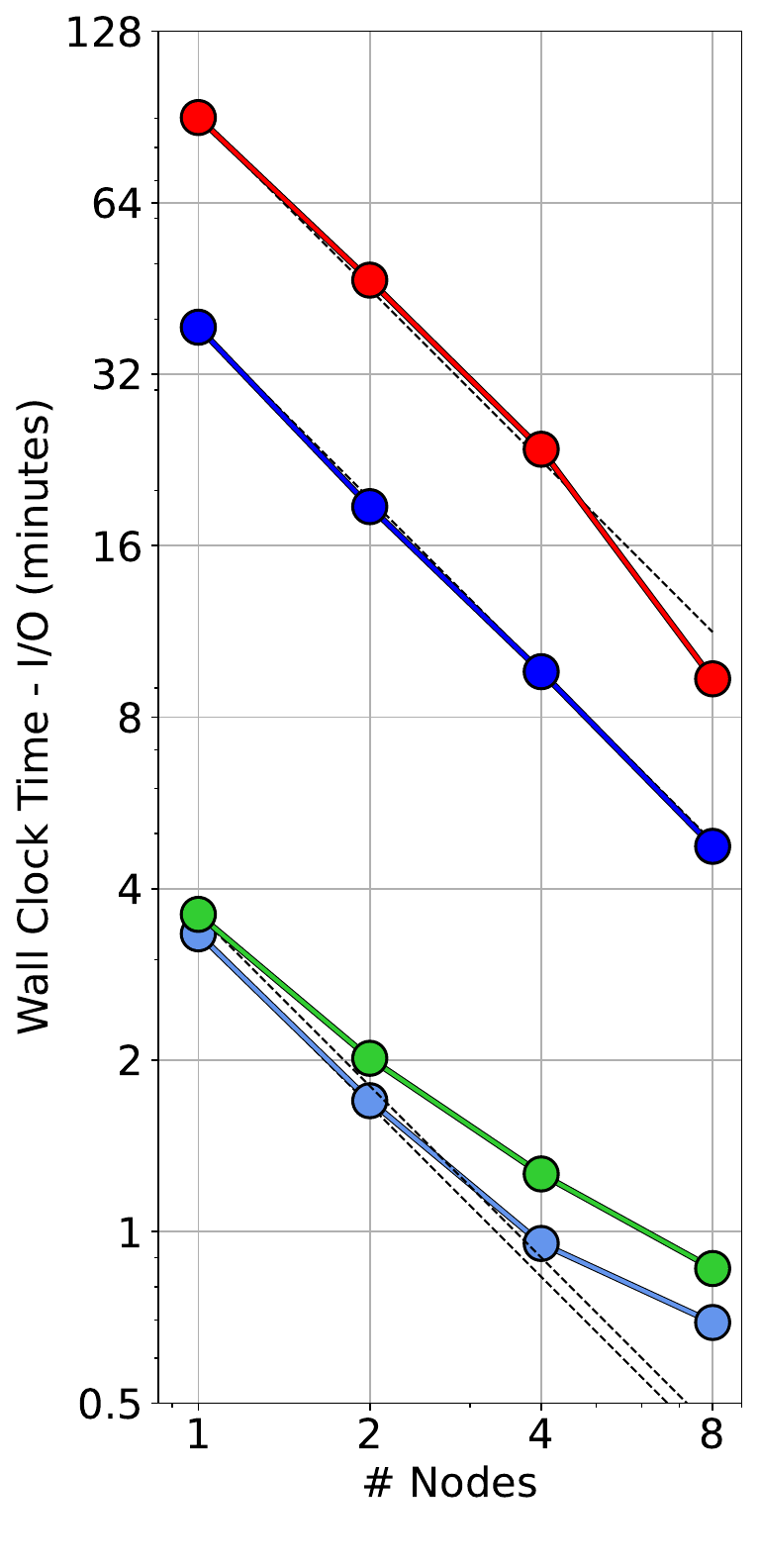}
\end{array}$
\caption{Timing results for running HipFT on the example case {\tt flux\_transport\_1rot\_flowAa\_diff\_r8}, but with a total of 128 realizations.  Due to differences in the file systems of the machines tested, we exclude the I/O time.  For CPUs, the realizations are spread optimally across the sockets and/or NUMA domains per node as indicated.  The most performative memory management method for each GPU is used. For the Intel GPUs, the realizations are spread across each GPUs two compute tiles (see \citet{waccpd24} for details). The scaling results for GPUs within a single node are shown on the left, and the results across CPU and GPU nodes is shown on the right. 
 The CPU and Intel GPU results used the Intel IFX compiler $2025.0.0$, while the NVIDIA GPU results used the nvfortran compiler $24.7/9$.\label{fig:scaling}}
\end{figure}

As described in Sec.~\ref{sec:code_par}, HipFT can use MPI to spread realizations across compute units.  To test how HipFT can scale for many realizations, we add redundant realizations to the example used above for a total of 128 realizations and run it across multiple GPUs within a compute node, and across GPU/CPU nodes.  The resulting timings for a variety of CPUs and GPUs are shown in Fig.~\ref{fig:scaling}.
We see that the code scales well across realizations for multiple GPUs (both NVIDIA and Intel) within a node, and across CPU and GPU nodes.  The scaling across GPU nodes starts to taper off at larger number of nodes, but the overall performance is much higher than the CPU nodes.

\begin{figure}[htb]
\centering
$\begin{array}{lll}
\includegraphics[width=0.33\textwidth]{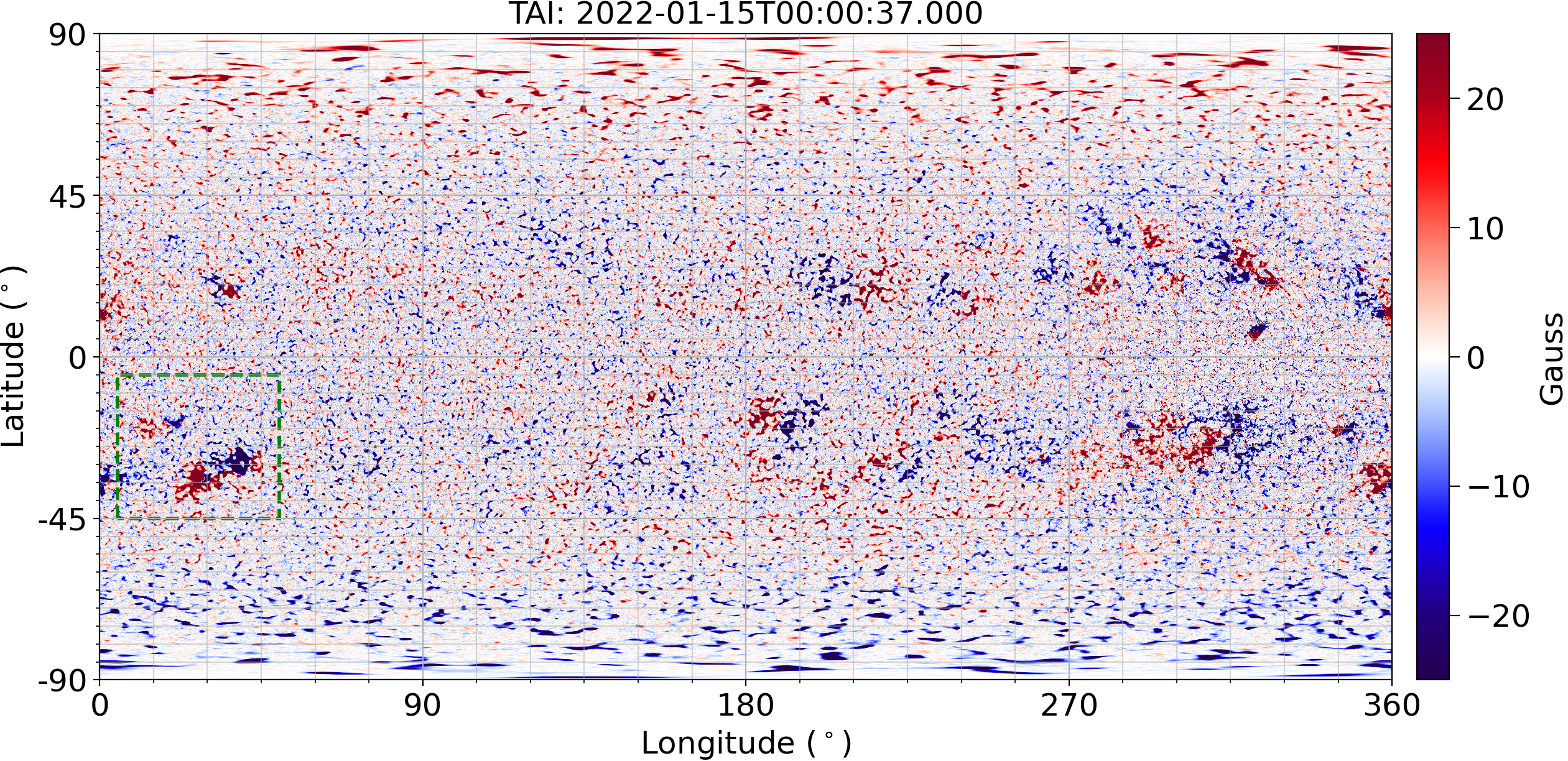} &
\includegraphics[width=0.33\textwidth]{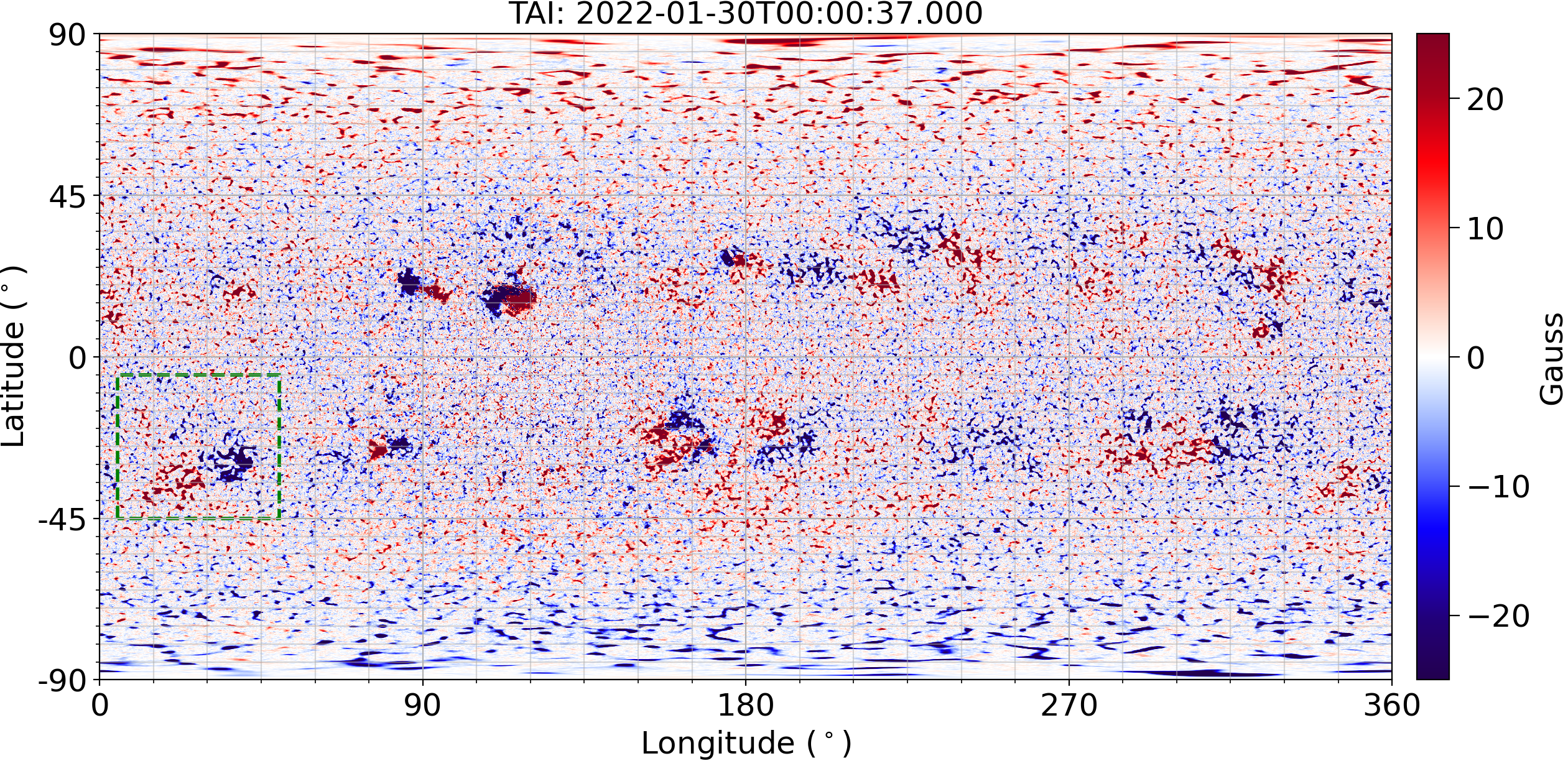} &
\includegraphics[width=0.33\textwidth]{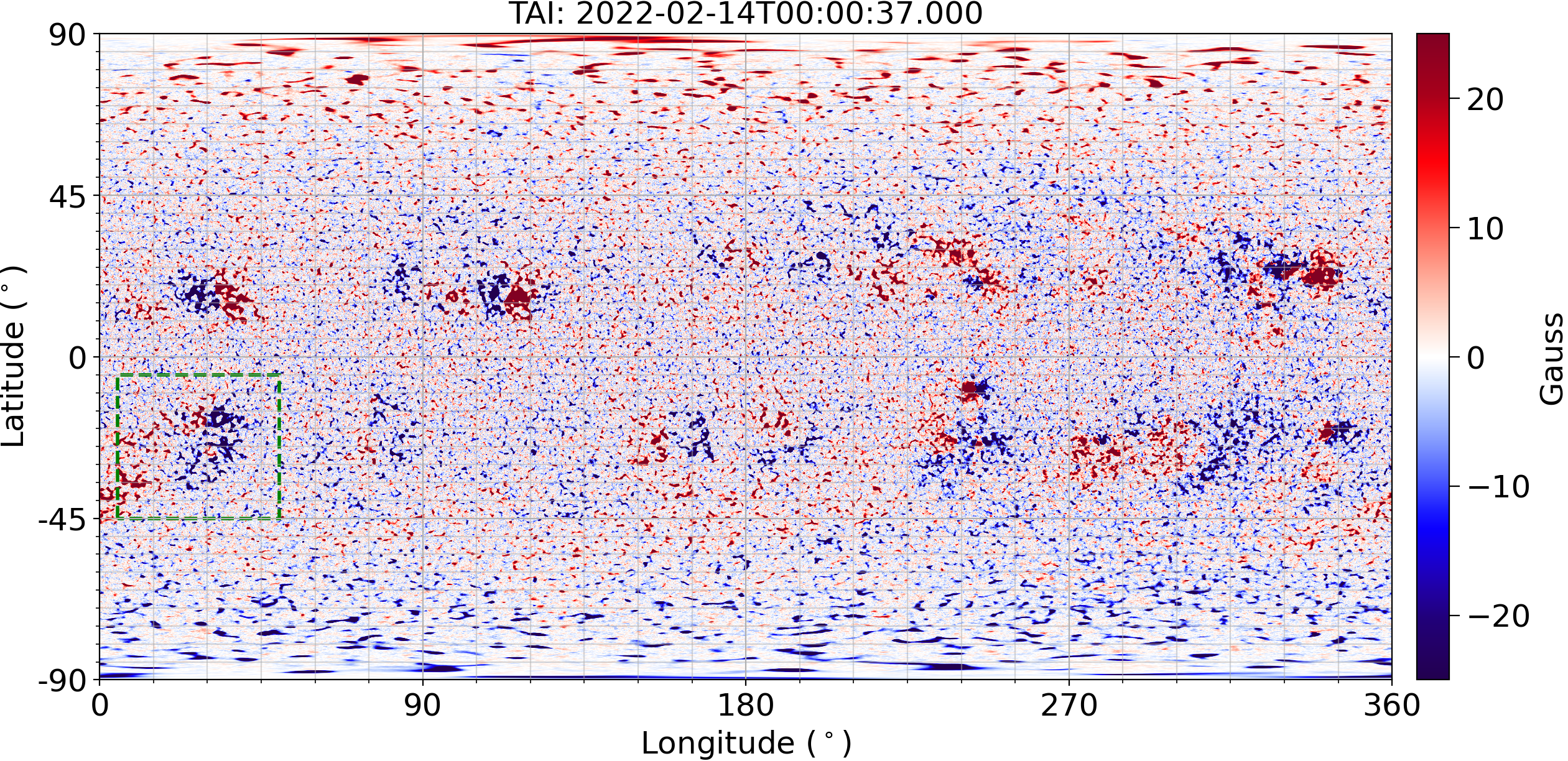}
\\
\includegraphics[width=0.3\textwidth]{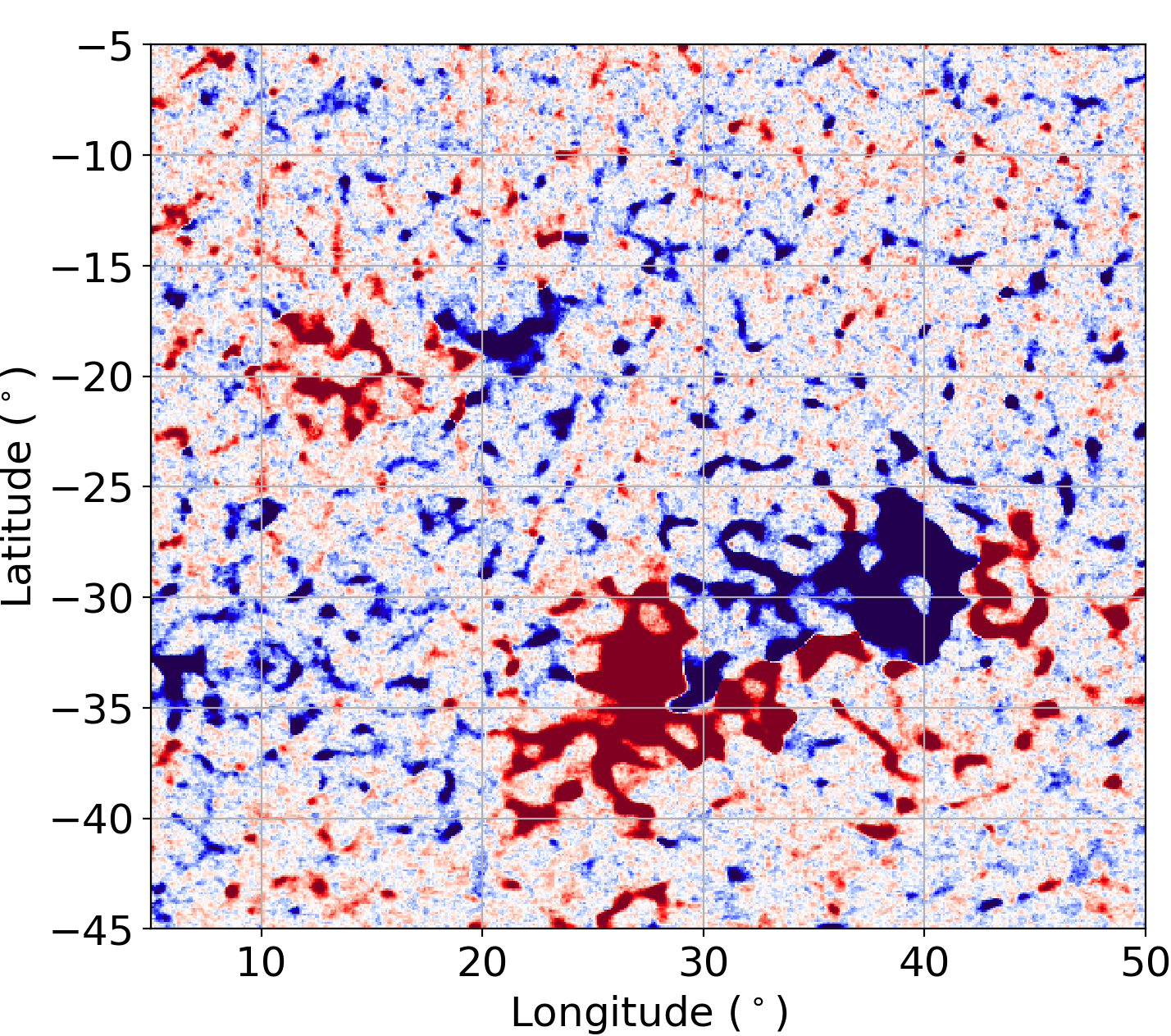} &
\includegraphics[width=0.3\textwidth]{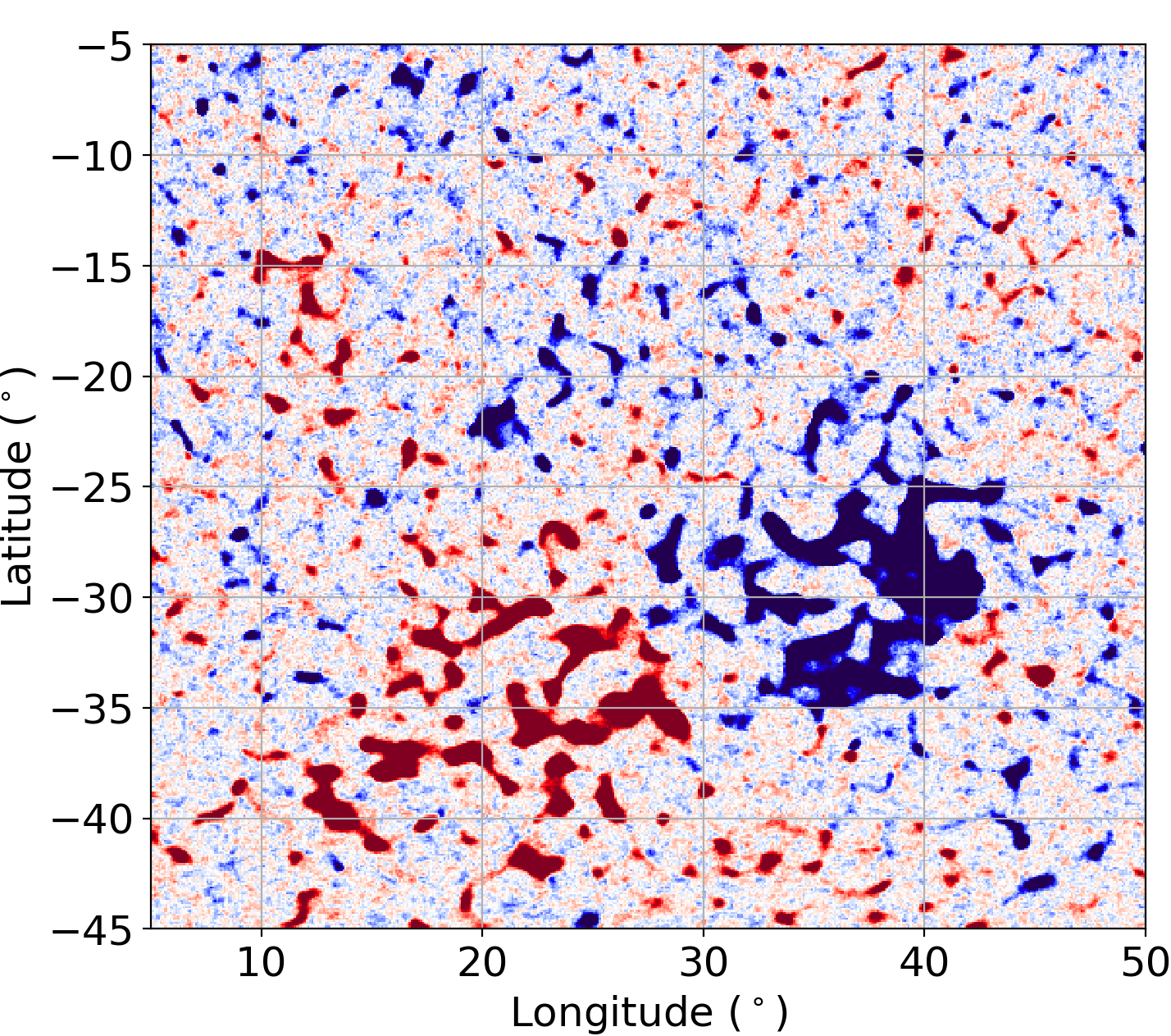} &
\includegraphics[width=0.3\textwidth]{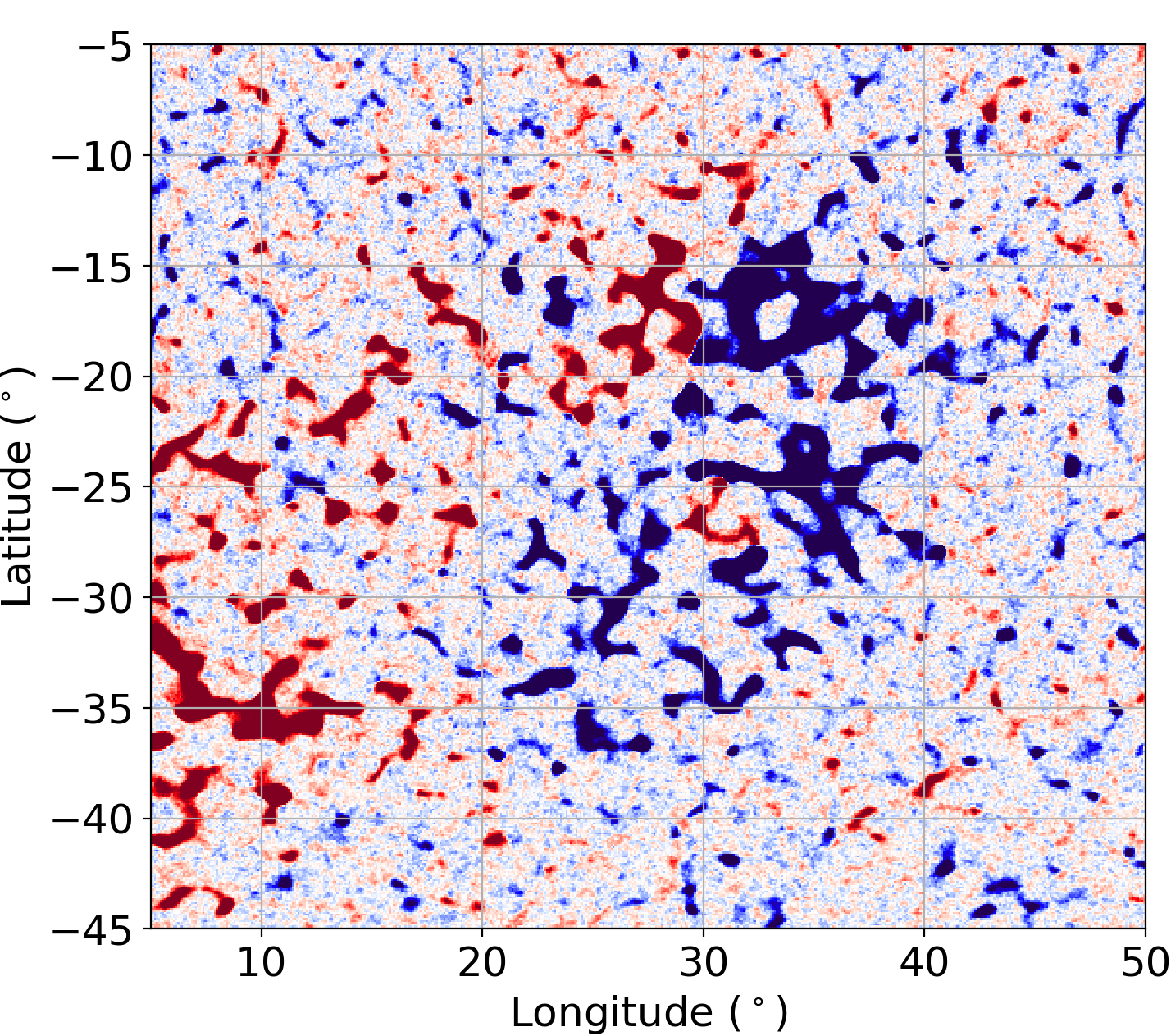}
\end{array}$
\caption{Map output of HipFT running at $4096\times 2048$ resolution.  The run is adapted from the example case {\tt flux\_transport\_1yr\_flowCAa\_diff300\_assimdata\_rfe} in the git repository, with the diffusion lowered to $\nu=200\,\mbox{km}^2\!/\mbox{s}$, the random flux amplitude risen to $\Phi/\mbox{hr}=650\times 10^{21} \mbox{Mx/hr}$, and the simulation time lowered to 56 days. High resolution versions of the required MagMAP database of assimilation data and convective flows from ConFlow were produced for the run. The maps of the run are shown at $\sim15$ (left), $\sim30$ (middle), and $\sim45$ (right) days into the simulation.  The active region highlighted by the green boxes is shown zoomed-in in the bottom row.  \label{fig:highres}}
\end{figure}

HipFT can be run at any practical resolution, but the higher the resolution, the more computationally expensive the integration steps become, and more importantly, the smaller the stable advection time step becomes, yielding many more computational steps.  The non-uniform grid capabilities of HipFT can be used to reduce the computational time, by coarsening the grid near the poles (increasing the stable flow time step) and/or focusing the high-resolution to limited sections of the map.  A full description and testing of these capabilities is beyond the scope of the current paper.  However, in order to get a feel on the computation time at very high resolutions in the worst case, we test HipFT at $4096\times 2048$ resolution.  We modify the HipFT git repository's example run {\tt flux\_transport\_1yr\_flowCAa\_diff300\_assimdata\_rfe} to run at $4096\times 2048$ resolution, set the added random flux amount $\Phi/\mbox{hr}=650\times 10^{21}\, \mbox{Mx}/\mbox{hr}$ (to compensate for the higher unsigned flux at higher resolution), set the diffusion to $\nu=200\,\mbox{km}^2\!/\mbox{s}$, and only run it for two Carrington rotations (56 days).  We also generated the required high-resolution MagMAP and ConFlow data.  The run completed in 24 hours on an NVIDIA A100 80GB PCIe GPU, which is a much larger computational time than that of the default resolution of HipFT ($\sim6$ minutes).  However, the run time is still over $50\times$ faster than real-time, opening the door for high resolution studies in the future.  In Fig.~\ref{fig:highres}, we show maps from the high resolution run with zoomed-in views of an active region.  
Exploration and analysis of high-resolution HipFT runs like these will be discussed in a future publication.

\section{Code use}
\label{sec:codeuse}

For full instructions on running the code and all input options, see the documentation included in the github repository.  Here, we provide a brief overview of the basic input, output, and post processing features of HipFT.

\subsection{Inputs}
\label{sec:code_input}

HipFT uses a Fortran `namelist' for input parameters.  This is a lightly formatted text file containing the desired input parameters and their values.  One key advantage of using a namelist is that only input parameters differing from the default values need to be specified, enabling concise, easy-to-read input files for most use cases.  Additionally, the code writes out the full namelist of all parameters to a file after reading the input, providing a complete record of every parameter used in the run.

Other possible inputs for HipFT include 2D HDF5 \citep{hdf5} map files for the initial map, as well as options for a custom spatially-dependent diffusion coefficient, source term, and/or flows. When using data assimilation and sequences of flows from file (such as when using MagMAP and/or ConFlow output), a CSV file is read in which directs HipFT to load the correct files at the desired times.  HipFT has the option to multiplicatively flux balance the input map and/or assimilated data.  Note that there are currently no interpolation/re-binning capabilities in the HipFT code, so any necessary map processing for inputs must be done externally (tools to do this are provided in the OFT repository).
 
\subsection{Outputs}
\label{sec:code_output}

Numerous outputs are provided by HipFT.  When the code is launched, it outputs the namelist of parameters that will be used in the run, as well as the initial map.  For validation runs, the analytic solution is also written out.  When using multiple realizations, a text file is outputted listing all parameters for each realization in a table for reference.

During the run, basic information about the current run time, time step, etc. are written to the terminal in order to monitor the run progress.  The analysis quantities described in Sec.~\ref{sec:model_analysis} are written to a history file for each realization.  Additionally, quantities related to the numerical methods (e.g. time step, stability limits, diffusion PTL cycles, etc.) are written to a separate history file across realizations.  All history files are appended and closed at each step, allowing them to be plotted (see Sec.~\ref{sec:code_postproc}) throughout the run.  If map output was activated, the maps are written at their chosen cadence and a list of the maps' output simulation time and filenames are written to a text file. For multiple realization runs, the maps are written as a 3D HDF5 file, with each slice corresponding to a realization.  If there is only 1 realization, the code can either write the maps as 3D HDF5 files with a unitary third dimension, or as 2D HDF5 files.  In addition to the maps, the code can also be set to output the flows at the same cadence as the maps.  This can be useful to see what the combined flows in the code were at specific times (for example when combining analytic flow profiles with flows from file and flow attenuation).  

When the run is complete, HipFT outputs a text file that shows timings for each part of the code for each MPI rank, including a summary of all ranks.  This summary is also outputted to the terminal at the end of a run.  Additionally, the code writes out the final map (the output of the initial and final maps are independent of whether or not there is a map output cadence set).  For all map output, an option to {\bf multiplicatively} balance the flux is available.

\subsection{Post processing and analysis}
\label{sec:code_postproc}

In the HipFT repository, we have provided several python scripts to help analyze and process the results of a HipFT run.  This includes plotting the quantities (and derived quantities) in the history files, plotting the output maps and making a movie of them, and generating and plotting a `butterfly diagram'.  Additional scripts are provided that add UTC/TAI dates and times to the run outputs and plots, extract realization slices, print history summaries, compare runs to each other, compare maps to each other, read map data into python {\tt numpy} arrays, and get map values of a series of $(\theta,\phi)$ in-situ points through interpolation. To facilitate comparisons to other models (or to analyze data derived maps), we provide a script that generates a HipFT history file from a sequence of HDF5 maps.  Scripts to run the validation tests of Sec.~\ref{sec:num_valid} are also included. For details on how to use the various tools, see the help documentation for each script by calling them with the `-h' flag.  

In addition to the HipFT tools described above, the OFT repository contains scripts and tools that can easily post-process HipFT input/output maps. These scripts allow for operations such as binning to a different resolution in a flux-balance preserving manner, flux balancing, and/or smoothing.  These tools can also be used to remesh HipFT output, which can be useful for models and analysis requiring lower resolution, smoothed maps, and/or alternative grids.  For example, once processed, the maps generated from HipFT can be directly used with the SWiG empirical solar wind generator model\footnote{\url{https://github.com/predsci/swig}} and/or the CORHEL-CME thermodynamic coronal mass ejection generator model\footnote{\url{https://ccmc.gsfc.nasa.gov/models/CORHEL-CME\~1}}.

%%%%%%%%%%%%%%%%%%%%%%%%%%%%%%%%%%%%%%%%%%%%%%%%%%%%%%%%%%%%
%%%%%%%%%%%   EXAMPLE PRODUCTION USE
%%%%%%%%%%%%%%%%%%%%%%%%%%%%%%%%%%%%%%%%%%%%%%%%%%%%%%%%%%%%

\section{Use Cases}
\label{sec:examples}

The use of HipFT in the context of the OFT model, including production runs and detailed analysis, will be described in a forthcoming publication \citep{oft3}.  In this section, we briefly highlight some use cases of HipFT.

\subsection{Flux transport with convective flows, data assimilation, and random flux emergence}
\label{sec:examples_1yr}

One primary use of HipFT is to run flux transport simulations in order to provide full-Sun magnetic maps.  To facilitate an example production run of this kind, we have provided a ready-to-use set of convective flow ConFlow simulation data (28 days at 15 minute cadence) as well as a full year of HMI $B_r$ data at 1 day cadence processed with MagMAP (see Sec.~\ref{sec:avail}).  The HipFT input file to run a full year simulation using this data is provided in the {\tt examples/flux\_transport\_1yr\_flowCAa\_diff300\_assimdata\_rfe} folder of the HipFT repository.  The run uses the default suggested values for this kind of simulation, including flow attenuation (Eq.~\ref{eq:flow_attenuation}), diffusion, data assimilation and random flux emergence.  

\begin{figure}[htb]
\centering
$\begin{array}{c}
\includegraphics[width=\textwidth]{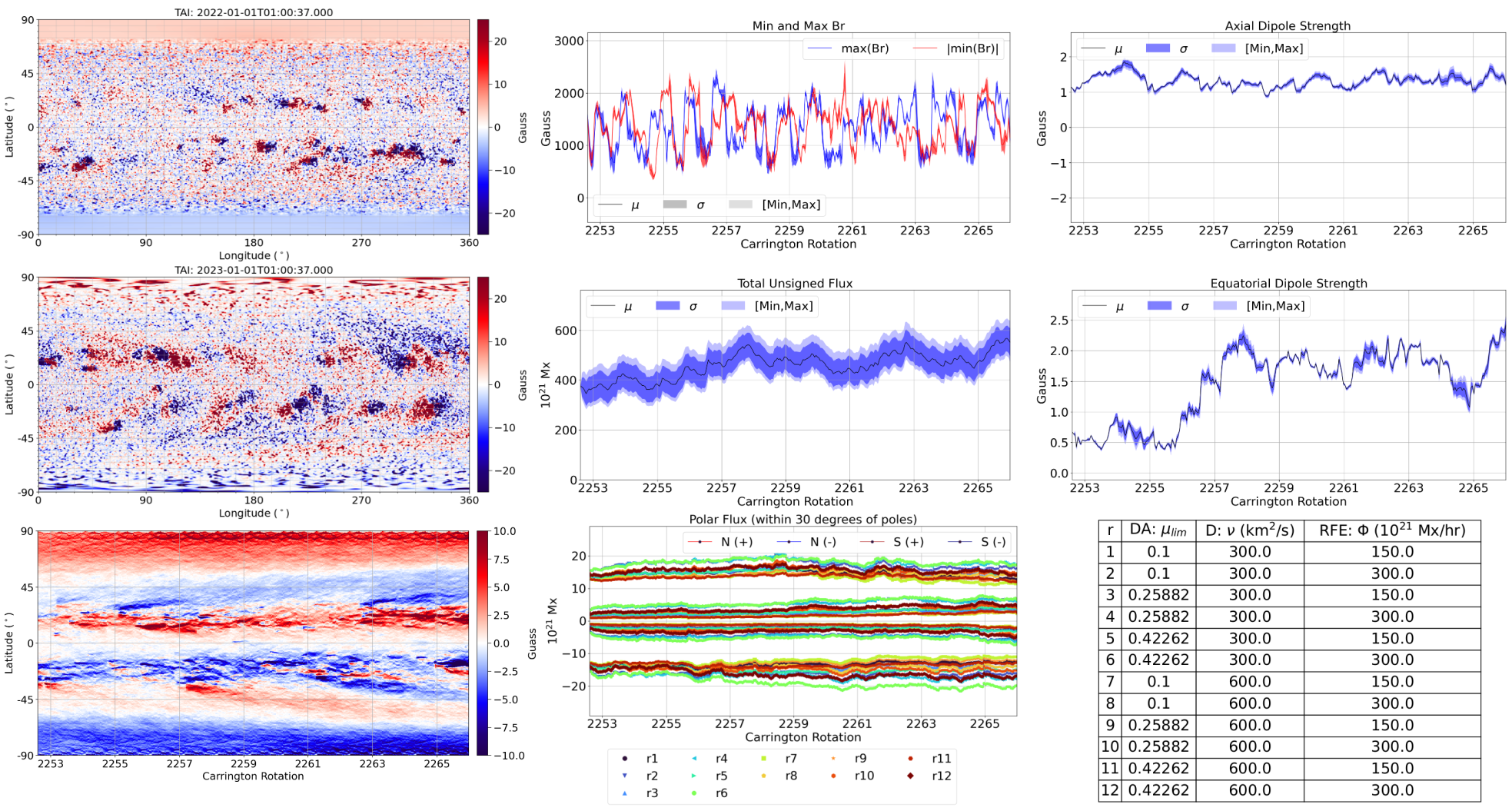}
\end{array}$
\caption{Samples of post processing output for the example case {\tt flux\_transport\_1yr\_flowCAa\_diff300\_assimdata\_rfe} included in the HipFT repository, but with 12 realizations spanning diffusion, data assimilation, and amount of added random flux quantities.  The initial and final $B_r$ maps for realization 1 are shown in the top and middle left respectively with the butterfly diagram shown in the bottom left.  Various derived quantities and metrics are shown in the center and right columns, with the center right showing all realizations, while the others show realization summary plots.  The bottom right shows the table of realization parameters generated by the post processing.\label{fig:hipft1yr}} 
\end{figure}

Here, we show results of running the one year example case, adding some multiple realization parameters.  Specifically, we set the data assimilation $\mu$ cutoff (see Eq.~\ref{eq:data_assim_f}) to $5.74^{\circ}$, $15^{\circ}$, and $25^{\circ}$ away from disk limb, the diffusivity $\nu$ to $300\,\mbox{km}^2\!/\mbox{s}$ and $600\,\mbox{km}^2\!/\mbox{s}$, and the random flux emergence total unsigned flux per hour to $150\times 10^{21}\,\mbox{Mx/hr}$ and $300\times 10^{21}\,\mbox{Mx/hr}$.  The combination of these parameters results in 12 realizations.  In Fig.~\ref{fig:hipft1yr}, we show some of the post processing outputs from the run including derived quantities, maps, and butterfly diagram plots. 
We see variations across realization combinations, with some combinations showing similarities while others can differ significantly.  The use of multi-realization runs like this is a useful tool in exploring the dynamics of flux transport.  A thorough analysis of a full solar cycle run (with a different set of realization parameters) will be described in a forthcoming publication \citep{oft3}.

\subsection{Generating time-dependent boundary conditions for an MHD model}
\label{sec:examples_tdc}

\begin{figure}[htb]
\centering
$\begin{array}{c}
\includegraphics[width=\textwidth]{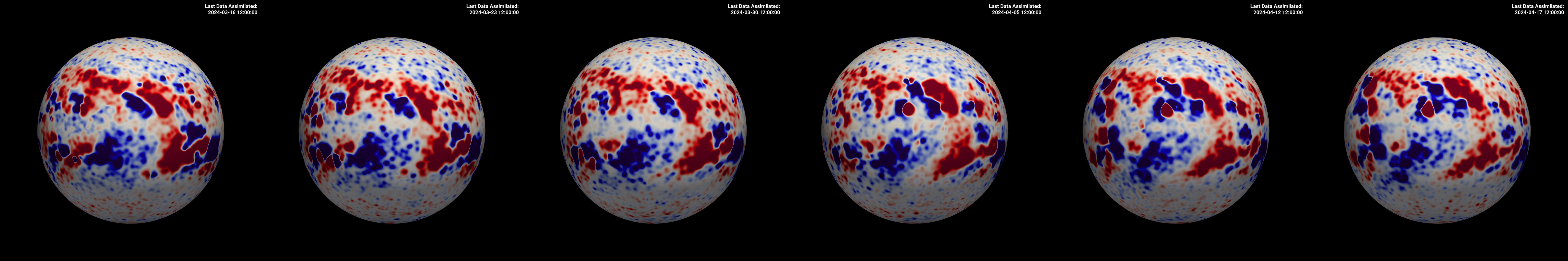}\\
\includegraphics[width=\textwidth]{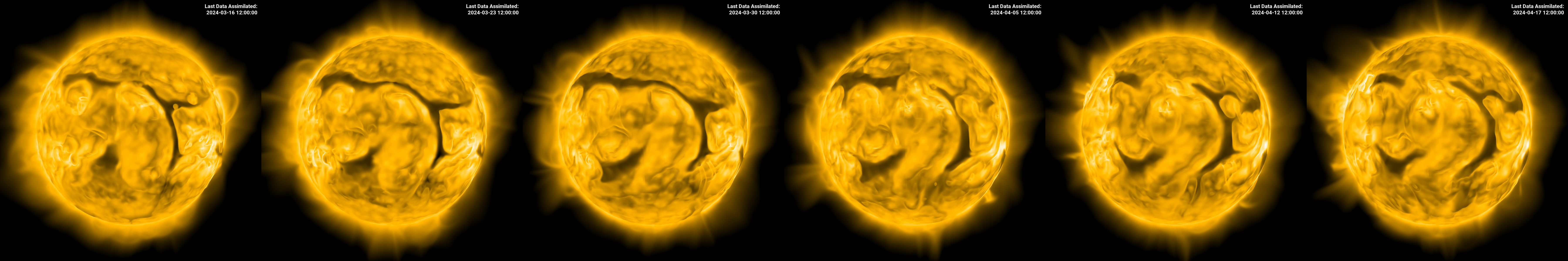}
\end{array}$
\caption{Example using HipFT to drive a time-dependent MHD model \citep{downs24}.  The top row shows the processed HipFT $B_r$ maps, while the bottom row shows the corresponding forward-modeled EUV images derived from the data-driven MHD model.  The view is from Earth's perspective on April 8th, 2024 18:42 UT, and the model snapshots have timestamps of 12 UT on 3/16, 3/23, 3/30, 4/5, 4/12, and 4/17 2024 respectively. The time-evolving model exhibits dynamics similar to those observed on the Sun that cannot be generated by non-evolving static relaxation models.  Images taken from \url{www.predsci.com/eclipse2024}.\label{fig:eclipse2024}} 
\end{figure}

The full Sun maps generated from a HipFT run of the kind described in Sec.~\ref{sec:examples_1yr} can be used to drive time-dependent MHD models of the solar corona and heliosphere \citep{Mason_2023,Lionello_2023,Feng_2013,downs24}.  For example, \citet{downs24} used HipFT to generate several months of full-Sun maps at one hour cadence, and used them to drive the lower boundary condition of a thermodynamic MHD model for 32 days, assimilating data in near real time.  The simulation was used to generate a running prediction of how the corona would look during the 2024 total solar eclipse\footnote{\url{https://www.predsci.com/eclipse2024}}.  In order to avoid strong jumps in the boundary from emerging active regions, a custom ramped weight map was used during data assimilation.  The maps were then processed using a flux-preserving re-meshing scheme as well as $B_r$-dependent smoothing to ensure the model could resolve the evolving map structures.  In Fig.~\ref{fig:eclipse2024} we show a time sequence of HipFT processed maps along with forward modeled EUV images of the time-dependent MHD model. These images use the same viewer position but each snapshot is separated by several days, illustrating how the solution evolves as new data from the sequence of HipFT maps is continuously assimilated.
 For full details and results of this novel simulation, including details on how HipFT was utilized to generate the boundary conditions, see \citet{downs24}.

\subsection{Flux-preserving smoothing tool}
\label{sec:examples_smooth}

\begin{figure}[htb]
\centering
$\begin{array}{c}
\includegraphics[width=0.25\textwidth]{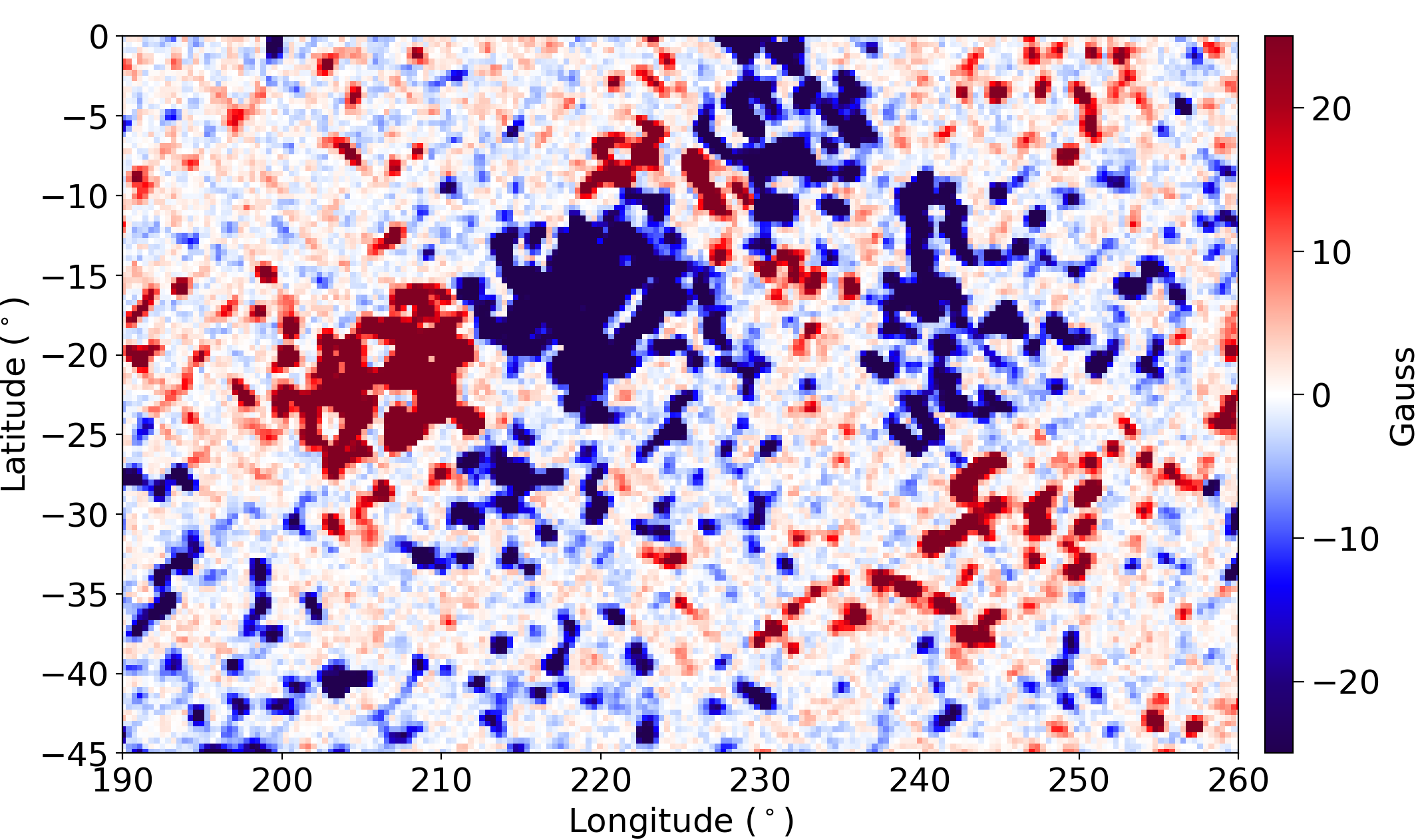}
\includegraphics[width=0.25\textwidth]{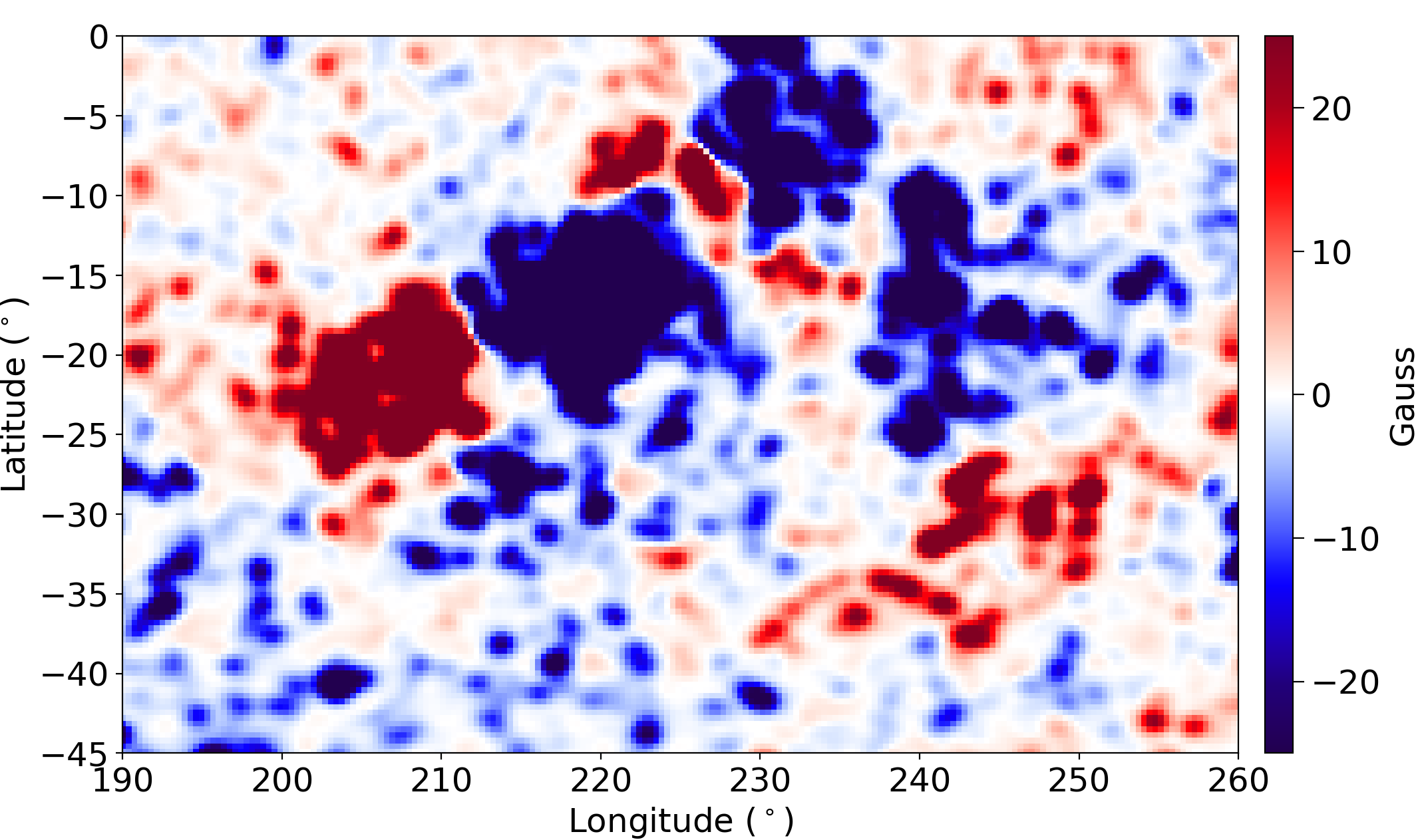}
\end{array}$
\caption{Example of using HipFT as a flux-preserving magnetogram smoother.  A zoomed-in portion of the last output map for realization 1 of the test used in Sec.~\ref{sec:examples_1yr} is shown on the left, while the same map is shown on the right after smoothing the map using the grid-based diffusivity of Eq.~\ref{eq:diffgrid} with $\alpha_{\nu}=0.5$.\label{fig:smoother}} 
\end{figure}

The efficiency of the diffusion advance in HipFT allows it to be used as a fast map smoothing tool.  Smoothing magnetic maps is often necessary for models to properly resolve the magnetic structure \citep{psi_map_processing_agu2019_rmc}.  While local filter algorithms can be used, they can suffer from aliasing artifacts and flux imbalance.  By using a surface diffusion advance, these difficulties are avoided.  The grid-based diffusivity shown in Sec.~\ref{sec:model_diffusion} allows the use of the minimum amount of smoothing necessary for different parts of the grid; however, a constant diffusivity can also be used.  This map smoothing capability of HipFT is used in OFT's map processing tools.  In Fig.~\ref{fig:smoother}, we show the result of using HipFT to smooth a map with grid-based diffusivity.
This smoothing is very fast (in this case, the $512\times 1024$ map took 0.25 seconds on a RTX 3090Ti GPU).  We note that the PTL time step limit described in Sec.~\ref{sec:num_diff_time} is critical in obtaining an accurate solution robustly in these runs as there is no stability limit on the time step (since the STS diffusion advance is unconditionally stable).  

%%%%%%%%%%%%%%%%%%%%%%%%%%%%%%%%%%%%%%%%%%%%%%%%%%%%%%%%%%%%
%%%%%%%%%%%   AVAILABILITY
%%%%%%%%%%%%%%%%%%%%%%%%%%%%%%%%%%%%%%%%%%%%%%%%%%%%%%%%%%%%

\section{Availability}
\label{sec:avail}

HipFT is open source and publicly available and developed on GitHub at \url{https://github.com/predsci/HipFT}.  The repository includes detailed instructions on building and running the code on multiple types of systems.  As the computational core of OFT, the HipFT repository is also included as a submodule of the OFT GitHub repository at \url{https://github.com/predsci/OFT}.   We have also set up a Zenodo data set at \url{https://zenodo.org/records/11205509} where a Carrington rotation of ConFlow convective flow maps and one full year of MagMAP HMI data for data assimilation is provided.  This allows testing HipFT in a true scientific context independent of installing and running ConFlow and MagMAP.  HipFT is continually developed along with the overall OFT code suite, with public suggestions, updates, and modifications being welcome.

%%%%%%%%%%%%%%%%%%%%%%%%%%%%%%%%%%%%%%%%%%%%%%%%%%%%%%%%%%%%
%%%%%%%%%%%   SUMMARY
%%%%%%%%%%%%%%%%%%%%%%%%%%%%%%%%%%%%%%%%%%%%%%%%%%%%%%%%%%%%

\section{Summary}
\label{sec:summary}

We have introduced the open source High-performance Flux Transport code HipFT, which serves as the computational core of the Open-source Flux Transport (OFT) model.  It is a Fortran code that implements highly accurate and efficient numerical methods to advance advection, diffusion, sources, and data assimilation.  It is designed to be modular, incorporating numerous model and numerical method options, and allows users to write extensions easily.  HipFT can compute multiple realizations across many model parameters within a single run to create ensembles of maps for uncertainty quantification and parameter studies.  By writing the code using Fortran standard language parallelism (using the {\tt do concurrent} construct), it can run in parallel on multi-core CPUs and on GPUs.  MPI is used to parallelize across realizations, allowing the code to run across multiple multi-CPU and multi-GPU compute nodes.  We have described the model, numerical methods, code implementation, and analysis tools of HipFT.  The numerical methods were validated on known solutions, and the performance on CPUs and GPUs was tested.  The code is available and developed on GitHub, and is included as a submodule in the OFT model suite.  

The OFT model suite contains two other important codes needed for realistic runs of HipFT, which are ConFlow and MagMAP.  ConFlow is used to generate time-dependent {\bf supergranular} convective tangential surface flows and will be {\bf described} in detail in a forthcoming paper \citep{oft2}.  MagMAP is a python library that is used to obtain, process, and prepare observations of the solar surface magnetic field for use in data assimilation in HipFT.  A full description of the OFT model and its use, including details of the MagMAP code will be described in a forthcoming paper \citep{oft3}. 

\acknowledgments{
Work at Predictive Science Inc.\ was supported by the NASA LWS Strategic Capabilities Program (grant 80NSSC22K0893), the NSF PREEVENTS program (grant ICER1854790), and the NSF/NASA SWQU program (grants AGS 2028154 and 80NSSC20K1582).  It also utilized the Cabeus system at NASA's High-End Computing Capability (HECC) through NASA grant's 80NSSC20K0192's NASA {\bf Advanced} Supercomputing division request SMD-24-72380598, as well as the Expanse system at the San Diego Supercomputing Center (SDSC), the Stampede3 system at the Texas Advanced Computing Center (TACC), and the Delta-AI system at the National Center for Supercomputing Applications (NCSA) through allocation TG-MCA03S014 from the Advanced the Advanced Cyberinfrastructure Coordination Ecosystem: Services \& Support (ACCESS) program, which is supported by National Science Foundation grants \#2138259, \#2138286, \#2138307, \#2137603, and \#2138296.

C.N.A.\ was supported in part by the NASA SWQU grant listed above and the NASA competed Heliophysics Internal Scientist Funding Model (ISFM). 

C.J.H.\ is partially supported by Air Force Office of Scientific Research (AFOSR) tasks 22RVCOR012 and 25RVCOR001. The views expressed are those of the authors and do not reflect the official guidance or position of the United States Government, the Department of Defense or of the United States Air Force. 
}

%%%%%%%%%%%%%%%%%%%%%%%%%%%%%%%%%%%%%%%%%%%%%%%%%%%%%%%%%%%%
%%%%%%%%%%%   APPENDECIES
%%%%%%%%%%%%%%%%%%%%%%%%%%%%%%%%%%%%%%%%%%%%%%%%%%%%%%%%%%%%

\newpage

\appendix
\section{Polar boundary conditions for advection and diffusion operators}
\label{a:polebc}

For the polar boundary condition for the advection and diffusion operator, we use a finite volume method.  We want to evaluate $\nabla\cdot F$ at the pole, where for advection, $F_a = \vec v\,B_r$, and for diffusion, $F_d = \nu\,\nabla B_r$.  From the divergence theorem, we have
\[
\int\int\nabla\cdot F\,dA = \int_C F\,ds.
\]
We treat the region at the pole out to the first internal half-mesh point ($\theta=\Delta\theta/2$) as a single cell.  $F$ is computed at the half-mesh points around the pole using the upwinding scheme for advection (see Sec.~\ref{sec:num_advect_space}), and central differencing for diffusion (see Sec.~\ref{sec:num_diff_space}).  Since only the flux in the theta direction contribute to the flux in and out of the polar cell, we ignore the $\phi$ direction in $F$.  The line integral above is evaluated as
\[
\int_C F\,ds = \int_{0}^{2\pi}F\,\sin\theta\,d\phi,
\]
which at $\theta=\Delta\theta/2$, numerically is integrated along $\phi$ as
\[
\int_{0}^{2\pi}F\,\sin\theta\,d\phi = \sin\left(\frac{\Delta\theta}{2}\right)\sum_{k=2}^{N_{\phi}-1}F_k\Delta\phi_k.
\]
For the surface integral, since the polar cap is treated here as a single cell, the $F$ is spatially independent within the cell, so it can be pulled out of the integral to yield
\[
{\int\int}_{\mbox{pole}}\nabla\cdot F\,dA \approx (\nabla\cdot F)_{\mbox{\scriptsize pole}}\,\int\int\,dA = (\nabla\cdot F)_{\mbox{\scriptsize pole}}\,\int_{0}^{2\pi}\int_{0}^{\Delta\theta/2}\sin\theta\,d\theta\,d\phi.
\]
Evaluating the integral yields
\[
\int_{0}^{2\pi}\int_{0}^{\Delta\theta/2}\sin\theta\,d\theta\,d\phi = 2\,\pi\,\left[-\cos\theta\right]_{0}^{\Delta\theta/2} = 2\,\pi\left[1-\cos\left(\frac{\Delta\theta}{2}\right)\right].
\]
Combining this with the line integral above, and using small angle approximations, we get the polar operator
\[
\nabla\cdot(\vec v\,B_r)_{\mbox{\scriptsize pole}}\, 
\approx 
\frac{\sin\left(\frac{\Delta\theta}{2}\right)}{2\,\pi\left[1-\cos\left(\frac{\Delta\theta}{2}\right)\right]}\,\sum_{k=2}^{N_{\phi}-1}F_k\Delta\phi_k 
\approx 
\frac{2}{\pi\,\Delta\theta}\,\sum_{k=2}^{N_{\phi}-1}F_k\Delta\phi_k 
\]
where $F_k$ is evaluated on the half mesh points next to the pole.

\section{A 3rd-order weighted essentially non-oscillatory scheme on a non-uniform grid}
\label{a:weno3}

\begin{figure}[htbp]
\centering
\includegraphics[width=0.25\textwidth]{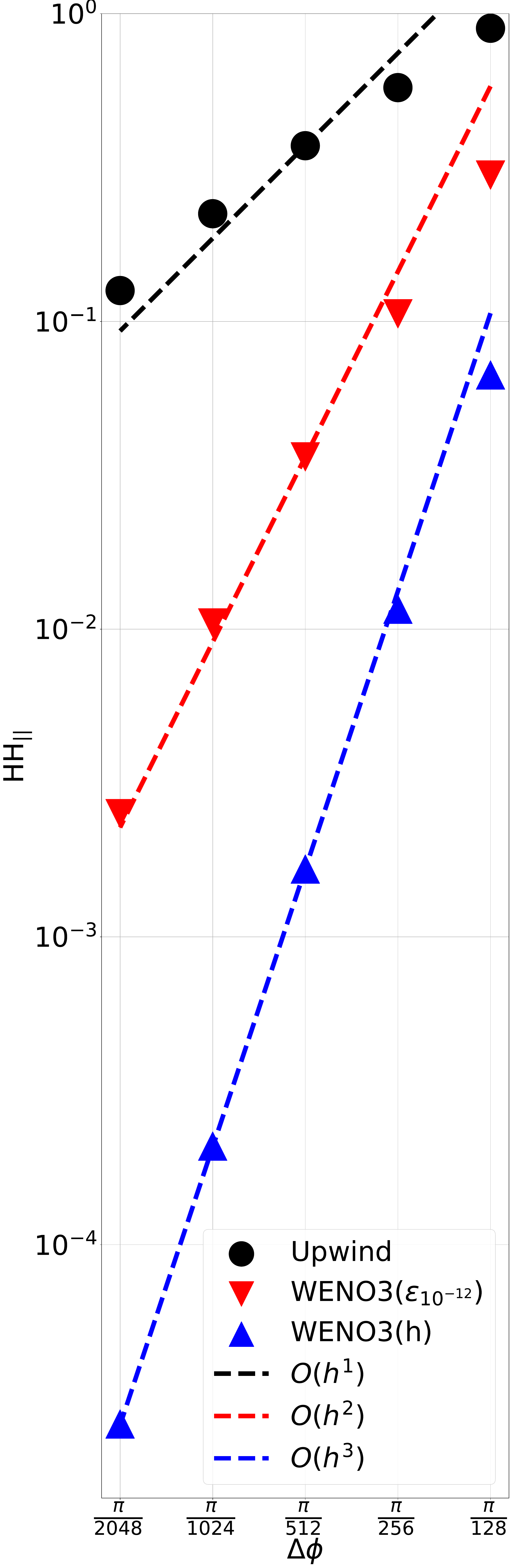}
\caption{Convergence of the $\phi$-rotation test case 1 (Eq.~\ref{eq:testblob}) with a fixed $\theta$-resolution of 512, and with $v_{\theta}=0$ for upwinding, WENO3 with a constant $\epsilon_w=10^{-12}$, and WENO3-CS(h).  We see that the WENO3($\epsilon_w$) exhibits 2nd-order accuracy, while the WENO3-CS(h) exhibits 3rd-order accuracy. \label{fig:wenoeps_vs_wenoh}} 
\end{figure}

The WENO3-CS(h) \citep{Cravero2015} is implemented using a left and right flux as in the upwinding scheme, but with different calculations of the flux $F$. Since $F_{i+1/2}$ is a right shifted version of $F_{i-1/2}$, only the calculation of $F_{i-1/2}$ is required to compute along each dimension over the whole domain. $F_{i-1/2}$ is calculated by first separating the left and right moving fluxes as:
\[
F_{i-1/2} = u_{i-1/2}^{+} + u_{i-1/2}^{-},
\]
where $u_{i-1/2}^{+}$ and $u_{i-1/2}^{-}$ are the the left and right moving numerical fluxes at the cell boundary, respectively. $u_{i-1/2}^{+}$ and $u_{i-1/2}^{-}$ are defined as follows:
\[
u_{i-1/2}^{\pm} = \dfrac{w_0^{\pm}}{w_0^{\pm} + w_1^{\pm}}\,p_0^{\pm} + \dfrac{w_1^{\pm}}{w_0^{\pm} + w_1^{\pm}}\,p_1^{\pm},
\]
where the $w$ fractions are the nonlinear weights, and the $p$ variables are the reconstruction polynomials. The scheme utilizes nonlinear weights based on the smoothness of the function for deciding how to add the reconstruction polynomials together to get $u_{i-1/2}^{+}$ and $u_{i-1/2}^{-}$. The WENO3-CS(h) scheme uses a first-order polynomial approximation, which gives us the following reconstruction polynomials:
\begin{align*}
p_0^{-} &= (1+D_{i-3/2}^{\mbox{c/cm}})\,\mbox{LM}_{i-1} - D_{i-3/2}^{\mbox{c/cm}}\mbox{LM}_{i-2},
&
p_0^{+} &= (1+D_{i    }^{\mbox{c/cp}})\,\mbox{LP}_{i  } - D_{i    }^{\mbox{c/cp}}\mbox{LP}_{i+1},
\\
p_1^{-} &=    D_{i    }^{\mbox{c/cm}}\,\mbox{LM}_{i-1} + D_{i-3/2}^{\mbox{c/cp}}\mbox{LM}_{i  },
&
p_1^{+} &=    D_{i-3/2}^{\mbox{c/cp}}\,\mbox{LP}_{i  } + D_{i    }^{\mbox{c/cm}}\mbox{LP}_{i-1},
\end{align*}
where the $D$ values are calculation constants and ${LP}_i$ and ${LM}_i$ are the left and right moving fluxes respectively. The components of the nonlinear weights are defined as:
\begin{equation}
w_0^{-} = \dfrac{D_{i-3/2}^{\mbox{p/t}}}{(\epsilon_w + \beta_0^{-})^2}, \qquad
w_1^{-} = \dfrac{D_{i-3/2}^{\mbox{cm/t}}}{(\epsilon_w + \beta_1^{-})^2}, \qquad
w_0^{+} = \dfrac{D_{i-1/2}^{\mbox{m/t}}}{(\epsilon_w + \beta_0^{+})^2}, \qquad
w_1^{+} = \dfrac{D_{i-1/2}^{\mbox{cp/t}}}{(\epsilon_w + \beta_1^{+})^2},
\end{equation}
where the $\beta$s are called `smoothness indicators' which describes the smoothness of the region, and $\epsilon_w$ is a small value to avoid division by zero. In other WENO3 schemes, the value of $\epsilon_w$ is often set to an arbitrary value smaller than the typical solution value \citep{shu2009high}.  However, the WENO3-CS(h) scheme used here sets $\epsilon_w=h$ where $h$ is the local cell spacing.  This modification avoids the drop in accuracy near critical points, keeping the scheme 3rd-order.  In Fig.~\ref{fig:wenoeps_vs_wenoh}, we show convergence plots for the Upwind, WENO3 with a constant $\epsilon_w=10^{-12}$, and WENO3-CS(h) schemes running test case 1 (Eq.~\ref{eq:testblob}) with $v_{\theta}=0$.  We see that the WENO3-CS(h) scheme displays third-order accuracy, while using the standard WENO3 scheme with $\epsilon_w=10^{-12}$ only exhibits second-order accuracy.

\begin{figure}[htbp]
\centering
$\begin{array}{c}
\includegraphics[width=0.45\textwidth]{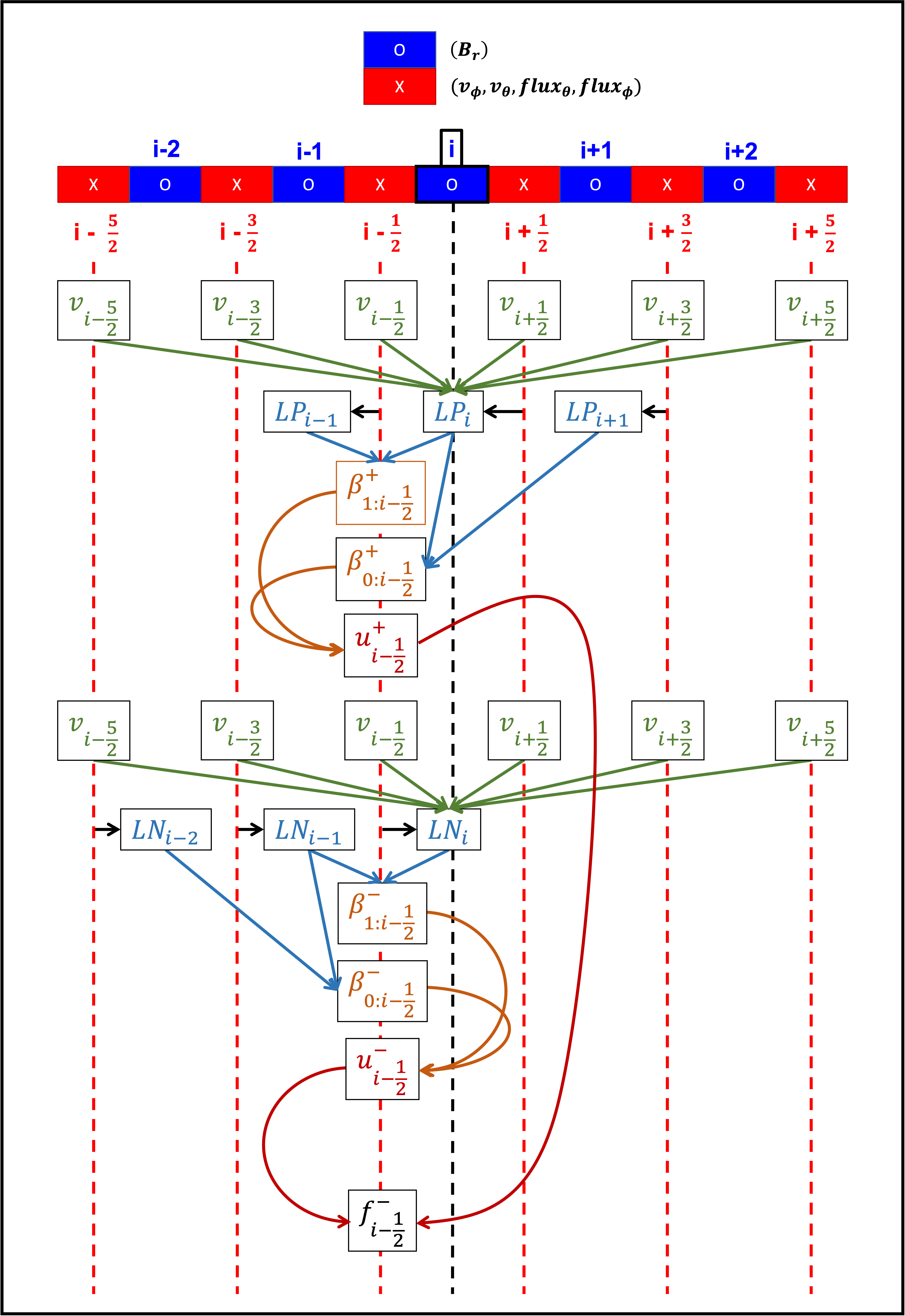}
\\
\includegraphics[width=0.45\textwidth]{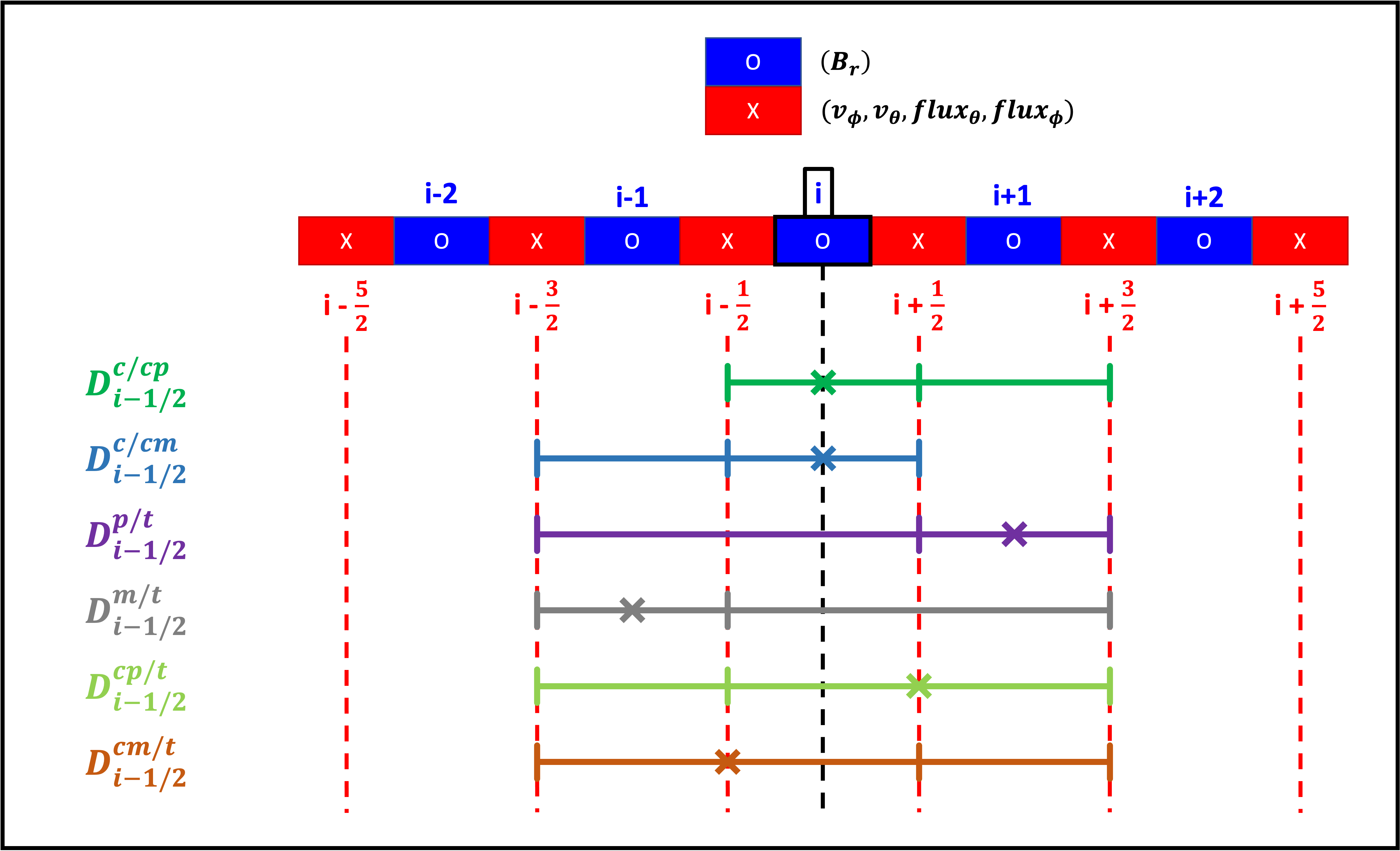}
\end{array}$
\caption{Top: Schematic of the WENO3-CS(h) scheme described in this section, showing the quantities used to calculate the left-side flux $f_{i-1}$ and where they reside on the grid.  Bottom: Schematic showing the values and locations of the non-uniform grid spacing used to compute the $D$ grid factors in the scheme.\label{fig:weno3}}
\end{figure}

The $\beta$s are defined as:
\begin{align*}
\beta_0^{-} &= 4\,(D_{i-3/2}^{\mbox{c/cm}}\,(\mbox{LM}_{i-1} - \mbox{LM}_{i-2}))^2,
&
\beta_0^{+} &= 4\,(D_{i}^{\mbox{c/cp}}\,(\mbox{LP}_{i+1} - \mbox{LP}_{i}))^2,
\\
\beta_1^{-} &= 4\,(D_{i-3/2}^{\mbox{c/cp}}\,(\mbox{LM}_{i} - \mbox{LM}_{i-1}))^2,
&
\beta_1^{+} &= 4\,(D_{i}^{\mbox{c/cm}}\,(\mbox{LP}_{i} - \mbox{LP}_{i-1}))^2,
\end{align*}
where because of the non-uniform grid the following equations are used for the values of $D$ instead of the usual constants \citep{smit2005grid}
\begin{align*}
D_{i-1/2}^{\mbox{c/cp}} &= \dfrac{\Delta x_i}{\Delta x_i + \Delta x_{i+1}},
&
D_{i-1/2}^{\mbox{p/t}} &= \dfrac{\Delta x_{i+1}}{\Delta x_{i-1} + \Delta x_i + \Delta x_{i+1}},
&
D_{i-1/2}^{\mbox{cp/t}} &= \dfrac{\Delta x_i + \Delta x_{i+1}}{\Delta x_{i-1} + \Delta x_i + \Delta x_{i+1}},
\\
D_{i-1/2}^{\mbox{c/cm}} &= \dfrac{\Delta x_i}{\Delta x_i + \Delta x_{i-1}},
&
D_{i-1/2}^{\mbox{m/t}} &= \dfrac{\Delta x_{i-1}}{\Delta x_{i-1} + \Delta x_i + \Delta x_{i+1}},
&
D_{i-1/2}^{\mbox{cm/t}} &= \dfrac{\Delta x_i + \Delta x_{i-1}}{\Delta x_{i-1} + \Delta x_i + \Delta x_{i+1}},
\end{align*}
Fig.~\ref{fig:weno3} shows the $D_i$ constants on the grid where the subsection of the line marked with an $x$ denotes the numerator. To obtain the values of $LP_i$ and $LM_i$, Local Lax-Friedrichs flux splitting (LLF) \citep{shu2009high} is used:
\begin{equation}
\mbox{LP}_i = \dfrac{1}{2}\,B_{r:i}\,(v_{i+1/2} - \alpha_i),
\qquad
\mbox{LM}_i = \dfrac{1}{2}\,B_{r:i}\,(v_{i-1/2} + \alpha_i),
\end{equation}
where $\alpha_i$ is the maximum velocity taken over a relevant range of the domain \citep{li2006weno}:
\[
\alpha_i = \mbox{max}\{|v_{i-5/2}|,|v_{i-3/2}|,|v_{i-1/2}|,|v_{i+1/2}|,|v_{i+3/2}|,|v_{i+5/2}|\}
\]
Fig.~\ref{fig:weno3} shows a flow chart schematic for each WENO3 component that resides on the grid and what previous calculation on the grid each subsequent component relies on. 


\section{Derivation of analytic theta solution}
\label{a:thetasol}

For advection in the $\theta$ direction in spherical coordinates, we have:
\[
\dfrac{\partial f}{\partial t} = -\nabla_s\cdot\left(v_{\theta}\,B_r\right) =  -\dfrac{1}{\sin\theta}\dfrac{\partial}{\partial\,\theta}(\sin\theta\,B_r\,v_{\theta}).
\]
If we choose $v_{\theta}$ to be constant this becomes
\begin{equation}
\label{eq:atheta1}
\dfrac{\partial f}{\partial t} =   -v_{\theta}\,\dfrac{1}{\sin\theta}\dfrac{\partial}{\partial\,\theta}(\sin\theta\,B_r)
\end{equation}
A translational function of the form $B_{r0}(\theta-v_{\theta}\,t)$ won't be a solution due to the terms involving $\sin\theta$.  Instead, we define 
\[
B_r(\theta,t) = \dfrac{1}{\sin\theta}\,B_{r0}(\theta-v_{\theta}\,t).
\]
Calling $g\equiv \theta-v_{\theta}\,t$, we have 
\[
B_r(\theta,t) = \dfrac{1}{\sin\theta}\,B_{r0}(g),
\]
which when inserted into Eq.~\ref{eq:atheta1} yields
\[
\dfrac{\partial Br}{\partial t} = -\dfrac{v_{\theta}}{\sin\theta}\,\dfrac{\partial B_{r0}(g)}{\partial g}.
\]
The spatial term can be written as 
\[
-v_{\theta}\,\dfrac{1}{\sin\theta}\dfrac{\partial}{\partial\,\theta}(B_{r0}(g)),
\]
where we have
\[
\dfrac{\partial B_{r0}(g)}{\partial\,\theta} = \dfrac{\partial B_{r0}(g)}{\partial\,g}.
\]
Inserting this into Eq.~\ref{eq:atheta1}, we see that both sides equate to $-(v_{\theta}/\sin\theta)\,\partial B_{r0}/\partial g$ so any profile $B_{r0}(\theta)$ makes $(1/\sin\theta)\,B_{r0}(\theta-v_{\theta}\,t)$ a solution.

%---------------------------------------------------------------------------------
% BIBLIOGRAPHY
%---------------------------------------------------------------------------------
\bibliography{main}
\bibliographystyle{yahapj}
\end{document}